\documentclass[a4paper,fleqn,usenatbib,useAMS]{mnras}

\usepackage{graphicx}	% Including figure files
\usepackage{amsmath}	% Advanced maths commands
\usepackage{amssymb}	% Extra maths symbols
\usepackage{multicol}        % Multi-column entries in tables
\usepackage{bm}		% Bold maths symbols, including upright Greek
\usepackage{pdflscape}	% Landscape pages

\newcommand{\oii}{$[\textnormal{OII}]$} 
\newcommand{\zspec}{z_{\rm spec}} % to be used in a $$ environment
\newcommand{\nser}{n_{\rm ser}}     % to be used in a $$ environment

\graphicspath{{}}

\usepackage[T1]{fontenc}
\usepackage{ae,aecompl}

\usepackage{newtxtext,newtxmath}
\usepackage{ulem}
\usepackage{multirow}
\usepackage{tablefootnote}

\usepackage[mode=multiuser,status=draft,lang=english]{fixme}
\fxsetup{theme=colorsig}
\FXRegisterAuthor{J}{aJ}{Johan}

\title[eBOSS ELG]{\textit{The SDSS-IV Extended Baryon Oscillation Spectroscopic Survey: final Emission Line Galaxy Target Selection}}

\author[A.~Raichoor et al.]{ 
\parbox{\textwidth}{
A.~Raichoor$^{1,2}$\thanks{e-mail: \texttt{\href{mailto:anand.raichoor@epfl.ch}{anand.raichoor@epfl.ch}}}, 
J.~Comparat$^{3,4,5}$,
T.~Delubac$^{1}$,
J.-P.~Kneib$^{1,6}$,
Ch~Y\`{e}che$^{2,7}$,
K.~S.~Dawson$^{8}$,
W.~J.~Percival$^{9}$,
A.~Dey$^{10}$,
D.~Lang$^{11}$,
D.~J.~Schlegel$^{7}$,
C.~Gorgoni$^{1}$,
J.~Bautista$^{8}$,
J.~R.~Brownstein$^{8}$,
V.~Mariappan$^{8}$,
H.-J.~Seo$^{12}$,
J.~L.~Tinker$^{13}$,
A.~J.~Ross$^{14,9}$,
Y.~Wang$^{9,15}$,
G.-B.~Zhao$^{9,15}$,
 J.~Moustakas$^{16}$,
N.~Palanque-Delabrouille$^{2}$,
 E.~Jullo$^{6}$,
 J.~A.~Newmann$^{17}$,
 F.~Prada$^{4,18,19}$,
 G.~B.~Zhu$^{20,\dagger}$
}\vspace{0.4cm}\\
\parbox{\textwidth}{
$^{1}$ Institute of Physics, Laboratory of Astrophysics, Ecole Polytechnique F\'ed\'erale de Lausanne (EPFL), Observatoire de Sauverny, 1290 Versoix, Switzerland\\
$^{2}$ CEA, Centre de Saclay, IRFU/SPP, F-91191 Gif-sur-Yvette, France\\
$^{3}$ Departamento de Fisica Teorica, Universidad Autonoma de Madrid, Cantoblanco E-28049, Madrid, Spain\\
$^{4}$ Instituto de F\'{\i}sica Te\'orica, (UAM/CSIC), Universidad Aut\'onoma de Madrid,  Cantoblanco, E-28049 Madrid, Spain\\
$^{5}$ Max-Planck-Institut f\"{u}r extraterrestrische Physik (MPE), Giessenbachstrasse 1, D-85748 Garching bei München, Germany\\
$^{6}$ Aix Marseille Universit\'e, CNRS, LAM (Laboratoire d'Astrophysique de Marseille), UMR 7326, 13388, Marseille, France\\
$^{7}$ Lawrence Berkeley National Labortatory, 1 Cyclotron Rd, Berkeley, CA 94720, USA\\
$^{8}$ Department of Physics and Astronomy, University of Utah, Salt Lake City, UT 84112, USA\\
$^{9}$ Institute of Cosmology \& Gravitation, Dennis Sciama Building, University of Portsmouth, Portsmouth, PO1 3FX, UK\\
$^{10}$ National Optical Astronomy Observatory, Tucson, AZ 85719, USA\\
$^{11}$ Bruce and Astrid McWilliams Center for Cosmology, Department of Physics, Carnegie Mellon University, 5000 Forbes Ave, Pittsburgh, PA 15213, USA\\
$^{12}$ Department of Physics and Astronomy, Ohio University, 251B Clippinger Labs, Athens, OH 45701\\
$^{13}$ Center for Cosmology and Particle Physics, Department of Physics, New York University, 4 Washington Place, New York, NY 10003, USA\\
$^{14}$ Center for Cosmology and Astro-Particle Physics, Ohio State University, Columbus, OH 43210\\
$^{15}$ National Astronomy Observatories, Chinese Academy of Science, Beijing, 100012, P. R. China\\
$^{16}$ Department of Physics and Astronomy, Siena College, 515 Loudon Road, Loudonville, NY 12211, USA\\
$^{17}$ Department of Physics and Astronomy and PITT PACC, University of Pittsburgh, Pittsburgh, PA 15260, USA\\
$^{18}$ Campus of International Excellence UAM+CSIC, Cantoblanco, E-28049 Madrid, Spain\\
$^{19}$ Instituto de Astrof\'{i}sica  de Andaluc\'{i}a (CSIC),Glorieta de la Astronom\'{i}a, E-18080 Granada, Spain\\
$^{20}$ Center for Astrophysical Sciences, Department of Physics and Astronomy, Johns Hopkins University, 3400 North Charles Street, Baltimore, MD 21218, USA\\
$^{\dagger}$ Hubble fellow \label{Hubble}\\
}}

\date{Last updated xxx; in original form xxx}

\pubyear{2017}

\begin{document}
\label{firstpage}
\pagerange{\pageref{firstpage}--\pageref{lastpage}}
\maketitle

\begin{abstract}
We describe the algorithm used to select the Emission Line Galaxy (ELG) sample at $z \sim 0.85$ for the extended Baryon Oscillation Spectroscopic Survey of the Sloan Digital Sky Survey IV, using photometric data from the DECam Legacy Survey.
Our selection is based on a selection box in the $g-r$ vs. $r-z$ colour-colour space and a cut on the $g$-band magnitude, to favour galaxies in the desired redshift range with strong \oii~ emission.
It provides a target density of 200 deg$^{-2}$ on the North Galactic Cap (NGC) and of 240 deg$^{-2}$ on the South Galactic Cap (SGC), where we use a larger selection box because of deeper imaging.
We demonstrate that this selection passes the eBOSS requirements in terms of homogeneity.
About 50,000 ELGs have been observed since the observations have started in 2016, September.
These roughly match the expected redshift distribution, though the measured efficiency
%, defined as the percentage of observed targets usable for the cosmological analysis, 
is slightly lower than expected.
The efficiency can be increased by enlarging the redshift range and with incoming pipeline improvement.
The cosmological forecast based on these first data predict $\sigma_{D_V}/D_V = 0.023$, in agreement with  previous forecasts.
Lastly, we present the stellar population properties of the ELG SGC sample.
Once observations are completed, this sample will be suited to provide a cosmological analysis at $z \sim 0.85$, and will pave the way for the next decade of massive spectroscopic cosmological surveys, which heavily rely on ELGs.
The  target catalogue over the SGC will be released along with DR14.
\end{abstract}


%%%%%%%%%%%%%%%%%%%%%%%%%%%%%%%%%%%%%%%%%%%%%%%%%%

%%%%%%%%%%%%%%%%% BODY OF PAPER %%%%%%%%%%%%%%%%%%

% The MNRAS class isn't designed to include a table of contents, but for this document one is useful.
%% I therefore have to do some kludging to make it work without masses of blank space.
%\begingroup
%\let\clearpage\relax
%\tableofcontents
%\endgroup
%\newpage

%=================================
% Introduction
%=================================
\section{Introduction} \label{sec:intro}

% ELGs
It is now well-established that the star-formation density of the Universe is increasing with redshift, from $z \sim 0$ to $z \sim 2$ \citep[e.g.,][]{lilly96,madau98,madau14}.
This implies that a typical galaxy at $z \sim 0.5$-2 will present emission lines in its spectrum \citep[e.g.,][]{moustakas06a}, the most characteristic ones being the \oii~ doublet (emitted at ($\lambda$3727, $\lambda$3729) \AA~ and observed in the optical) and the H$\alpha$ emission line (emitted at $\lambda6563$ \AA~ and observed in the near-infrared).
These emission lines allow one to measure the spectroscopic redshift $\zspec$ of a $z \sim 0.5$-2 galaxy within an optimised amount of observing time, without requiring a detection of the continuum at a signicant level.
Those two observational facts -- a high density and a rapid $\zspec$ measurement -- make such star-forming galaxies the ideal tracer at $0.5 \lesssim \zspec \lesssim 2$ for massive spectroscopic cosmological surveys, which aim to measure a large number (10$^5$-10$^7$) of $\zspec$ within the minimum of observing time.

% surveys with ELGs
The WiggleZ experiment \citep[2006--2011,][]{drinkwater10} was the first survey to use such Emission Line Galaxies (ELGs), with the observation of $\sim$ 200,000 ELGs at $z \sim 0.6$ over 800 deg$^2$.
Future massive spectroscopic cosmological surveys heavily rely on the ELGs in the $0.5 \lesssim z \lesssim 2$ range:
PFS\footnote{Prime Focus Spectrograph: \href{http://sumire.ipmu.jp/en/2652/}{http://sumire.ipmu.jp/en/2652/}} \citep[2019,][]{sugai12, takada14} will target 4 millions ELGs at $0.8<z<2.4$ over 1400 deg$^2$,
DESI\footnote{Dark Energy Spectroscopic Instrument: \href{http://desi.lbl.gov/ cdr/}{http://desi.lbl.gov/ cdr/}} \citep[2019,][]{desi-collaboration16a,desi-collaboration16b} will target 28 millions ELGs at $0.6<z<1.6$ over 14,000 deg$^2$,
4MOST\footnote{4-meter Multi-Object Spectroscopic Telescope: \href{https://www.4most.eu/}{https://www.4most.eu/}} \citep[2021,][]{de-jong14} will target 7 millions ELGs at $0.7<z<1.1$ over 5000 deg$^2$; 
and \textit{Euclid} \citep[2020,][]{laureijs11} will obtain 50 millions redshifts form ELGs at $0.8<z<1.8$ over 15,000 deg$^2$, based on their H$\alpha$ emission line.

\begin{figure}
\begin{center}
\includegraphics[width=0.95\columnwidth]{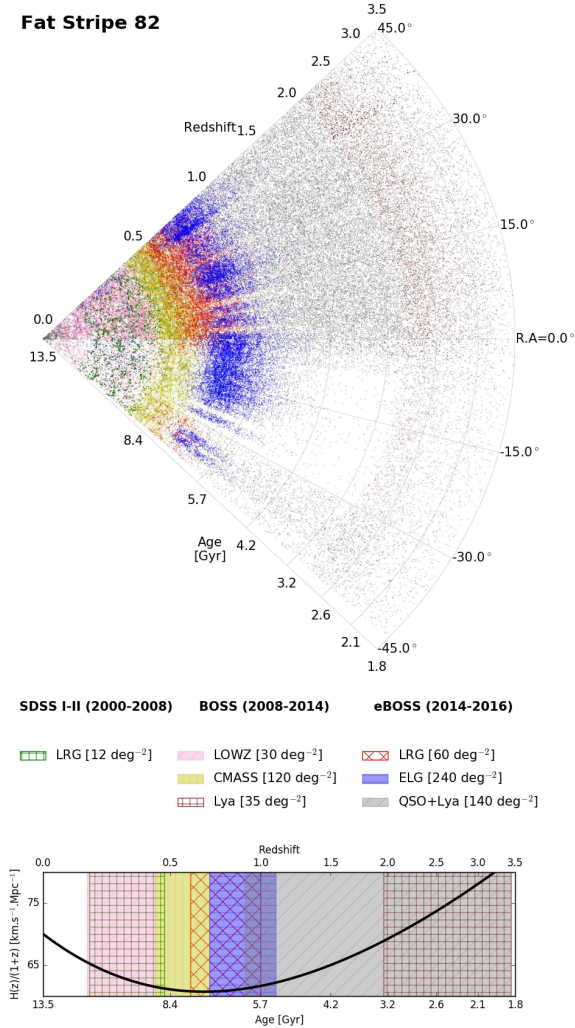}
\caption{
Compilation of the SDSS BAO surveys in the Fat Stripe 82 (see Section \ref{sec:footprint}).
The LRG program from SDSS I-II is in green.
The LOWZ, CMASS, and Lyman-$\alpha$ programs from BOSS are in pink, yellow, and brown, respectively.
The LRG, ELG, QSO+Lyman-$\alpha$ programs from eBOSS are in red, blue, and gray, respectively; for eBOSS, we report the data observed between 2014 and 2016.
For BOSS and eBOSS, the density in the $0^\circ < \textnormal{R.A.} < 45^\circ$ is higher than in $-45^\circ < \textnormal{R.A.} < 0^\circ$, because the included range in declination is higher ($-5^\circ < \textnormal{Dec} < 5^\circ$ vs. $-2^\circ < \textnormal{Dec} < 2^\circ$).
\textit{Top panel}: sector view of observed targets.
\textit{Bottom panel}: 
Universe rate of expansion as a function of redshift, along with the redshift range used for each tracer for the BAO measurement.}
\label{fig:FS82_sector}
\end{center}
\end{figure}

% SDSS and BAO
The Extended Baryon Oscillation Spectroscopic Survey \citep[eBOSS, 2014--2020,][]{dawson16} program of the fourth generation of the Sloan Digital Sky Survey \citep[SDSS,][]{york00} experiment will pave the way for those next generations surveys, as it will also use ELGs as tracers at $z \sim 0.8$.
Since its beginning, the SDSS experiment has been constraining the nature of dark energy through the measurement of the Baryon Acoustic Oscillations (BAOs) signal in the two-point clustering of galaxies at different redshifts.
SDSS I-II \citep{abazajian09} measured at $z=0.35$ the distance-redshift relation with a 5\% precision \citep{eisenstein05} through the observation of 45,000 Luminous Red Galaxies \citep[LRGs,][]{eisenstein01}, making the first BAO detection along with the 2dF Galaxy Redshift Survey \citep{colless03,cole05}.
The Baryon Oscillation Spectroscopic Survey \citep[BOSS, 2008--2014, ][]{dawson13} from the SDSS-III \citep{eisenstein11} observed 1.5 million galaxies and 160,000 quasars, which led to 1\%-2\% precision measurements of the cosmological distance scale for redshifts $z<0.6$ and $z=2.5$ \citep{ross17,delubac15}.

% eBOSS, eBOSS/ELG
eBOSS is one of three surveys from the SDSS-IV experiment \citep{blanton17}: it will use one million objects divided in four different tracers to expand the volume covered by BOSS focussing on the redshift range $0.6<z<2.2$.
The four eBOSS tracers are LRGs at $z\sim0.7$ \citep{prakash16}, ELGs at $z\sim0.8$,
"CORE" quasars at $0.9<z<3.54$ \citep{myers15} supplemented by variability-selected ($2.1<z<3.5$) quasars \citep{palanque-delabrouille16}; the quasars at $z<2.2$ are used as direct probes of large-scale structure, and those at $2.1<z<3.5$ are used to trace Lyman-$\alpha$ absorbers along their line-of-sight.
Regarding the ELGs, 300 plates are dedicated to the observation of 255,000 ELG targets with the BOSS spectrograph \citep{smee13}, in order to produce a 2\% precision distance estimate \citep{zhao16}.
% FS82 sector plot
The Figure \ref{fig:FS82_sector} illustrates how the eBOSS programs combine with the previous SDSS BAO programs, thus probing the universe over more than 70\% of its history.
The top panel shows the observed tracers, color-coded by programs, while the bottom panel displays the Universe rate of expansion as a function of redshift, along with the redshift range used for each tracer for the BAO measurement

% Figure: FS82 sector plot

% ELG TS
We present in this paper the final target selection for the eBOSS/ELG program, for which observations started in 2016, September.
The target selection is a key step in such cosmological experiments, as it will define the data sample used for the cosmological analysis.
It has to fulfill requirements with respect to the subsequent cosmological analysis (number of redshifts, area, homogeneity), but also with respect to the observational constraints (availability of sufficiently wide and deep imaging survey, reliable $\zspec$ measurement with the instrument in the available observing time).
According to previous experience with BOSS and cosmological forecast, the ELG target selection should fulfill the following criteria \citep{dawson16}: 
(1) a surface density $>$170 deg$^{-2}$;
(2) an absolute variation in expected density $<$15\% with respect to imaging depth, Galactic extinction, and stellar density;
(3) an absolute variation in expected density $<$15\% with respect to the estimated uncertainties in the imaging zeropoint;
(4) reliable $\zspec$ measurements, i.e. with a precision better than 300 km.s$^{-1}$;
(5) an ELG sample used for cosmology at $z \sim 0.85$ $>$190,000, i.e. $>$74\% of the observed targets with a reliable $\zspec$ measurement with $0.7<\zspec<1.1$;
(6) $<$1\% of this sample with a catastrophic $\zspec$ measurement (redshift error exceeding 1000 km.s$^{-1}$).

% eBOSS/ELG history...
Though the SDSS has more than a decade experience in target selection for cosmological surveys, the ELG target selection required a significant testing phase before the start of the program, because it is the very first time this tracer is used in the SDSS and because the targeted redshift is challenging for a 1-1.5 hour observation with the BOSS spectrograph.
In a preliminary work, \citet{comparat13} studied the feasibility of an ELG program with the BOSS spectrograph, concluding positively.
They demonstrated that a sufficiently high density of ELGs could be selected from optical photometry and that their $\zspec$ could be efficiently measured with the BOSS spectrograph.
\citet{comparat16a} led a detailed study of the $\zspec$ measurement of ELGs with the BOSS spectrograph, using pilot surveys of different target selections.
They demonstrated the reliability of the $\zspec$ measurement, and developed a posteriori flags which ensure the rate of catastrophic measurement is $<$1\%.
Then \citet{raichoor16} presented a possible target selection based on the SDSS imaging, with a target density of 180 deg$^{-2}$: thanks to public data and dedicated test plates observed with the BOSS spectrograph, they showed that $\sim$70\% of this selection has $0.6<\zspec<1.0$, with an expected catastrophic rate of $\sim$1\%.
\citet{delubac17} demonstrated that this selection passed the density homogeneity requirements.

% and we go for DECaLS!
However, the recent advent of the DECam Legacy Survey\footnote{\href{http://legacysurvey.org/}{http://legacysurvey.org/}} (Dey et al., in preparation) provided an opportunity to design a further ELG target selection, which presented the noteworthy advantage of being at higher redshift than the one presented in \citet{raichoor16} and \citet{delubac17}.
This one, based on the DECaLS imaging, was chosen by the eBOSS team, and is the one we present hereafter.
We present in Section \ref{sec:photometry} the DECam observations and DECaLS photometric catalogs, which are used to define the target selection.
Section \ref{sec:TSalgorithm} details the target selection algorithm.
We demonstrate in Section \ref{sec:systematics} that the selection passes the homogeneity requirements.
Section \ref{sec:firstobs} presents the first months of observations, which have started in 2016, September, the redshift distribution and the cosmological forecast based on these observed plates.
Those first observations are then used to present in Section \ref{sec:meanproperties} the mean photometric, spectroscopic, and structural properties of the ELG sample.
We conclude in Section \ref{sec:conclusion}.

We consider a standard cosmology with $H_0=70$ km.s$^{-1}$.Mpc$^{-1}$, $\Omega_{\rm m}=0.30$, and $\Omega_\Lambda=0.70$, except in Section \ref{sec:cosmo_forecast}, where the \textit{Planck} cosmology \citep{planck-collaboration16} is assumed for the cosmological forecast.
All magnitudes are in the AB system \citep{oke83} and have been corrected for Galactic extinction using the maps of \citet{schlegel98}.

%=================================
% Photometry: DES and DECaLS
%=================================
\section{Imaging and photometry} \label{sec:photometry}

% TS based on DECaLS
The eBOSS/ELG target selection is based on the DECaLS data, processed independently by us, but using the same pipeline and software.
Thorough tests have been conducted to design the eBOSS/ELG target selection.
\citet{raichoor16} and \citet{delubac17} have presented a target selection based on the SDSS imaging, which passed the eBOSS requirements.
Though the SDSS imaging has the great advantage of being highly homogeneous and understood, its shallowness limits the target selection performance, degrading the median redshift and the efficiency.
The deeper imaging of the DECaLS survey allows one to design a target selection with higher redshift and higher efficiency, even with current observations not reaching yet the nominal depth.

% DECaLS
The DECam Legacy Survey (DECaLS; co-P.I.s: A.Dey \& D.J.Schlegel) is an on-going imaging survey covering 6700 deg$^2$ of the extragalactic sky that lies in the region $-20^\circ < \textnormal{Dec.} < +30^\circ$ to depths of $g=24.7$, $r=23.9$, and $z=23.0$ mag (5$\sigma$ point-source).
In addition, the DECaLS program also processes any public DECam observations in the DESI footprint, the largest component being the DES survey.
The DECaLS data and processing pipeline are public, with a release every six months.
DECaLS will be used as the main imaging for the DESI target selection, which will observe $\sim$35 millions targets, of which $\sim$28 millions will be ELGs \citep{desi-collaboration16a}.

% footprint
\subsection{ELG footprint over the SGC and the NGC} \label{sec:footprint}

\begin{figure}
	\includegraphics[width=0.95\columnwidth]{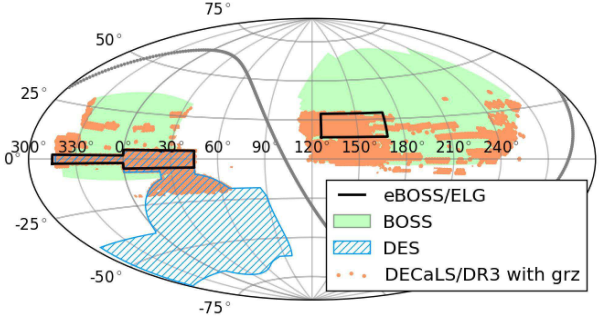}
	\caption{Survey footprints: eBOSS/ELG footprint (black lines; SGC is on the left, NGC is on the right), SDSS/BOSS footprint (light green shaded regions),  DES footprint (blue hatched), and DECaLS/DR3 regions with $grz$-imaging (orange dots).}
	\label{fig:footprint}
\end{figure}

The eBOSS/ELG footprint is divided in two regions (see Figure \ref{fig:footprint}):
\begin{itemize}
\item $\sim$620 deg$^2$ over the Fat Stripe 82 in the SGC, covered by DES observations; the boundaries are: 
($317^\circ<\textnormal{R.A.}<360^\circ$ and $-2^\circ<\textnormal{Dec.}<2^\circ$)
or
($0^\circ<\textnormal{R.A.}<45^\circ$ and $-5^\circ<\textnormal{Dec.}<5^\circ$);
\item $\sim$600 deg$^2$ over the NGC, covered by DECaLS observations; the boundaries are:
$126^\circ<\textnormal{R.A.}<169^\circ$ and $14^\circ<\textnormal{Dec.}<29^\circ$.
\end{itemize}

The exact footprint over the NGC may be updated when the plates position and targets fiber assignment are set (tiling), as the DECaLS observations in this region are not yet complete (see Section \ref{sec:imaging} below).

% Figure: footprints

% Imaging observations
\subsection{Imaging observations} \label{sec:imaging}

% imaging origin
The eBOSS/ELG target selection is based on the imaging included in the Data Release 3 (DR3), the third public data release of images and catalogues for the DECaLS.
DECaLS/DR3 includes DECaLS $grz$-band observations (co-P.I.s: A.Dey \& D.J.Schlegel; NOAO Proposal \# 2014B-0404) from August 2014 through March 2016, and DECam data from a range of non-DECaLS surveys, including observations that were conducted from September 2012 to March 2016.
For the currently tiled part of the eBOSS/ELG NGC footprint, we also included 28 DECaLS exposures from 2016, April 08-10.
Figure \ref{fig:footprint} displays the 4200 deg$^2$ of DECaLS/DR3 observed in all three $grz$-bands.

% DECam
All observations included in DECaLS/DR3  are made with the DECam camera \citep{flaugher15} mounted at the prime focus of the Victor M. Blanco 4m telescope on Cerro Tololo near La Serena, Chile.
The DECam camera has a 3 deg$^2$ field-of-view covered by 62 2k$\times$4k CCDs for imaging, with a resolution of 0.263 arcsec/pixel.

% DECaLS obs. / NGC
DECaLS observations follow a 3-pass tiling strategy, with nominal exposure times of $[t_g,t_r,t_z]=[70,50, 100]$ sec; those nominal exposure times are dynamically adjusted during the observations in order to optimise the observing efficiency (see Burleigh et al. 2017 for details).
DECaLS observations over the eBOSS/ELG NGC footprint are still on-going: the part of eBOSS/ELG NGC footprint not covered with $grz$-bands in DR3 ($\sim$170 deg$^2$, see Figure \ref{fig:footprint}) will be observed in the first semester of 2017; we will include all DECaLS/post-DR3 observations available over the footprint before the final tiling.
Note that those data will be processed with the very same DR3 pipeline, which has been used already for the data presented in this paper.

% DES obs. / SGC
The Dark Energy Survey\footnote{\href{http://www.darkenergysurvey.org}{http://www.darkenergysurvey.org}} (DES, 2013-2018; P.I.: J. Frieman) is an on-going $grizY$-imaging survey over 5000 deg$^2$, down to $i \sim 4.1$ mag (10$\sigma$, extended source).
At the end of the 5 years of observations, DES will have observed each region with ten individual exposures, with typical exposure times of 90 s for the $grz$-bands.

\begin{figure}
	\includegraphics[width=0.95\columnwidth]{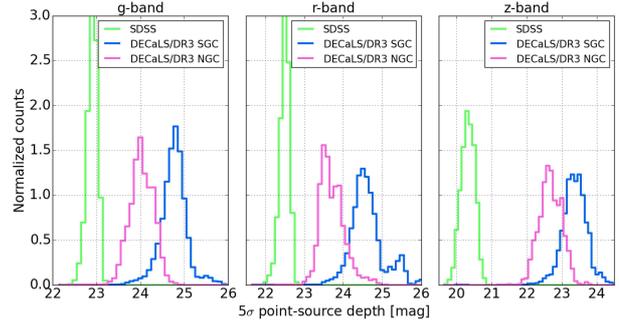}
	\caption{DECam imaging depths over our ELG footprint (SGC in cyan, NGC in pink). We report the SDSS (green) to illustrate the improvement brought by DECaLS.}
	\label{fig:depths}
\end{figure}

% table+figure comments
Imaging properties over the eBOSS/ELG footprint are summarised in Table \ref{tab:DECamImg}, and the imaging depths over this footprint are displayed in Figure \ref{fig:depths}.
Over the NGC, the DECam imaging is about one magnitude deeper than the SDSS imaging in the $g$- and $r$-band, and about two magnitudes deeper in the $z$-band.
Over the SGC, the DECam imaging is about two magnitudes deeper than the SDSS imaging in the $g$- and $r$-band, and about three magnitudes deeper in the $z$-band; when compared to the DECam NGC imaging, the DECam SGC imaging is about 0.8 magnitude deeper.

% Figure: imaging depth

% Table: DECam imaging properties
\begin{table}
	\centering
	\begin{tabular}{lccccccc}
		\hline
		\hline
		Region & Area          &Observing & filter & nexp & texp & seeing & depth\\
		        	    & [deg$^2$]  & program    &        &         & [s]    & [arcsec] & [mag]\\
		\hline
		\multirow{3}{*}{SGC} & \multirow{3}{*}{$\sim$620} & \multirow{3}{*}{DES} & $g$ & 4 & 360 & 1.6 & 24.7\\
						&				    &				      & $r$  & 3 & 270 & 1.2 & 24.5\\
						&				    &				      & $z$ & 3 & 270 & 1.1 & 23.3\\
		\hline
		\multirow{3}{*}{NGC} & \multirow{3}{*}{$\sim$600} & \multirow{3}{*}{DECaLS} & $g$ & 2 & 140 & 1.4 & 24.0\\
						&					  &			 	   & $r$  & 2 & 100 & 1.3 & 23.6\\
						&					  &				   & $z$ & 1 & 120 & 1.2 & 22.6\\
		\hline
	\end{tabular}
	\caption{Imaging properties over the eBOSS/ELG footprint. We report the median values. The depth is 5$\sigma$ point-source depth.
The NGC numbers are for the DECaLS/DR3 imaging covering our footprint in the $grz$-band, described in Section \ref{sec:imaging}.}
	\label{tab:DECamImg}
\end{table}

% Data processing and photometry
\subsection{Data processing and photometry}

We processed the images with the DECaLS/DR3 pipeline\footnote{for a more detailed description, please see: \href{http://legacysurvey.org/dr3/description/}{http://legacysurvey.org/dr3/description/}}, which is based on \textit{the Tractor}\footnote{\href{https://github.com/dstndstn/tractor}{https://github.com/dstndstn/tractor}} (Lang et al. in prep.).
Apart from source detection that uses on stacked images, all measurements are based on individual exposures.
Astrometric calibration is tied to Pan-Starrs1 measurements \citep{kaiser10}, as is the photometric calibration, with the use of a color correction term to transform the PS1 magnitudes into a DECam-based system.
Each source is modeled with a simple analytic profile (point-source, exponential, de Vaucouleurs, or composite) and a model image is generated for each exposure.
Increasingly more complex profiles are allowed for sources detected with higher signal-to-noise ratio.
The source properties (position, shape, flux) are measured through a likelihood optimisation ($\chi^2$ minimisation) of the set of model images covering the considered region.
For our purpose, this approach has the advantage to provide accurate colours, based on the same profile for all bands and accounting for the Point-Spread Function (PSF).

We processed the data independently of the DECaLS team for the two following reasons.
First, because the target catalogues were required for the tiling before the public release of DECaLS/DR3.
Second, as the DECaLS imaging over our ELG NGC footprint is not yet finished, we will later on process any new DECaLS imaging in our ELG NGC footprint, thus using the very same pipeline version.
A posteriori comparison with the publicly released DECaLS/DR3 catalogues has shown that source detections and flux measurements in regions processing the same imaging dataset were virtually similar, resulting in identical target catalogues.

%=================================
% Algorithm
%=================================
\section{Target selection algorithm} \label{sec:TSalgorithm}

The eBOSS/ELG target selection is based on three criteria: 
i) clean photometry, using catalogue flags,
%minimum depth requirement, 
and masking bright stars/objects neighbourhoods;
ii) favouring \oii~ emitters, through a cut in the $g$-band magnitude \citep[see][]{comparat15};
iii) selecting galaxies in the desired redshift range, through a box cut in the $g-r$ vs. $r-z$ colour-colour diagram ($grz$-diagram, hereafter).
The corresponding cuts, detailed in Table \ref{tab:TScriteria}, have been fine-tuned with thorough tests, using various photometric and spectroscopic public catalogues and dedicated observations with the BOSS spectrographs.
 % TS bug...
A coding error in the target selection scripts
\footnote{At the step of rejecting areas with shallow depths in the NGC, we read the depth brick images ($0.25^\circ\times0.25^\circ$) to obtain the depth at a target position: we took the (x,y) position instead of the (y,x) position.
This results for $\sim$15\% of the targets in reading the depth at a position without imaging coverage because of the CCD gaps, hence a rejection.
The main consequence is that objects which should have been observed are not observed.
The $\sim$0.5\% of observed objects, which should have not been observed, can be \textit{a posteriori} rejected when making the cosmological analysis, using minimum \texttt{depth\_ivar} values of 62.79, 30.05, and 11.00 for the $g$-, $r$-, and $z$-band respectively.
This error is fully reproducible, hence can be accounted for in the clustering analysis, especially at the step of random generation.}
translates to an additional rejection mask for the $\sim$400 deg$^2$ already tiled in the NGC (chunk \texttt{eboss23} chunk, see Section \ref{sec:obs_design}).
The effect of this mask is to reject $\sim$15\% of the targets in the NGC, which explains why we select objects slightly fainter in the $g$-band in the NGC to reach our target density of 200 deg$^{-2}$.

% Table: TS criteria
\begin{table*}
	\centering
	\begin{tabular}{lcc}
		\hline
		\hline
		Criterion & eBOSS/ELG SGC & eBOSS/ELG NGC\\
			      & [240 deg$^{-2}$] & [200 deg$^{-2}$]\\
		\hline
		\multirow{3}{*}{Clean photometry} & \multicolumn{2}{c}{SDSS bright object mask\tablefootnote{\href{http://data.sdss3.org/sas/dr10/boss/lss/reject\_mask/}{http://data.sdss3.org/sas/dr10/boss/lss/reject\_mask/}} and $0 \; \textnormal{mag} < V < 11.5 \; \textnormal{mag}$ Tycho2 stars mask}\\
		& \multicolumn{2}{c}{\texttt{BRICK\_PRIMARY} and \texttt{decam\_anymask$[grz]$=0} and \texttt{tycho2inblob==False}}\\
		&&Custom mask$^\dagger$ [chunk \texttt{eboss23} only]\\
		%& $\texttt{decam\_depth}[g,r,z] > 62.79,30.05, 12.75$ &  $\texttt{decam\_depth}[g,r,z] > 62.79,30.05, 11.00$ \\
		&&\\
		 \oii~ emitters & $21.825 < g < 22.825$ & $21.825 < g < 22.9$\\
		&&\\
		\multirow{2}{*}{Redshift range}	& $-0.068 \times (r-z)+0.457 < g-r <  0.112 \times (r-z)+0.773$ & $-0.068 \times (r-z) +0.457 < g-r < 0.112 \times (r-z)+0.773$\\
				 	& $0.218  \times (g-r)+0.571 < r-z < -0.555 \times (g-r)+1.901$ & $0.637 \times (g-r)+0.399 < r-z < -0.555 \times (g-r)+1.901$\\
		\hline
	\end{tabular}
	\caption{
eBOSS/ELG target selection over the SGC and the NGC footprints.
The reported densities are computed with the plate covered areas (i.e. not accounting for the masked regions).
The DECaLS catalogue quantities \texttt{BRICK\_PRIMARY}, \texttt{decam\_anymask}, and \texttt{tycho2inblob} are described here: \href{http://legacysurvey.org/dr3/catalogs/}{http://legacysurvey.org/dr3/catalogs/}.
$^\dagger$: this custom mask is used because of a bug in the target selection scripts, and is relevant for the chunk \texttt{eboss23} only.
}
	\label{tab:TScriteria}
\end{table*}

In order to profit from the DES imaging depth over the eBOSS/ELG SGC footprint, our selection has a higher density in the SGC than in the NGC.
The deeper, hence less scattered, DES photometry allows us to enlarge our selection box in the $grz$-space towards blue $r-z$ colour, with keeping a low contamination level from lower-redshift galaxies.

\begin{figure*}
	\begin{tabular}{lr}
		\includegraphics[width=0.95\columnwidth]{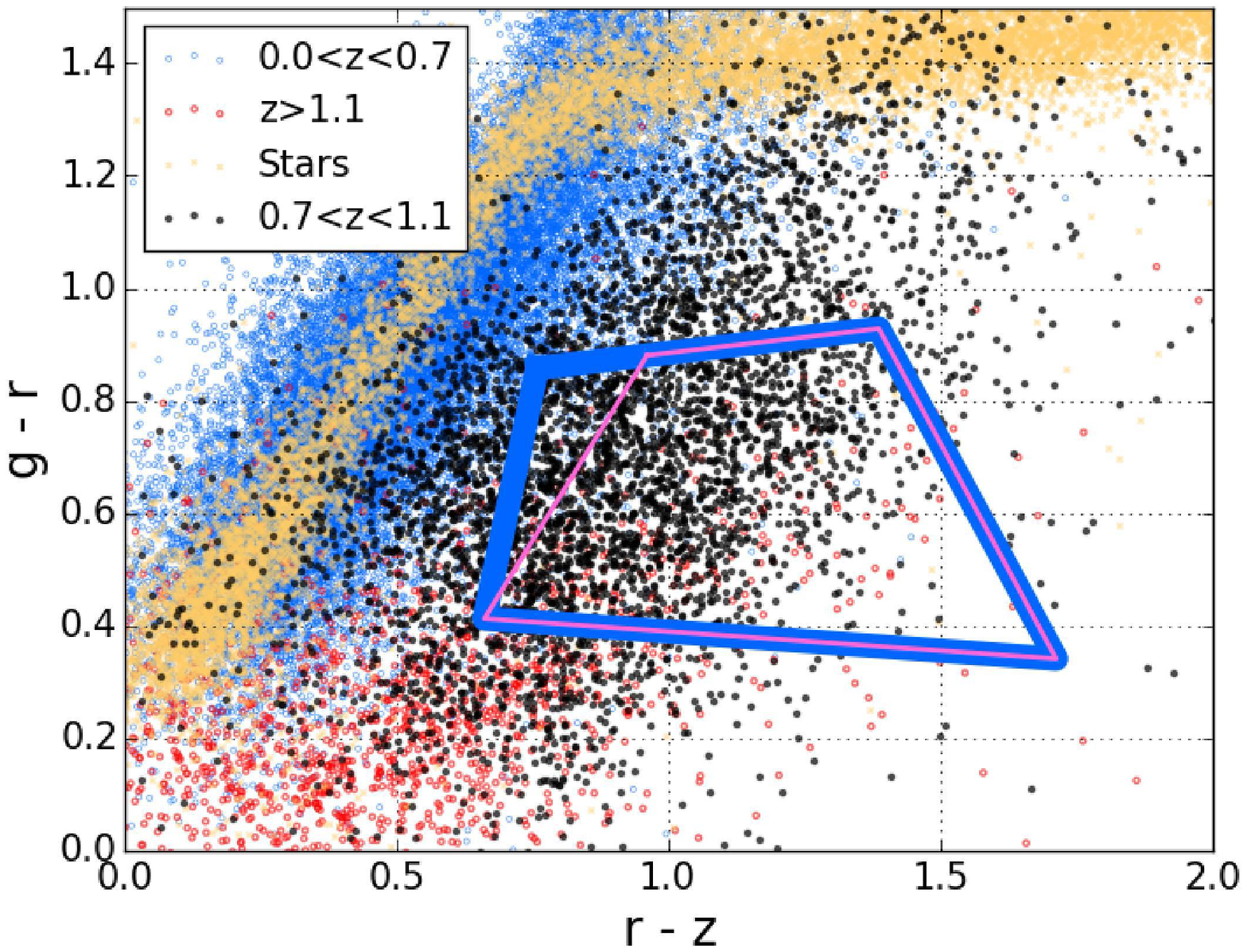} &
		\includegraphics[width=0.95\columnwidth]{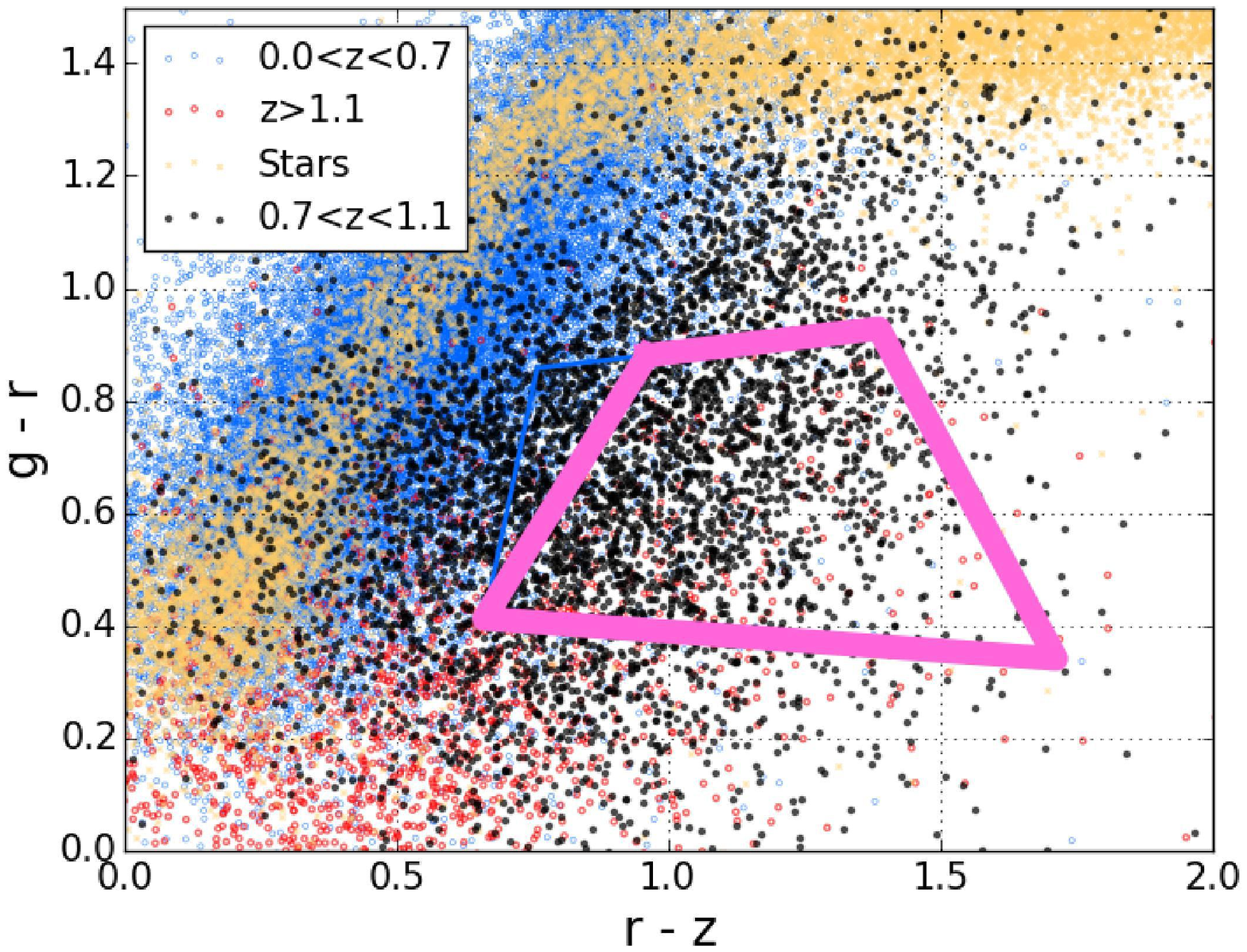}\\
		\includegraphics[width=0.95\columnwidth]{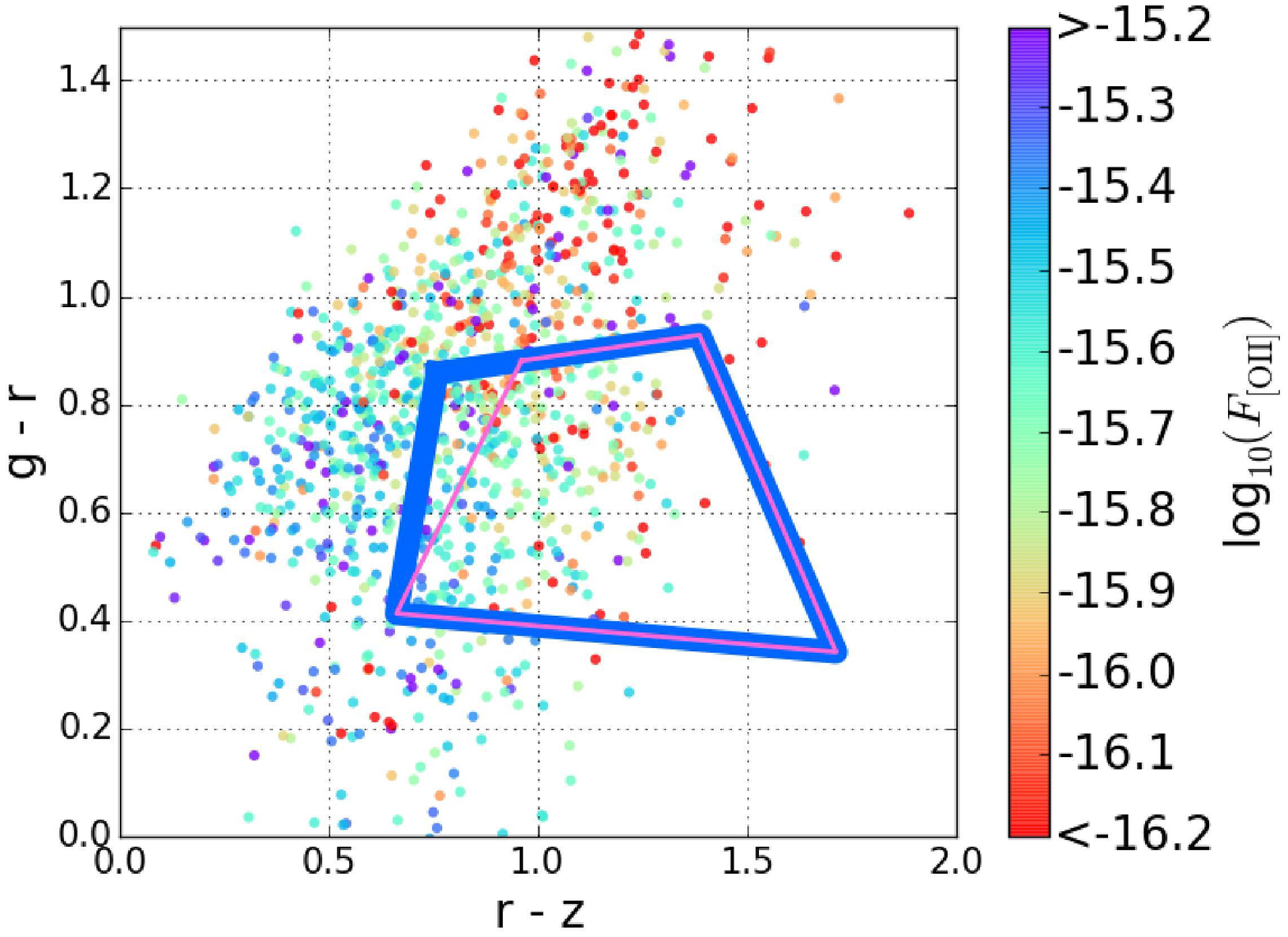} &
		\includegraphics[width=0.95\columnwidth]{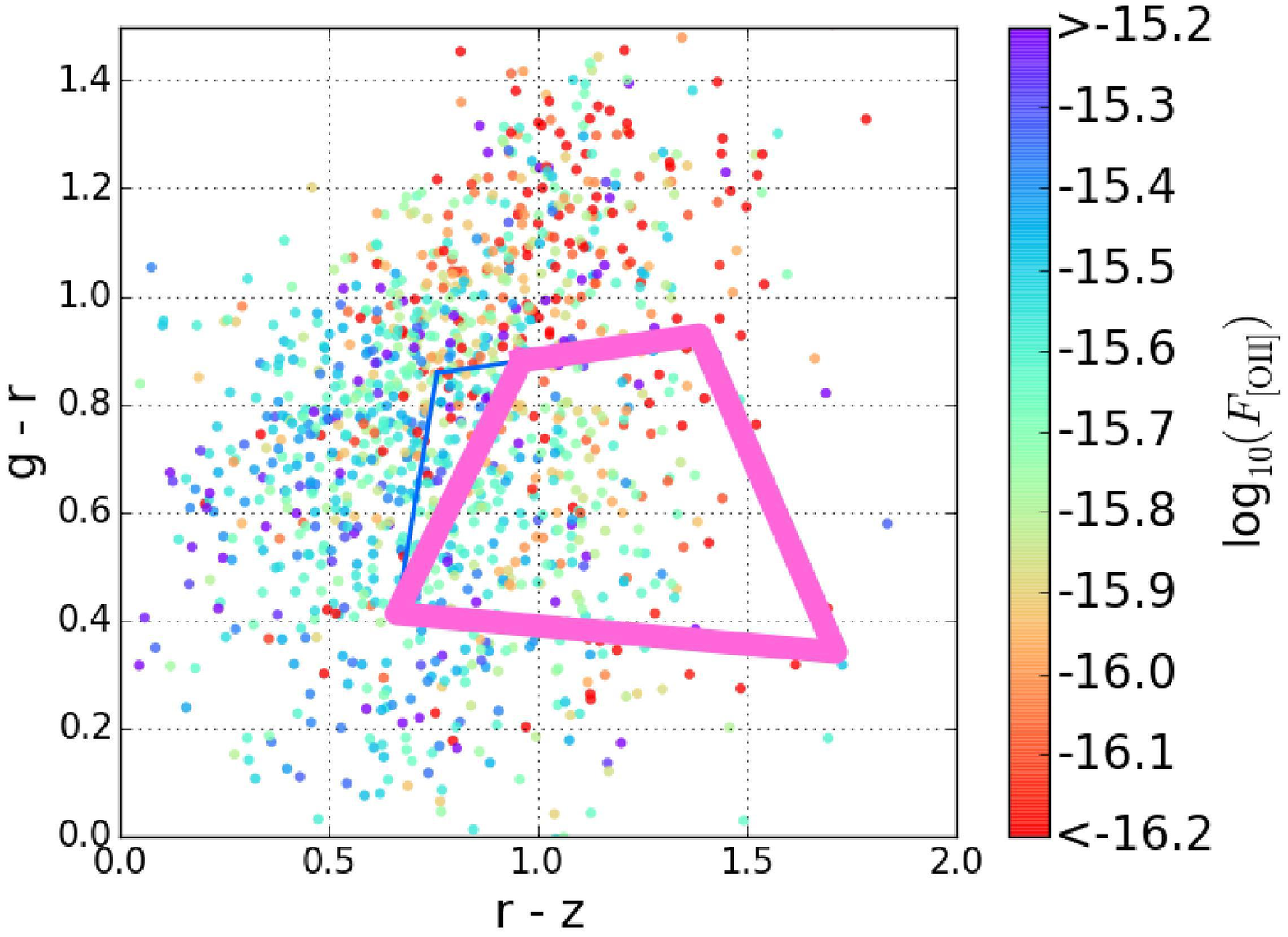}\\
	\end{tabular}
	\caption{eBOSS/ELG selection in the $grz$-colour space for the CFHTLS/W4 field.
Our $grz$-selection over the eBOSS/ELG SGC area (NGC area, respectively) is displayed as the cyan (pink, respectively) box. 
We display objects after applying the "clean photometry" and "\oii~ emitters" cuts (see Table \ref{tab:TScriteria}).
\textit{Left:} photometry from DECaLS/DR3 over the CFHTLS/W4 (DES observations), degraded down to DECaLS/DR3 depths over the eBOSS/ELG SGC footprint.
\textit{Right}: photometry from DECaLS/DR3 over the CFHTLS/W4 (DES observations), degraded down to DECaLS/DR3 depths over the eBOSS/ELG NGC footprint.
\textit{Top panels}: Photometric redshifts are taken from the CFHTLS survey: stars are displayed as beige crosses, $0<z_{\rm phot}<0.7$ objects as blue circles, $0.7<z_{\rm phot}<1.1$ objects as black circles, and $1.1<z_{\rm phot}$ objects as red circles.
\textit{Bottom Panels}: we only display objects in common with the VIPERS survey with $0.6<\zspec<1.1$; \oii~fluxes are measured from the VIPERS survey spectra.}
	\label{fig:grz}
\end{figure*}

Figure \ref{fig:grz} illustrates the cuts in the $grz$-space.
This figure is for illustration purpose only: it allows one to visually identify the expected redshift and \oii~ loci over the eBOSS/ELG SGC and NGC areas, showing the impact of the imaging depth on the $grz$-band.
We consider the CFHTLS/W4 region \citep[$\sim$20 deg$^2$ at R.A=333$^\circ$ and Dec.=2$^\circ$,][]{gwyn12}, which is covered by DES deep observations.
The redshift information is taken from the CFHTLS photometric redshifts \citep{ilbert06,coupon09} and the \oii~ flux is measured from the VIPERS survey spectra \citep{guzzo14,scodeggio16}.
The displayed photometry is the original DECaLS/DR3 photometry, degraded to the depths over the eBOSS/ELG SGC and NGC footprints.
To do so, for each of the $grz$-band, we add in quadrature random noise to the original photometry in order to reproduce the magnitude error-magnitude relation \citep[see Sect.3.1 of][]{raichoor16}.
Also, as our selection is based on a cut on the $g$-band magnitude, we randomly remove objects to reproduce the incompleteness in source detection for faint sources in the typical NGC/SGC data.
The eBOSS/ELG NGC area has an imaging depth $\sim$0.8 magnitude shallower than the eBOSS/ELG SGC area: this implies a larger scatter on the photometry and, as a consequence, less well-defined loci in terms of redshift.
Restricting to a smaller colour-colour box in the $grz$-diagram for the eBOSS/ELG selection over the NGC prevents our target selection to be contaminated by $z<0.7$ objects: as those objects have a higher density than the $0.7<z<1.1$ objects, the scattering due to shallow photometry would imply more $z<0.7$ objects entering our selection than $0.7<z<1.1$ objects exiting our selection.

% Figure: grz

%=================================
% Systematics
%=================================
\section{Systematics} \label{sec:systematics}

To ensure that our clustering analysis and cosmological measurements will be limited by statistical uncertainties, the target density should pass a list of homogeneity requirements, mainly derived from knowledge of the BOSS survey.
As stated in the Section \ref{sec:intro}, the target density should have a $<$15\% variation with respect to imaging PSF and depth, Galactic extinction, and stellar density; and it also should have a $<$15\% variation with respect to the estimated uncertainties in the imaging zeropoint.

In this Section we present a thorough analysis on the density variation over the footprint, similar to that presented in \citet{ross11}, \citet{myers15}, \citet{prakash16}, and \citet{delubac17} for other samples, to which we refer the reader for the details.

% subsection: Dependency over the observational parameters
\subsection{Dependency over the observational parameters} \label{sec:syst_obs}
\begin{figure*}
\begin{center}
\includegraphics[width=1.05\columnwidth]{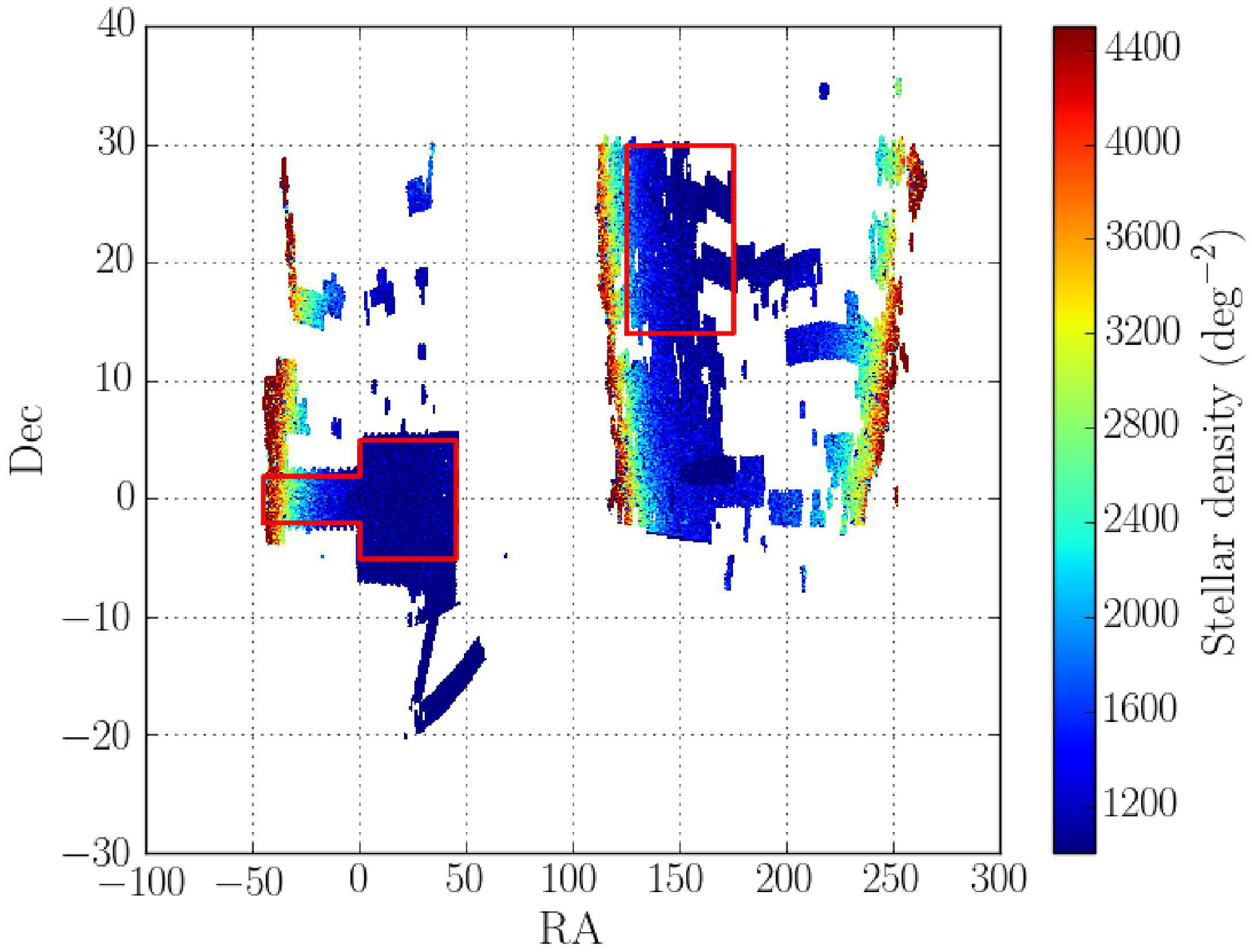}
\includegraphics[width=0.75\columnwidth]{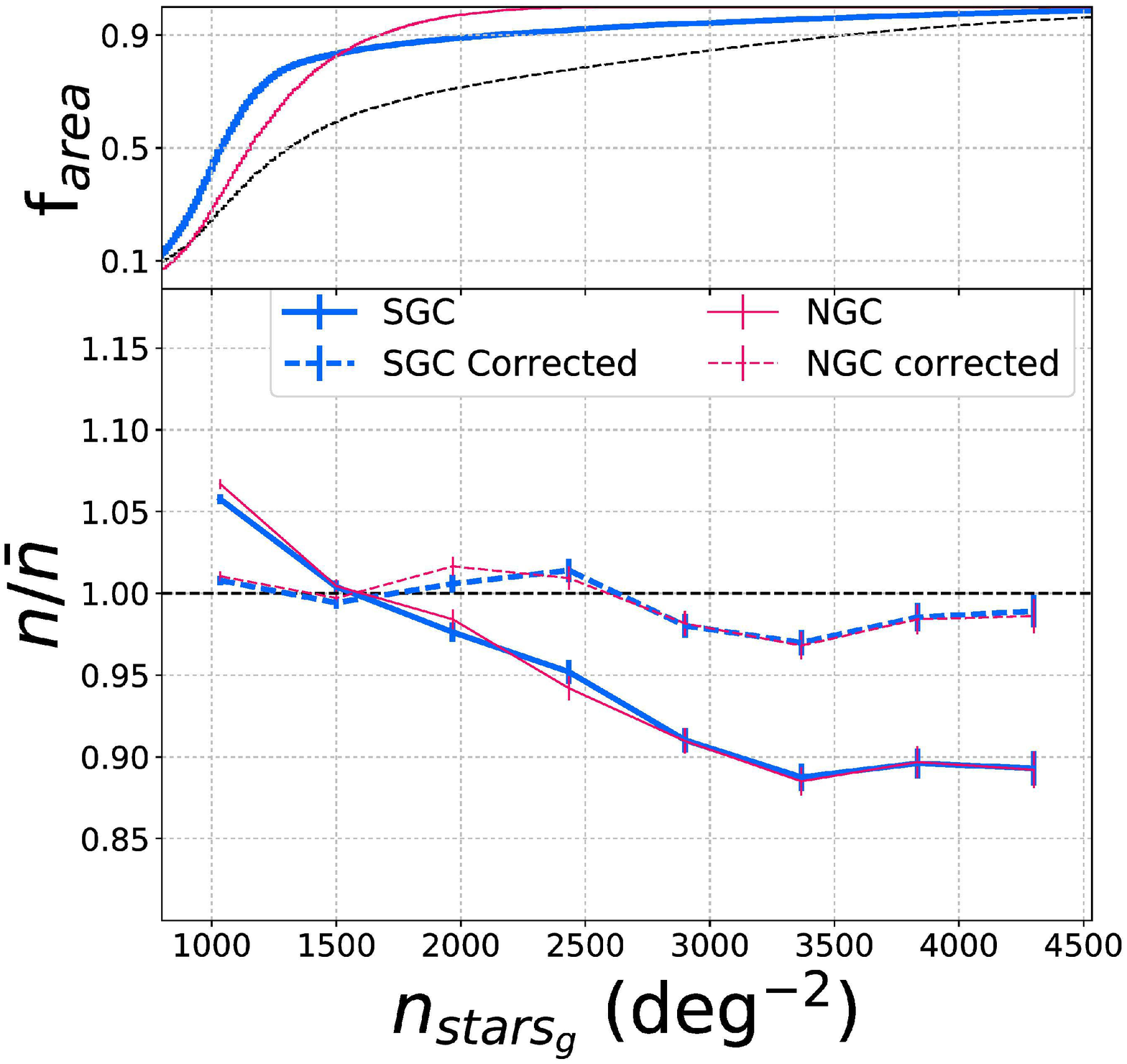}
\includegraphics[width=1.05\columnwidth]{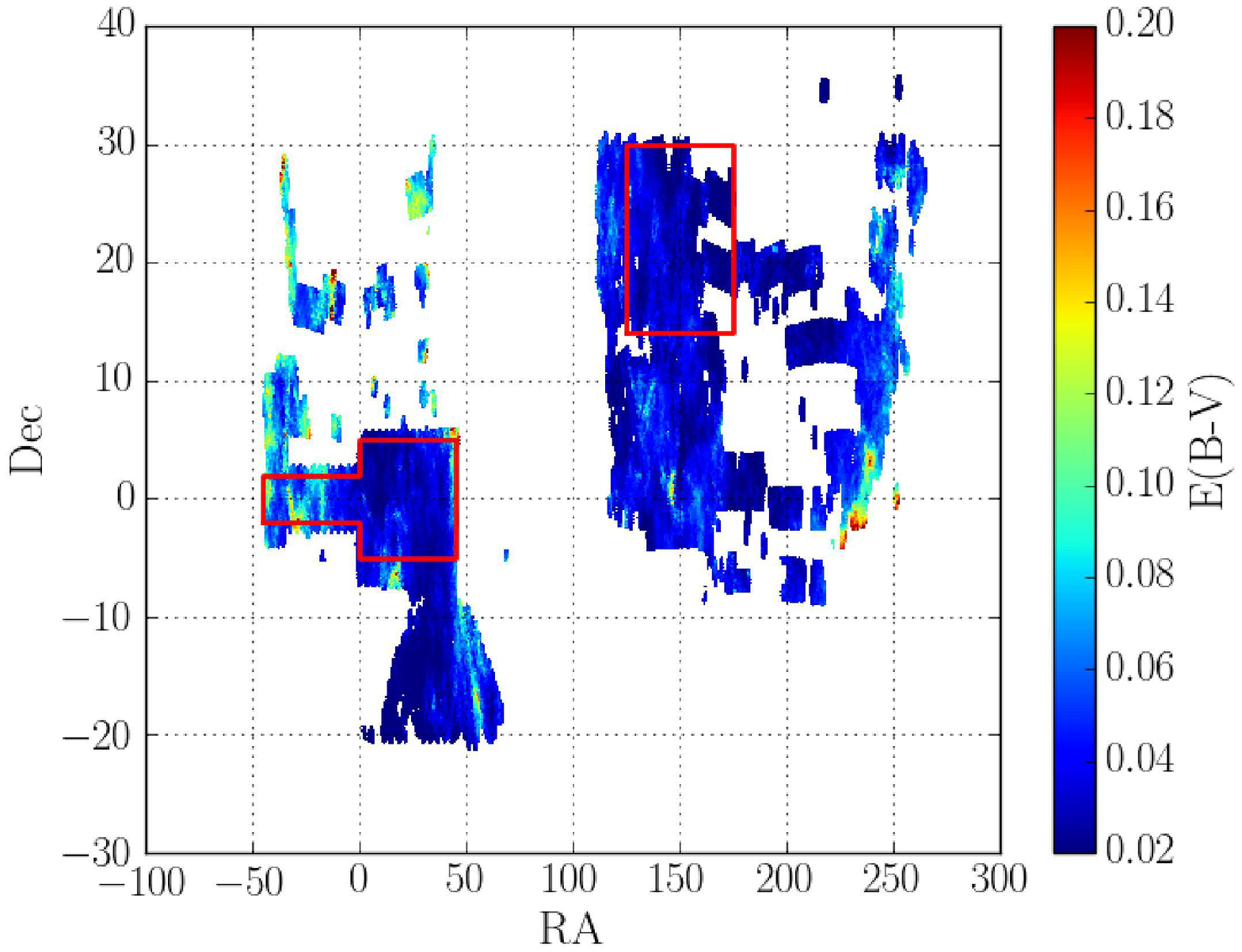}
\includegraphics[width=0.75\columnwidth]{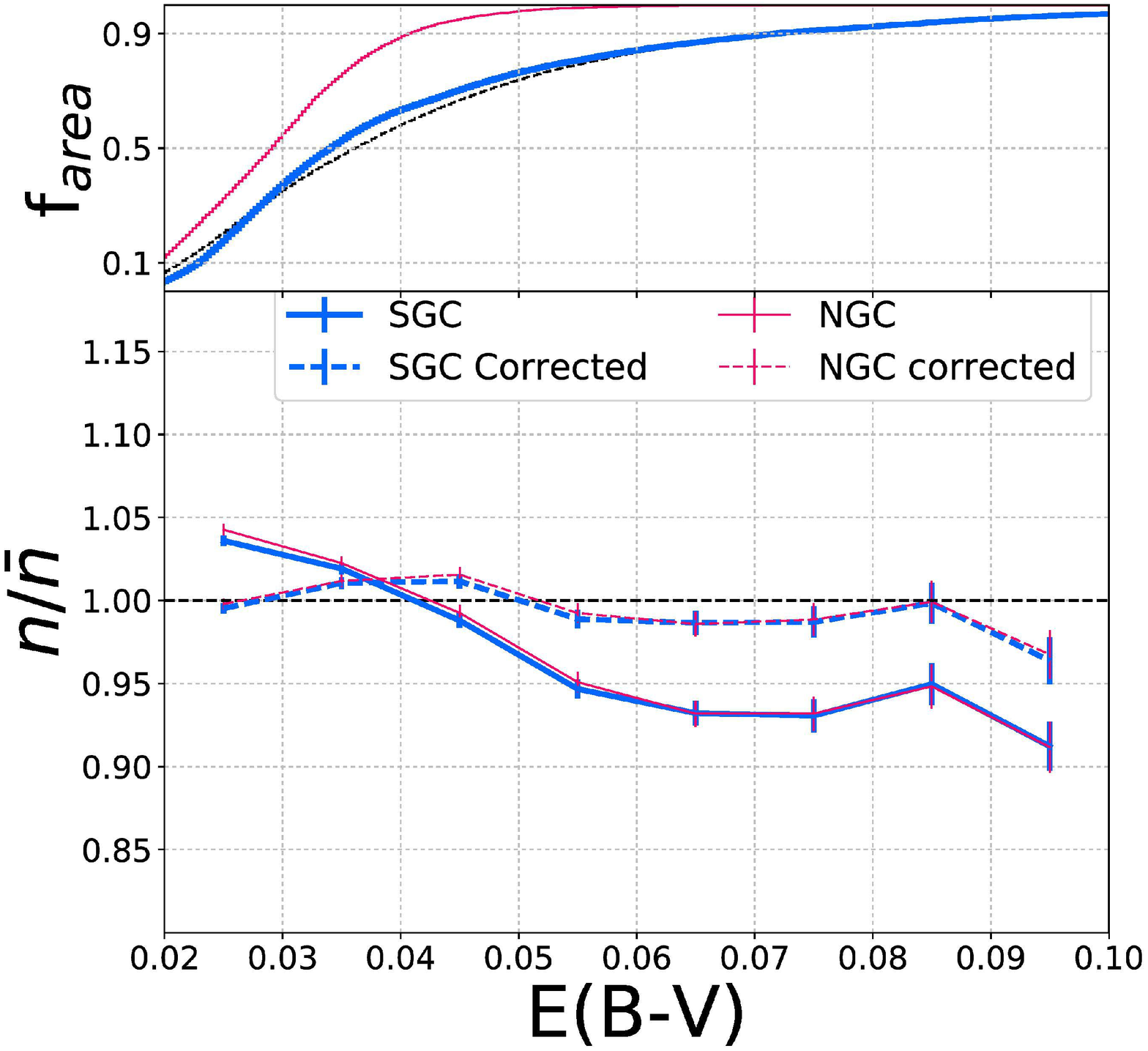}
\caption{Target selection dependency on the stellar density (\textit{top}) and the Galactic extinction (\textit{bottom}).
\textit{Left}: maps of the observational parameters
\textit{Right}: evolution of the normalized average number density ($n/\overline{n}$) as a function of each observational parameter.
Cyan curves are for the eBOSS/ELG SGC selection and pink curves are for the eBOSS/ELG NGC selection.
In each panel on the right, the top curve shows the fractional area of the survey which has a value of the parameter lower than or equal to the x-axis value; the black dashed line stands for the full DECaLS/DR3 footprint.
In the bottom panels, the solid lines correspond to the uncorrected density fluctuations, while the dashed curves represent the fluctuations remaining after applying the weights defined in Section \label{sec:syst_model}.}
\label{fig:syst_starebv}
\end{center}
\end{figure*}

\begin{figure*}
\begin{center}
\includegraphics[width=1.05\columnwidth]{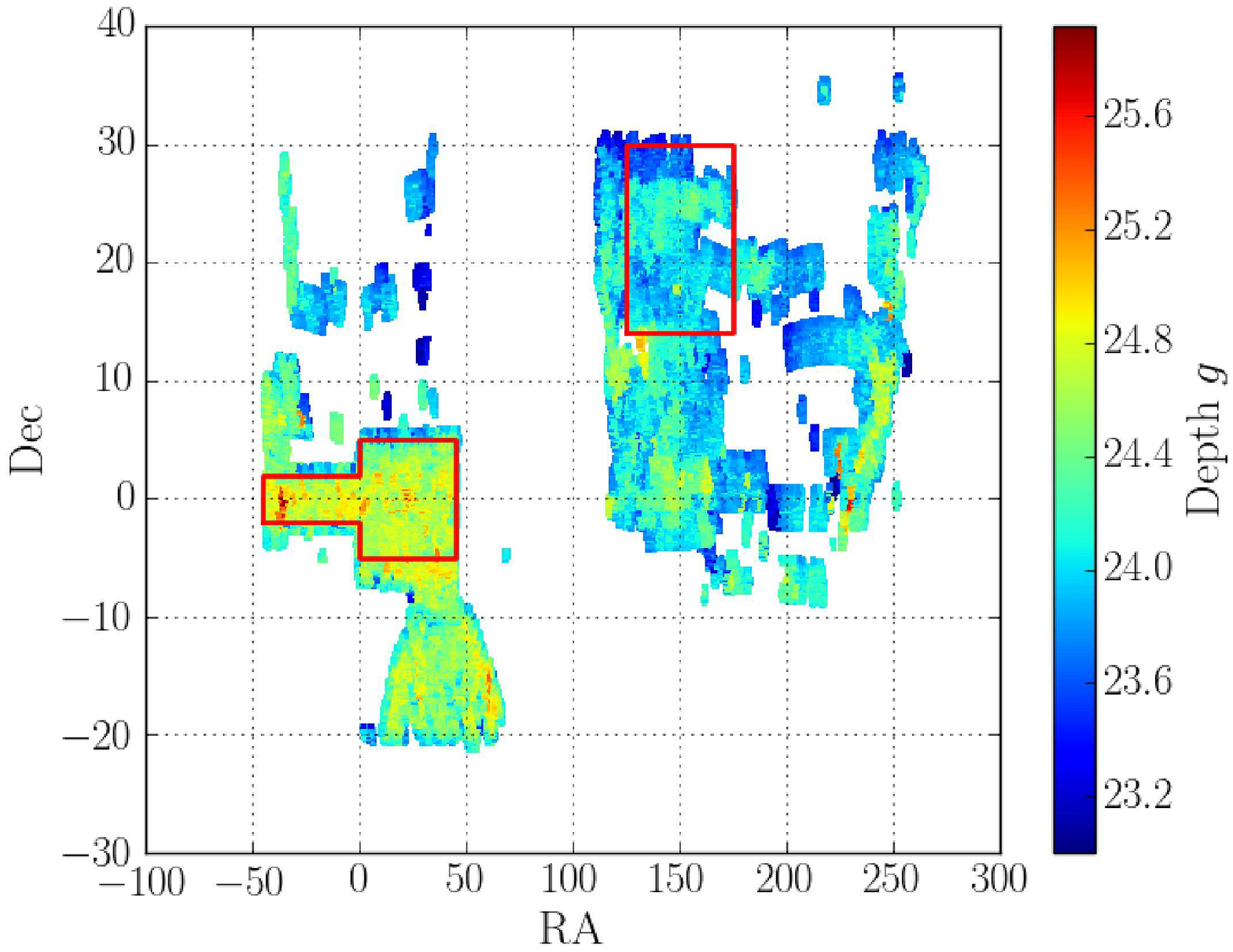}
\includegraphics[width=0.75\columnwidth]{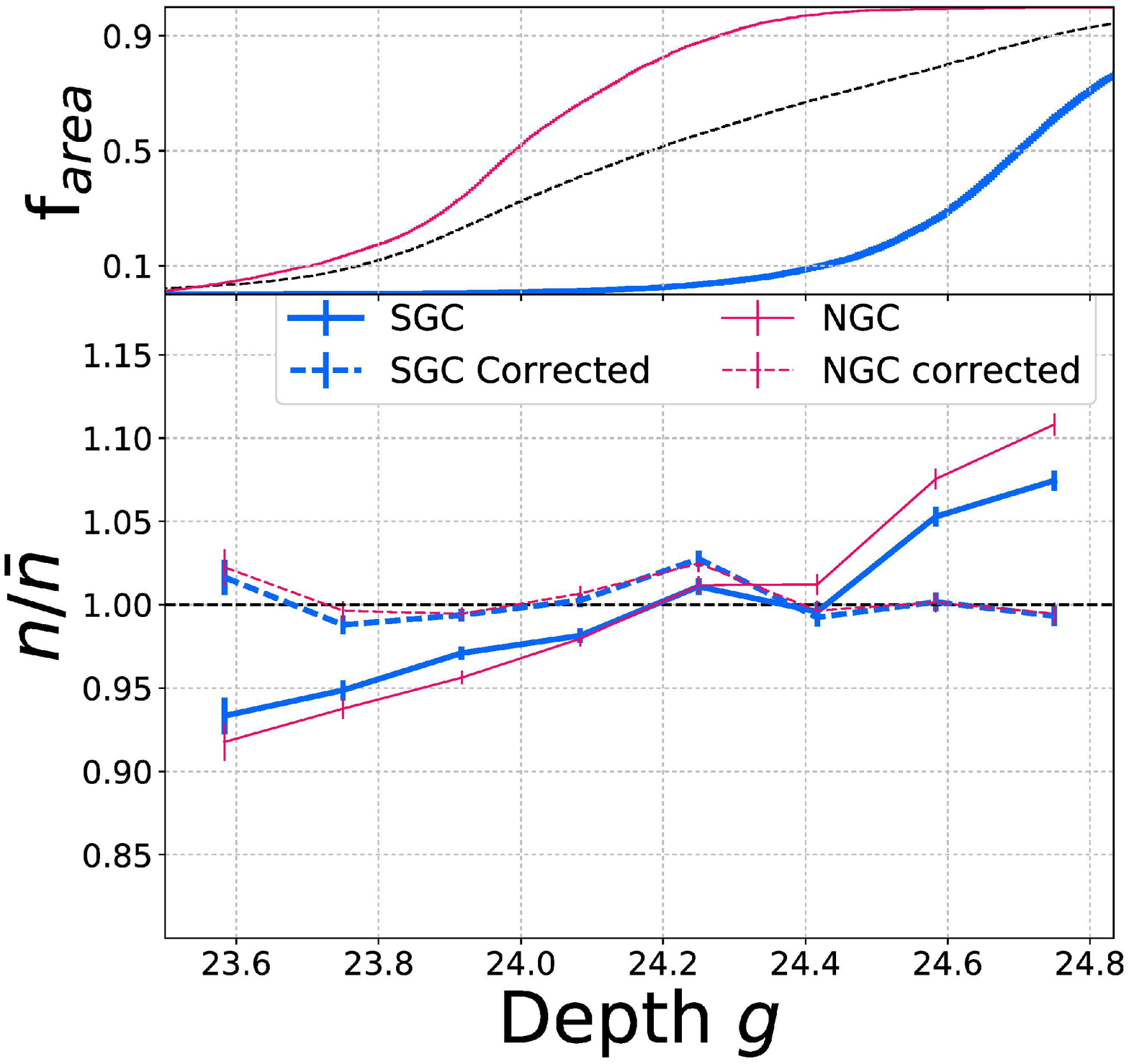}
\includegraphics[width=1.05\columnwidth]{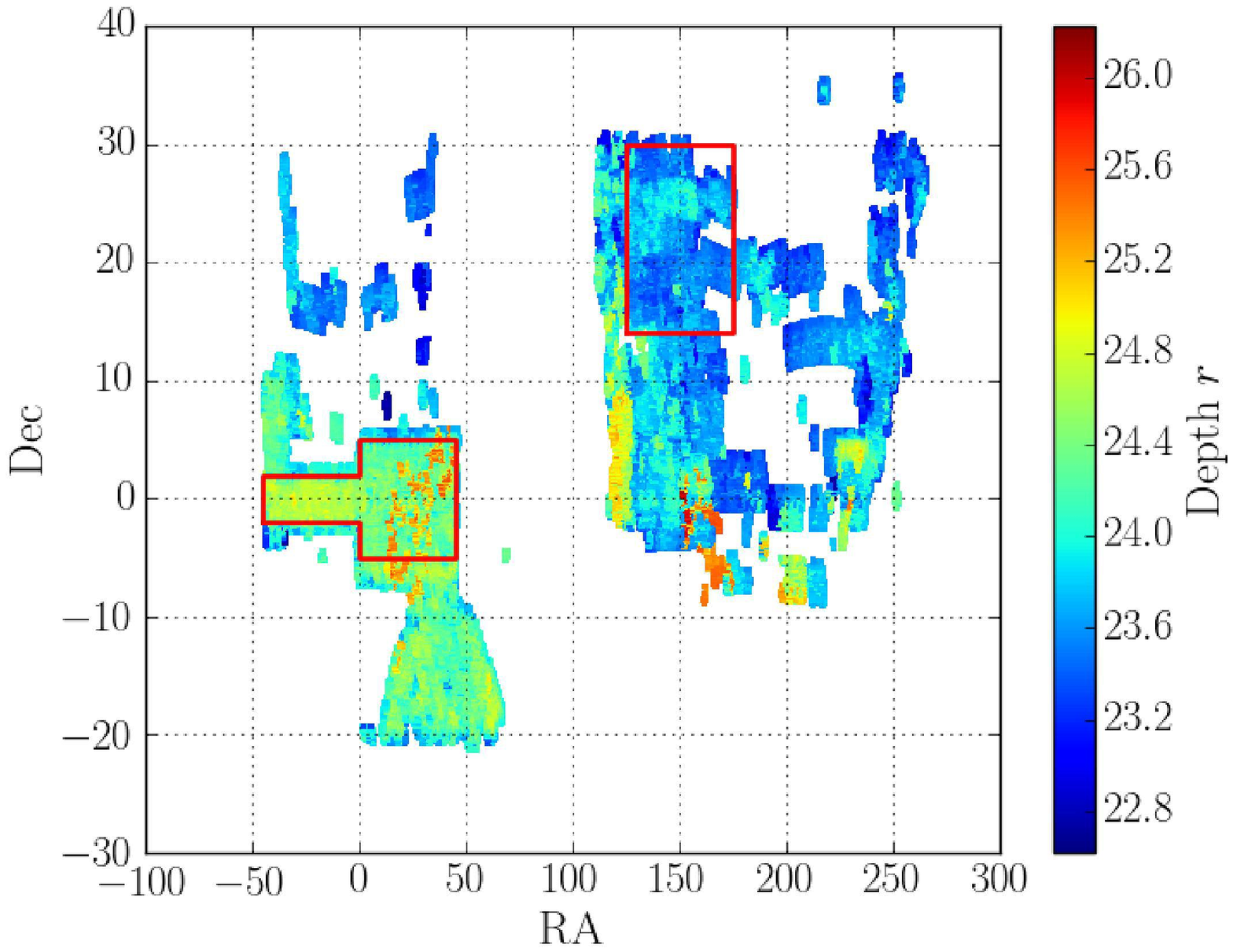}
\includegraphics[width=0.75\columnwidth]{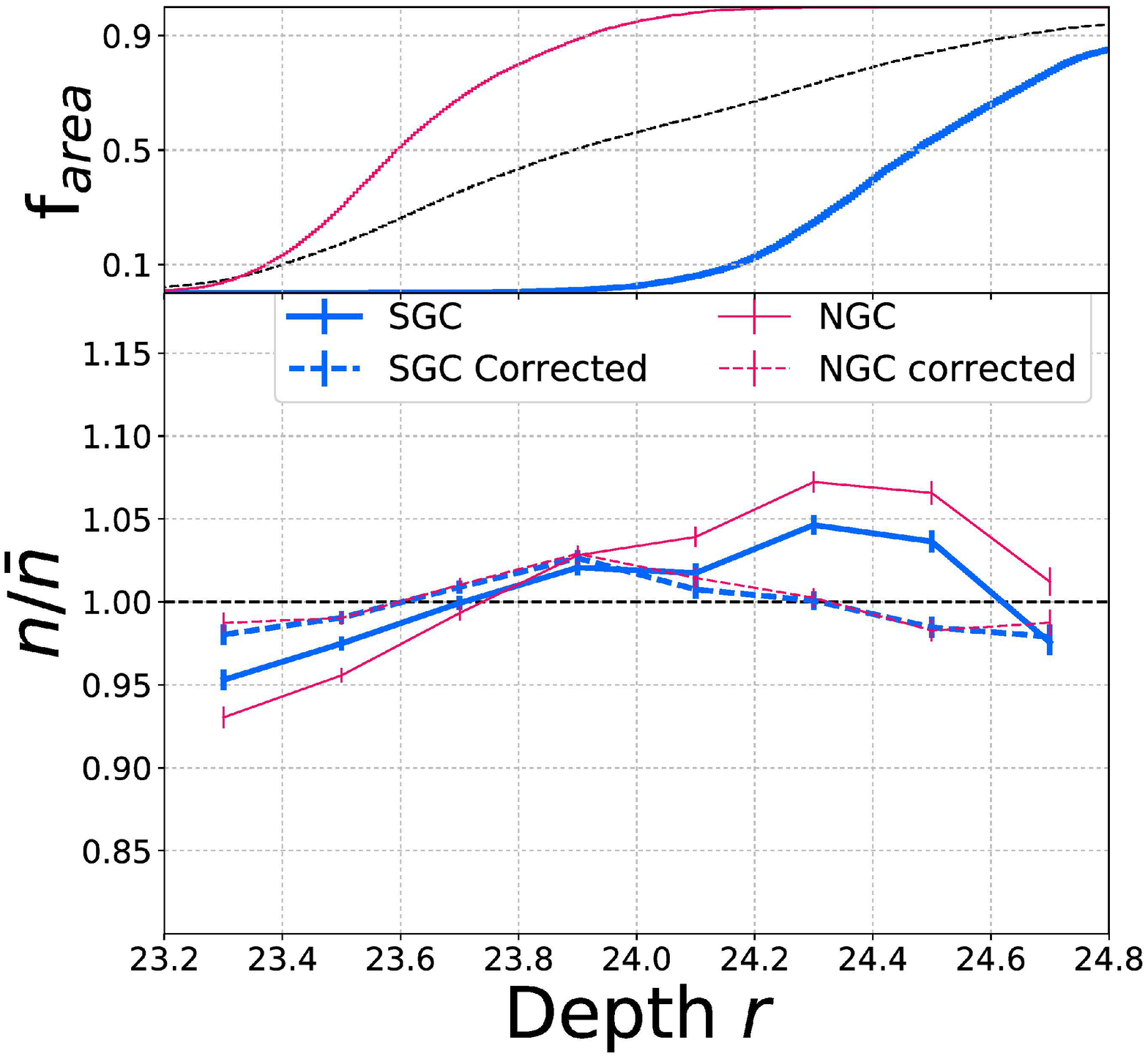}
\includegraphics[width=1.05\columnwidth]{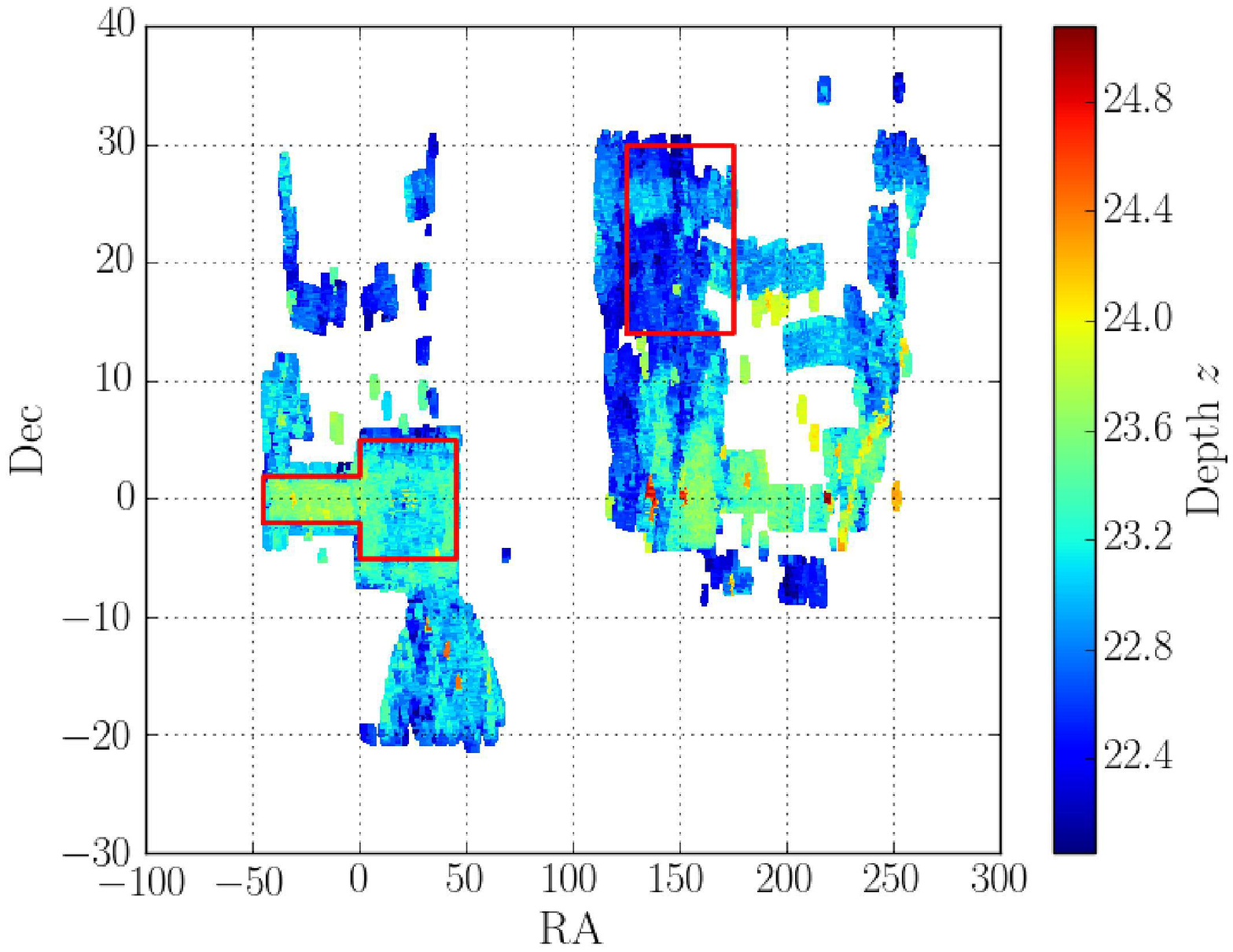}
\includegraphics[width=0.75\columnwidth]{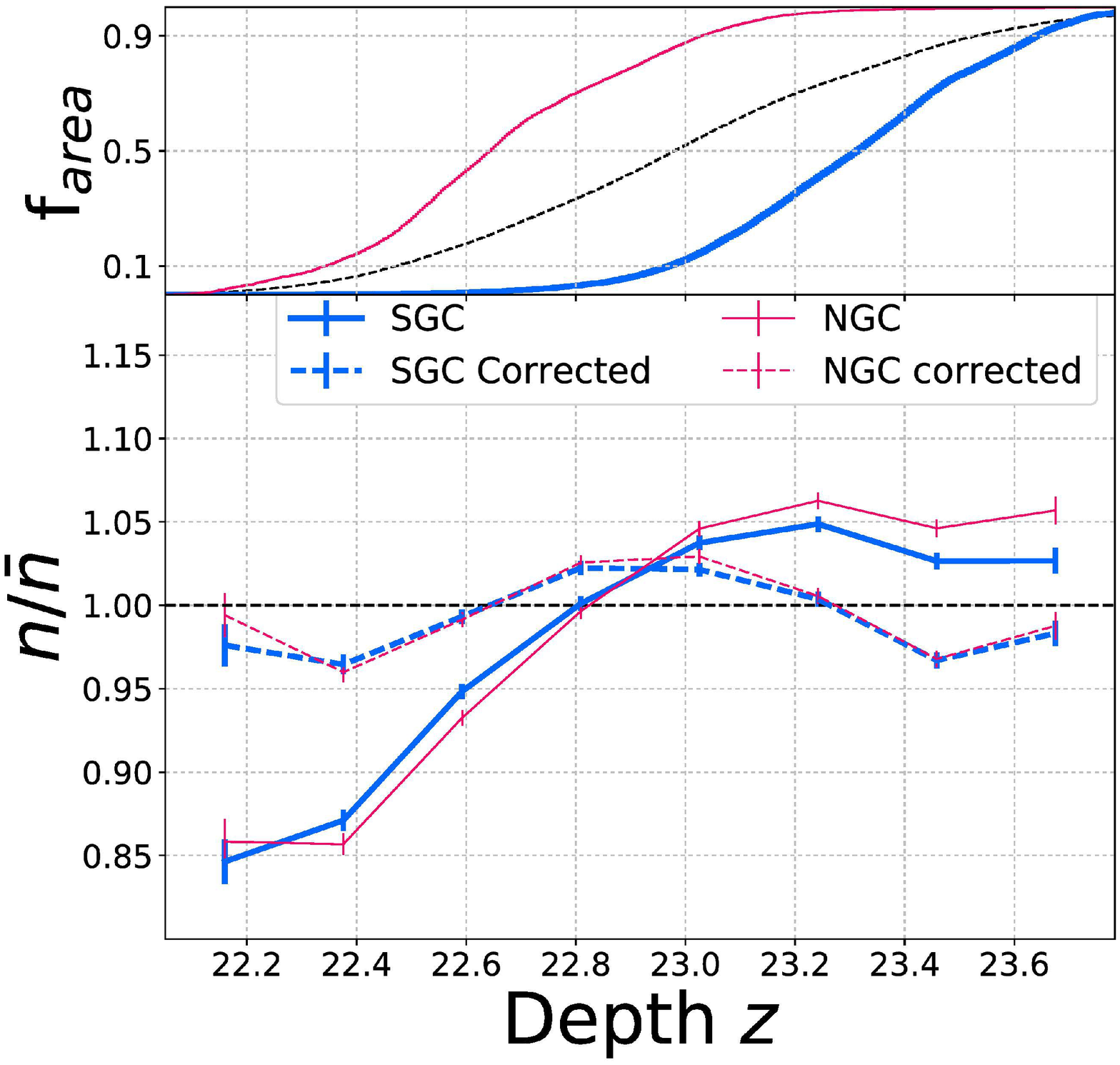}
\caption{As Figure \ref{fig:syst_starebv} but for the DECaLS imaging depth in the $g$-, $r$-, and $z$-band (from top to bottom.).
}
\label{fig:syst_depth}
\end{center}
\end{figure*}

\begin{figure*}
\begin{center}
\includegraphics[width=1.05\columnwidth]{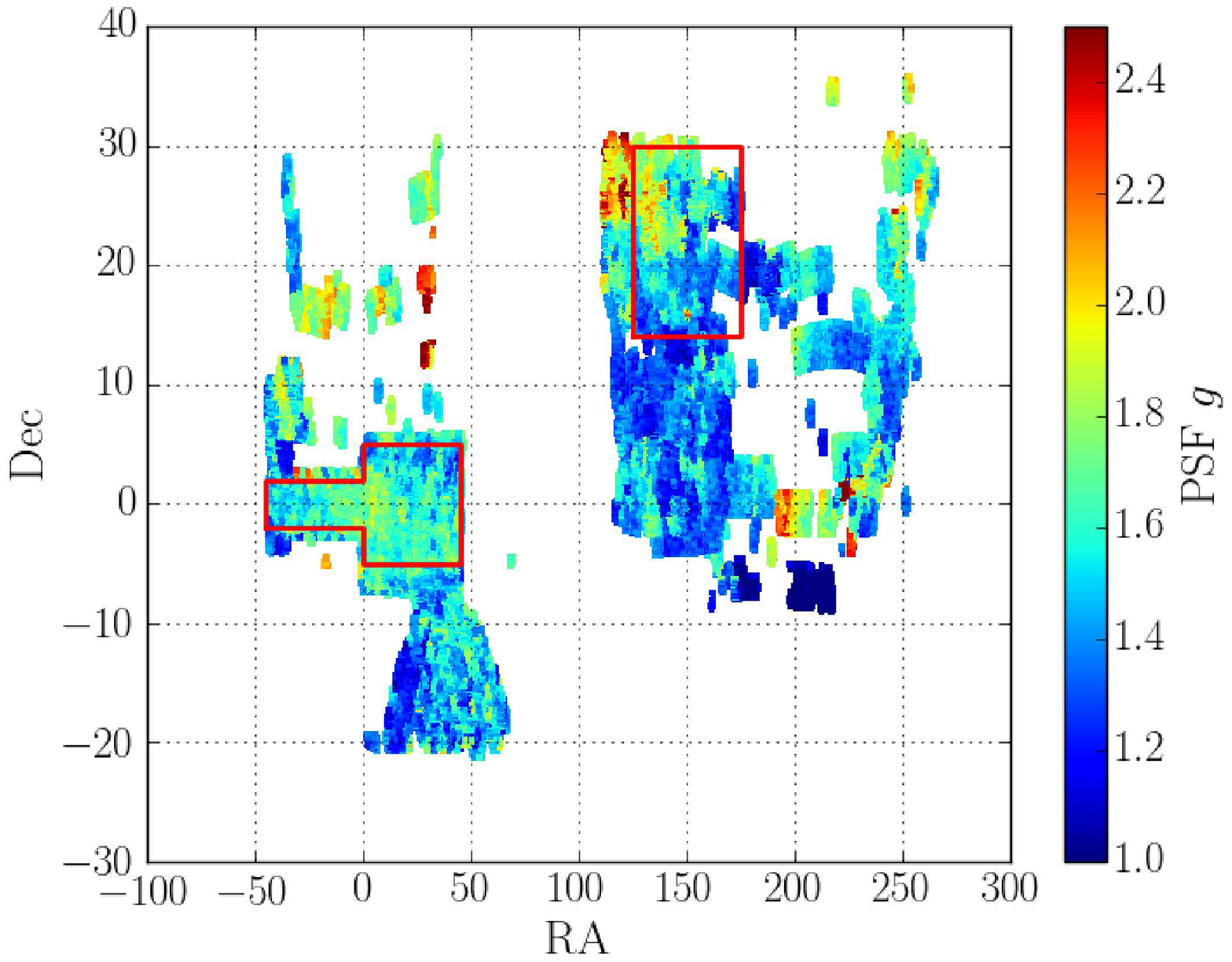}
\includegraphics[width=0.75\columnwidth]{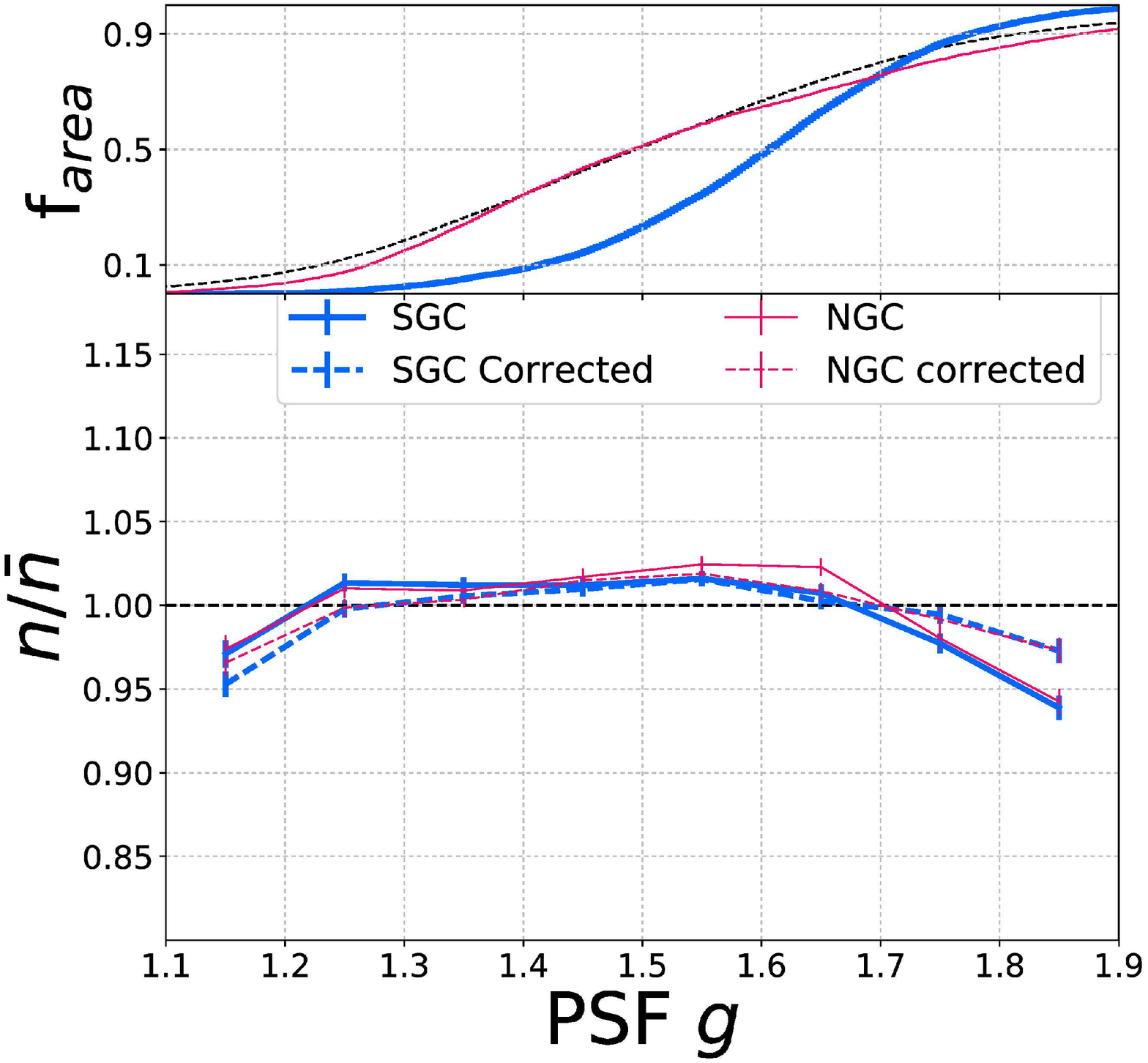}
\includegraphics[width=1.05\columnwidth]{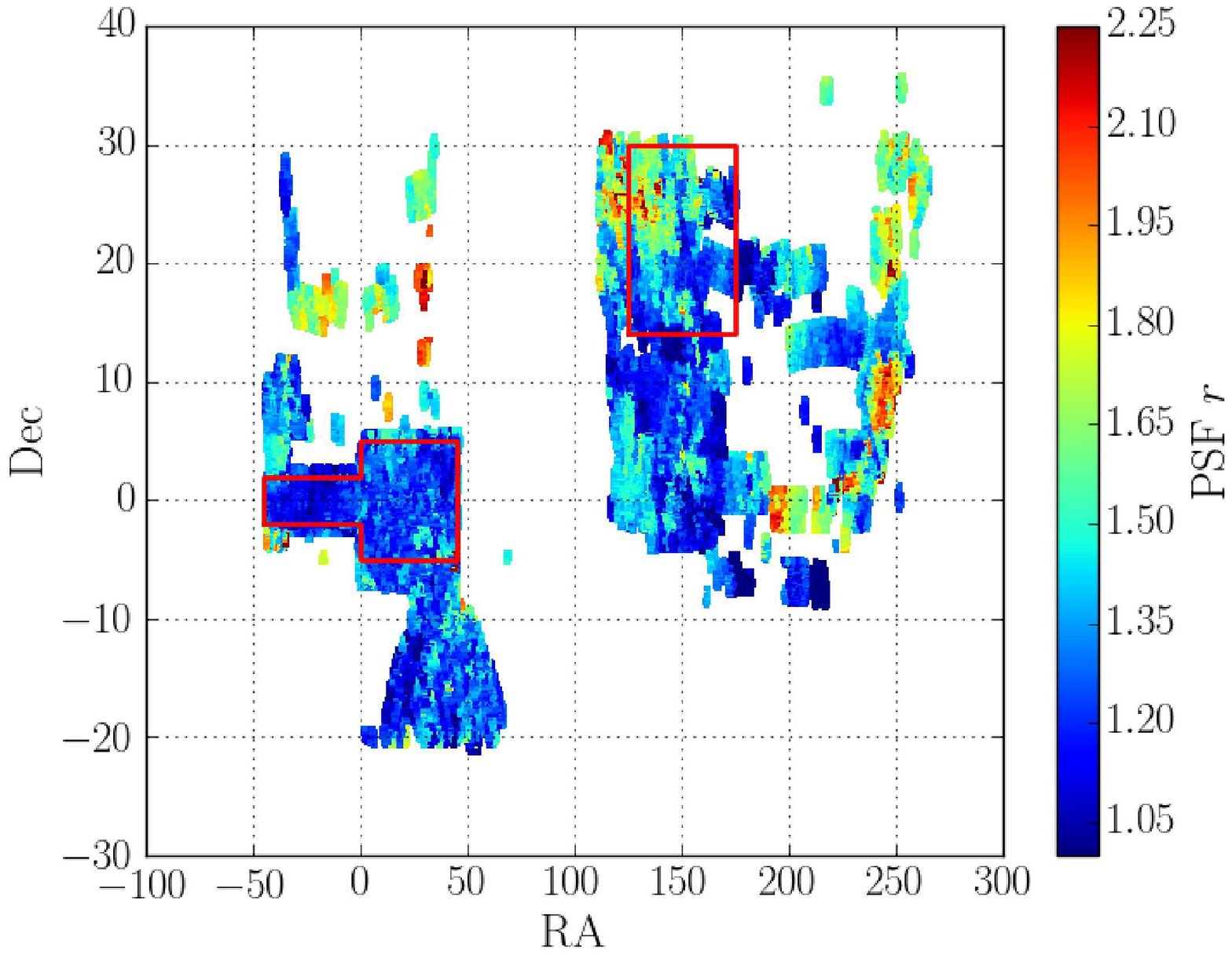}
\includegraphics[width=0.75\columnwidth]{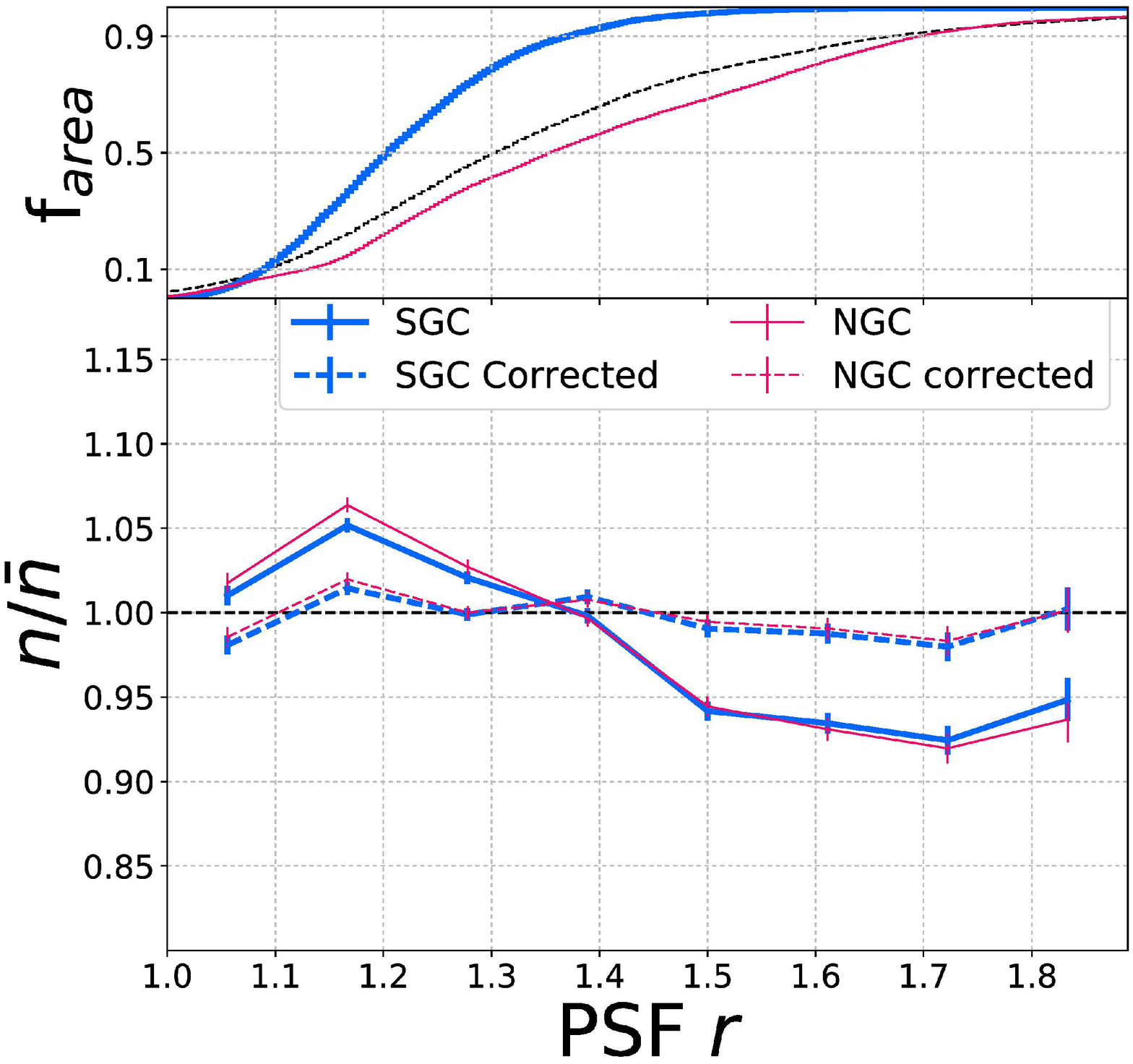}
\includegraphics[width=1.05\columnwidth]{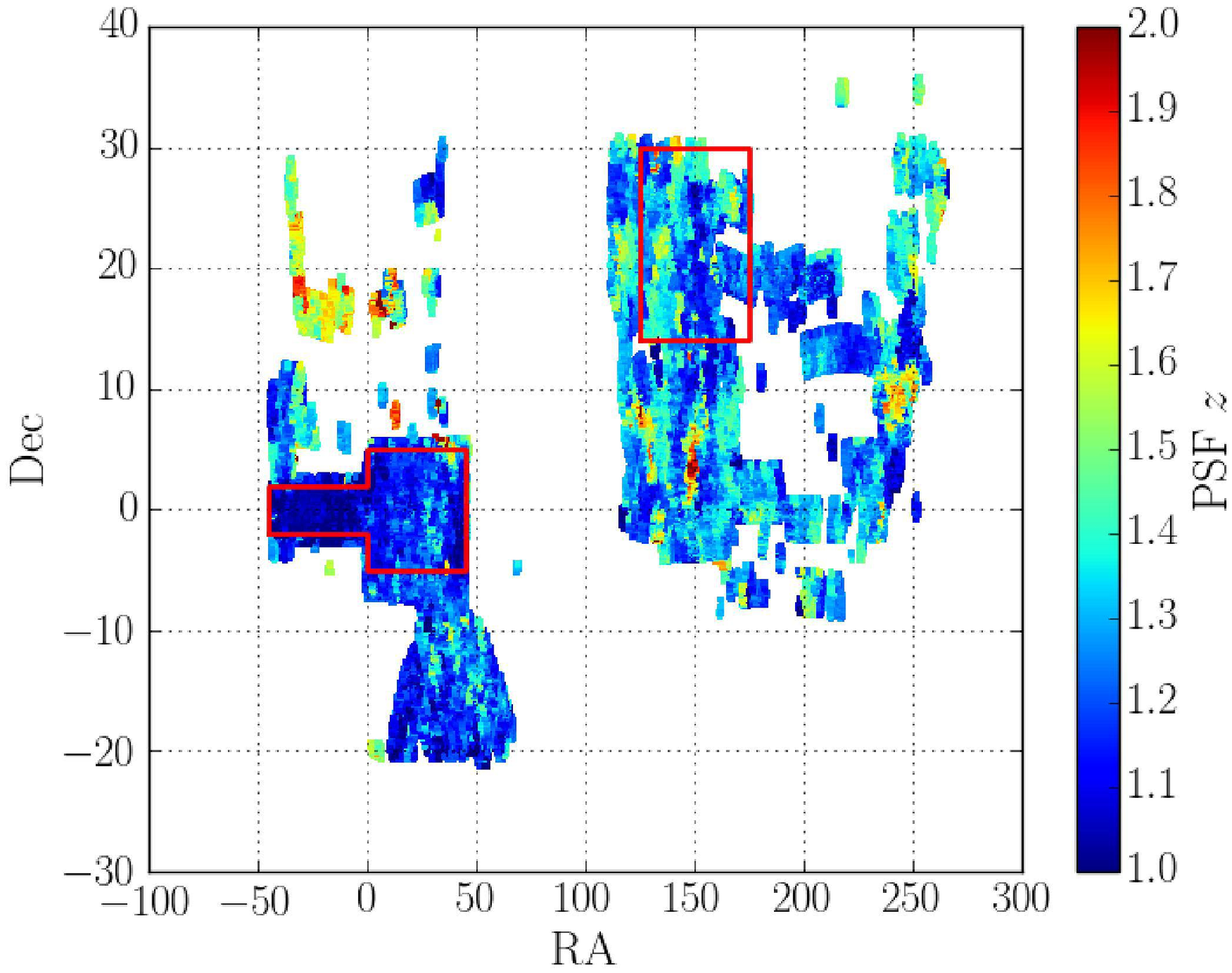}
\includegraphics[width=0.75\columnwidth]{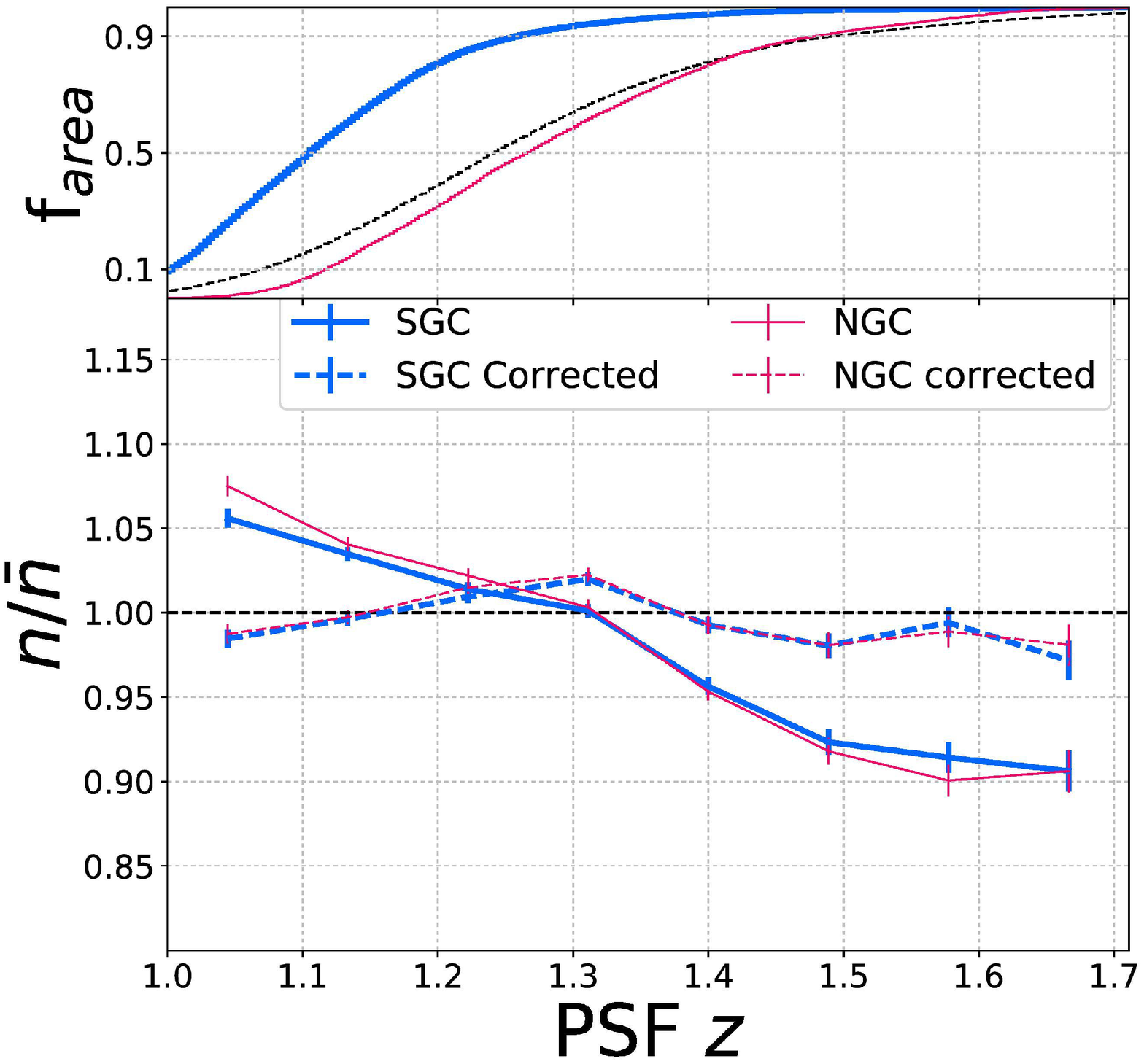}
\caption{As Figure \ref{fig:syst_starebv} but for the DECaLS imaging PSF in the $g$-, $r$-, and $z$-band (from top to bottom.).
}
\label{fig:syst_psf}
\end{center}
\end{figure*}

As a first step, we use the whole DECaLS/DR3 footprint to construct a map of the potential sources of systematic error in target selection due to variations in stellar density, Galactic extinction, $grz$ PSF, and depth: those maps are displayed on the left column panels in Figures \ref{fig:syst_starebv}-\ref{fig:syst_psf}.
We divide our maps into equal-area pixels of $\sim$0.05 deg$^2$ (corresponding to \texttt{HEALPIX}\footnote{\href{http://healpix.jpl.nasa.gov/}{http://healpix.jpl.nasa.gov/}} \texttt{nside} = 256), and compute the target density in each pixel, for both SGC and NGC selections.
For both footprints (SGC and NGC), we measure how the target density varies with each of these sytematics.

% plots description
The lower panels in the right column of Figures \ref{fig:syst_starebv}-\ref{fig:syst_psf} show those dependencies.
The selections over the SGC and the NGC have very similar behaviour, which is expected given that they are based on very similar cuts.
These curves are computed for the full DECaLS/DR3 footprint: the upper panels in the right column plots of Figures \ref{fig:syst_starebv}-\ref{fig:syst_psf} display the cumulative histograms of the systematics over the eBOSS/ELG SGC and NGC footprint.
The histograms thus indicate the relevant systematics range for each selection.
Cumulative histograms over the full DECaLS/DR3 footprint are also reported in black.
For instance, the $g$-band depth over the eBOSS/ELG SGC and the NGC footprints are distributed differently: 90\% of the SGC footprint has a $g$-band imaging deeper than 24.4 mag, whereas the corresponding depth for the NGC footprint is 23.7 mag.
% variations are ok!
Overall, the target density variations over the SGC and NGC footprints are satisfactory.
For each observational parameter, the variation in target density is always smaller than 12\% over the range including the 10\%-90\% of the cumulative histograms; the only exception being for the $z$-band depth for the NGC selection, where the variation in the target density is $\sim$20\%.

Some of the dependencies can be explained as follows.
% stellar density
The target density decreases with increasing stellar density: it has been shown in \citet{ross11} and \citet{delubac17} that this can be understood as resulting from the low stellar contamination of our selections plus the fact that each star masks a small area of the sky, preventing the selection of targets in that area.
However, we underline that the stellar density over the ELG footprint has limitedimpact as the density is generally low: for instance, 90\% of the ELG footprint have $n_{\rm stars_g}<2000$ deg$^2$.
% ebv
The target density also decreases with increasing Galactic extinction: this is expected because of the luminosity function shape, as the faintest objects in $g$-band, removed from our selection by a high Galactic extinction value are overwhelmingly more numerous than the brightest ones which enter our selection.
Additionaly, there is also a correlation between the stellar density and the Galactic extinction.
% depth
The target density has a clear dependence on the $g$-band depth, with an increase for deeper imaging.
This can be explained by the fact that our selections are based on a $g$-band magnitude cut and that deep imaging will provide more detected objects.

% Figure: syst nstar & ebv

% Figure: syst depth

% Figure: syst PSF

% subsection: Modelling the systematic effects
\subsection{Modelling the systematic effects} \label{sec:syst_model}
We then simultaneously model the effect of those systematics using a multivariate regression,as in \citet{delubac17}.
We use a quadratic dependence as a function of the Galactic extinction, the $g$- and $z$-band PSF, the $r$- and $z$-band depth, which exhibit explicit non-linear behaviour, and a linear dependence as a function of all other systematics.

% correction is working
Figures \ref{fig:syst_starebv}-\ref{fig:syst_psf} show that applying the correction defined in \citet{delubac17} to the pixel densities reduces the systematic effects (dashed lines): the fluctuations of the reduced average number density are now consistent with zero given the uncertainties computed as the root mean square in the bin.
Note that, once the spectroscopic ELG observations are complete, that weights will be re-computed using the ELG cosmological sample.

\begin{figure}
	\includegraphics[width=0.95\columnwidth]{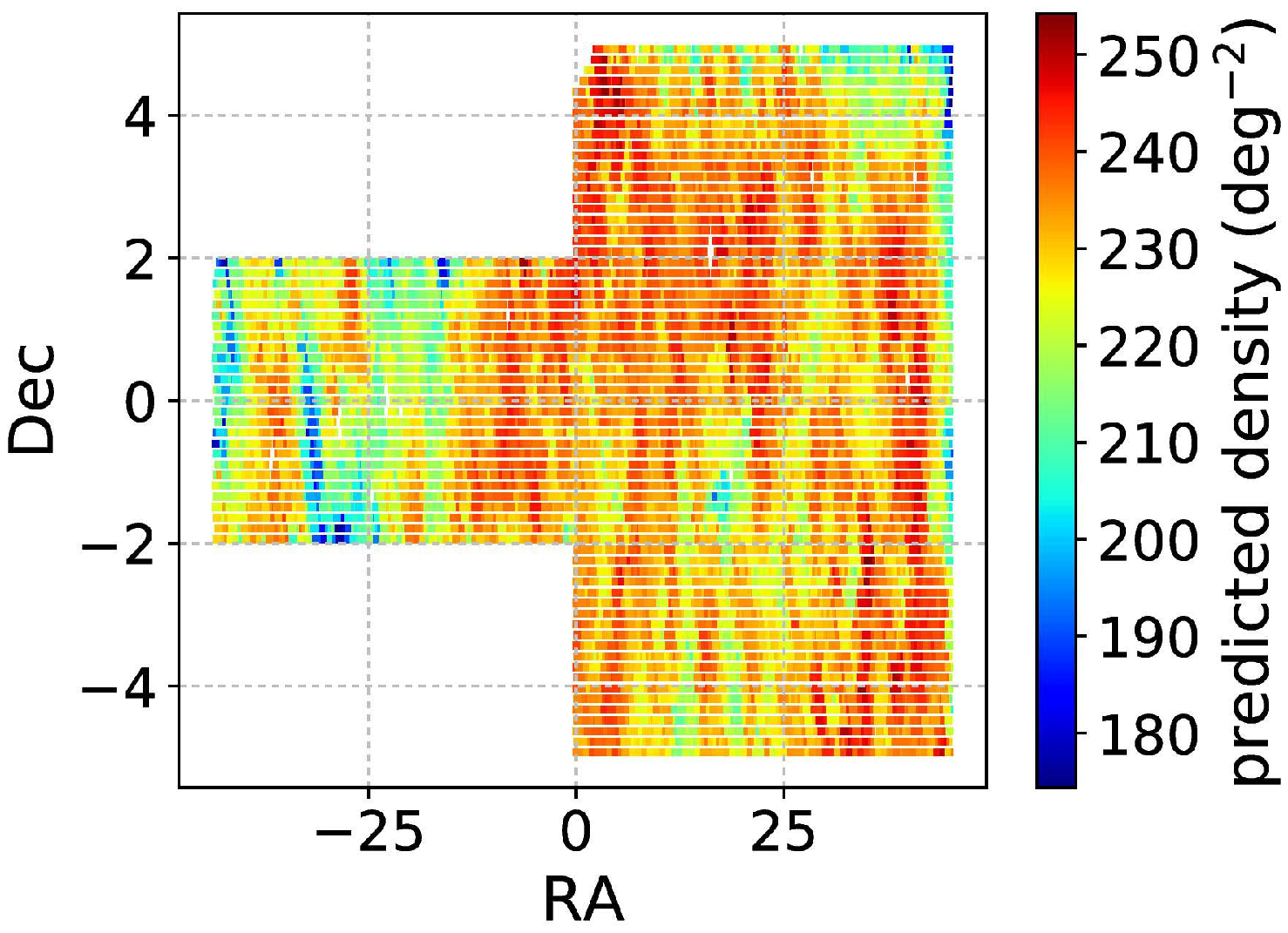}
	\includegraphics[width=0.95\columnwidth]{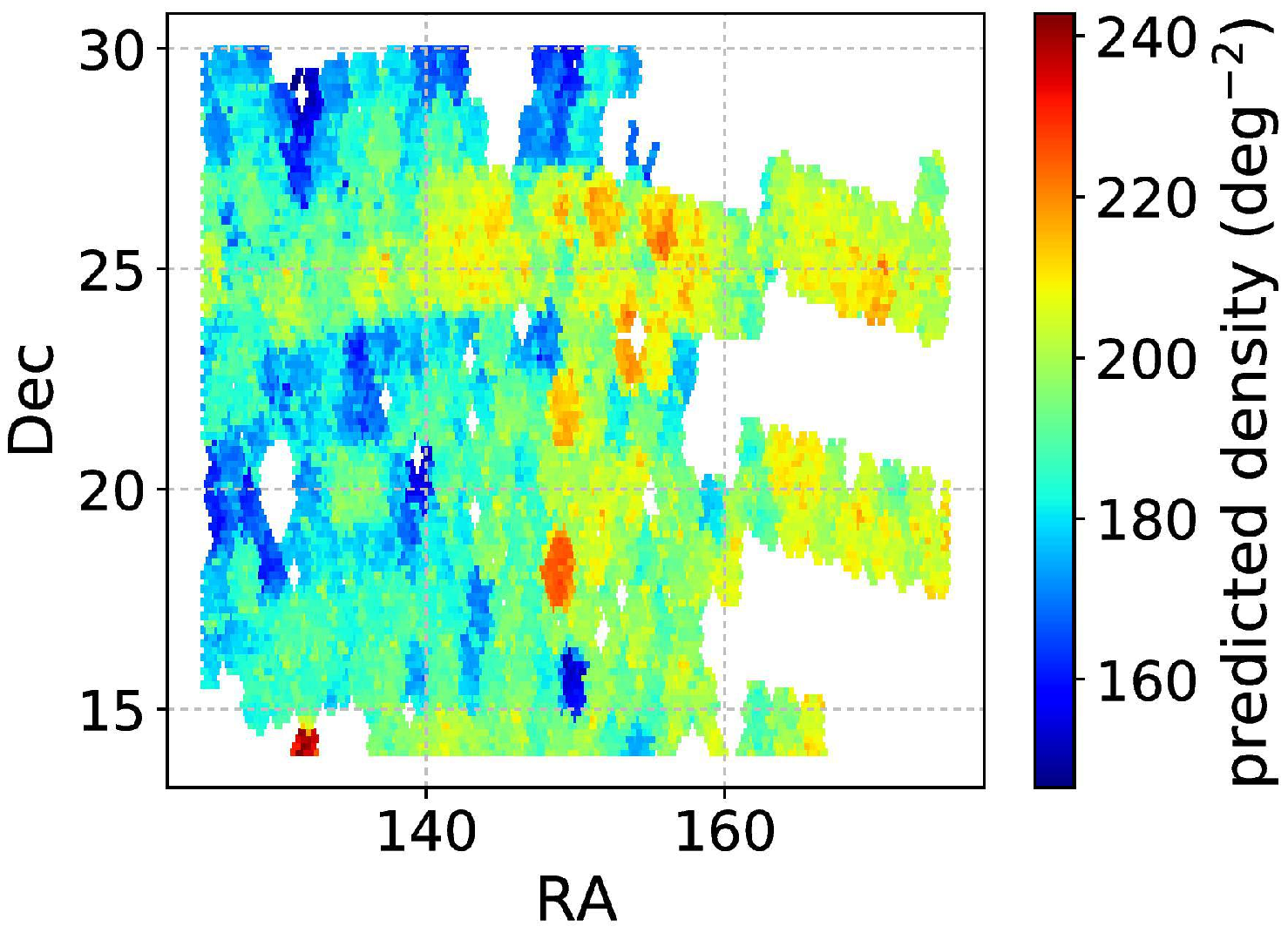}
	\includegraphics[width=0.95\columnwidth]{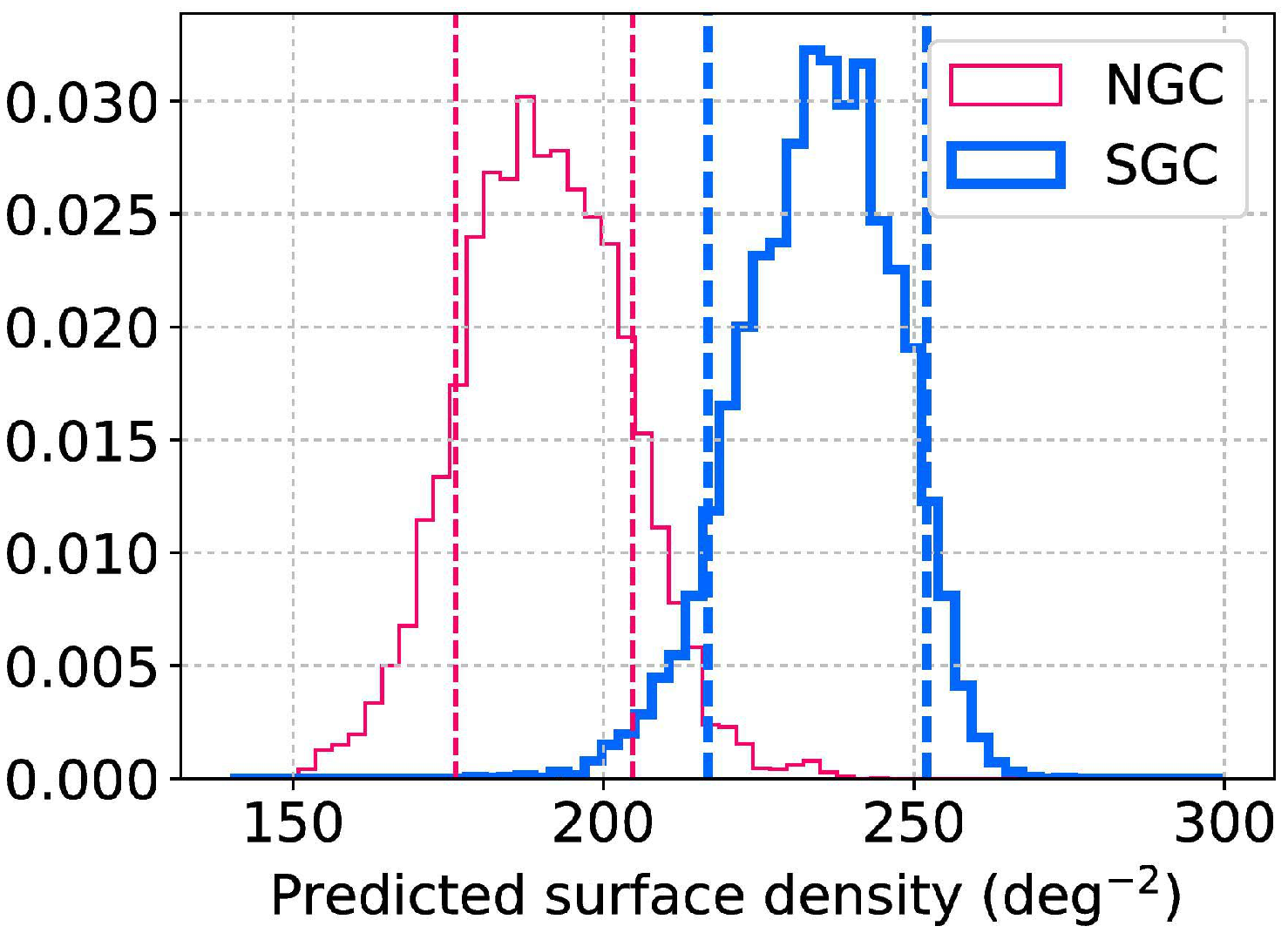}
\caption{Predicted pixel densities for both selections.
\textit{Top panel}:      map of the predicted densities for the SGC.
\textit{Middle panel}: map of the predicted densities for the NGC.
\textit{Bottom panel}: histogram of the predicted densities; the dashed vertical lines mark a $\pm$7.5\% fluctuation around 234 and 190 deg$^{-2}$, broadly corresponding to the maximum of the distribution of the two selections.}
\label{fig:pred_dens}
\end{figure}

% predicted densities
Once we have modelled how the target selection depends on the systematics, we can compute the predicted density of the selections in the absence of shot noise and cosmic variance given the value of the systematics.
We present Figure \ref{fig:pred_dens} the maps of this predicted densities (top and middle panels), along with their distribution (bottom panels); note this quantity is different from the measured pixel densities.
In the histogram panel, we emphasize two regions corresponding to a $\pm$7.5\% fluctuation around 234 and 190 deg$^{-2}$.
The choice of the central values is somewhat arbitrary, but they broadly correspond to the maximum of the distribution of the two selections, i.e. they are close to maximizing the surface of the footprint passing the homogeneity requirements.
With these central values, 84.5\% deg$^2$ of the SGC footprint and 72.2\% deg$^2$ of the NGC footprint pass the homogeneity requirement.

% Figure: predicted density

% subsection: Zero-point fluctuations
\subsection{Zero-point fluctuations} \label{sec:syst_zp}

% syst zeropoints
We test the zeropoint-requirement for each photometric band individually, following an approach similar to the one presented in \citet{dawson16} and \citet{myers15} and \citet{prakash16}.
We add $\pm0.01$ mag to each photometric band used and then re-run the target selection algorithm to estimate $\delta N_{0.01} = \frac{|\Delta N|}{N}$, the change in target density due to this $\pm$0.01 mag shift.
In the $g$-, $r$-, and $z$-bands, we find for $\delta N_{0.01}$ values of 3.6\%, 4.3\%, and 2.9\%, respectively for the SGC, and of 3.4\%, 4.0\%, and 2.9\%, respectively for the NGC.

% Table: syst zeropoints
\begin{table*}
	\centering
	\begin{tabular}{cccccccc}
		\hline
		\hline
				& 		& \multicolumn{3}{c}{SGC selection} &  \multicolumn{3}{c}{NGC selection} \\
		Band	 	& $\sigma_{\rm zp}$ & $\delta N_{0.01}$  & Fraction & Fluctuation over & $\delta N_{0.01}$  & Fraction &  Fluctuation over\\
				& 				&   & passing  $\pm$7.5\% &  95\% of the area &  & passing  $\pm$7.5\% &  95\% of the area\\
				& [mmag] &	[\%] 	&  [\%] & [\%] &  [\%] & [\%] & [\%] \\
		\hline
		$g$		& 9		& 3.6		& 97.9	& 13.0	& 3.4		& 98.6	& 12.2\\
		$r$		& 7		& 4.3		& 98.7	& 12.0	& 4.0		& 99.3 	& 11.2\\
		$z$		& 8		& 2.9		& 99.9	& 9.3		& 2.9		& 99.9	& 9.3	\\
		\hline
	\end{tabular}
\caption{
Impact of fluctuations in imaging zeropoints on the number densities of the selections.
The 2$^{\rm nd}$ column is the expected rms error in the photometric calibration ($\sigma_{\rm zp}$).
The 3$^{\rm rd}$  and 6$^{\rm th}$ columns are the change in target density due to a $\pm$0.01 mag shift in the zeropoint ($\delta N_{0.01} = \frac{|\Delta N|}{N}$).
The 4$^{\rm th}$ and 7$^{\rm th}$ columns are the fraction of the footprint passing the $\pm$7.5\% uniformity requirement.
The 5$^{\rm th}$ and 8$^{\rm th}$ columns are the fluctuations in density over 95\% of the footprint.}
	\label{tab:zeropoints}
\end{table*}

We quantify the fraction of the footprint area which passes the $\pm$7.5\% uniformity requirement as in \citet{dawson16}.
Let it be $\sigma_{\rm zp}$, the expected rms error in the photometric calibration of the considered band.
We use the $\sigma_{\rm zp}$ error estimates of \citet{finkbeiner16} for the SDSS bands (summarised in Table \ref{tab:zeropoints}): those have been estimated for the SDSS with Pan-Starrs1, which is the very same survey used by DECaLS for its zeropoint calibration.
For a given band, uniformity with 15\% peak-to-peak amplitude occurs in regions where the zeropoint is in error by less than $\pm 0.01 \times \frac{7.5}{\delta N_{0.01}}$.
Using the $\sigma_{\rm zp}$ error estimates of \citet{finkbeiner16} and assuming Gaussian errors for the zeropoints, this happens for $>$97.9\% of the footprint in all bands (see Table \ref{tab:zeropoints}).
The selections are robust against variation of the imaging zero-points. 

Another way to consider this is that the expected rms variation in target density due to shifts of the imaging zero-point is $\frac{\delta N_{0.01}}{0.01} \times \sigma_{\rm zp}$ \citep{myers15,prakash16}.
Assuming Gaussian errors for the zeropoints, 95\% of our footprint lies within a $\pm2\sigma_{\rm zp}$ variation from the expected zero-point of any given photometric band, meaning that 95\% of our footprint has a variation in target number density lower than $4\times \sigma_{zp} \times \frac{\delta N_{0.01}}{0.01}$.
The resulting fluctuations for each photometric band are given in Table \ref{tab:zeropoints}.
Both selections have fluctuations of 9\%-13\%, below the 15\% requirements, thus confirming that the selections pass the zeropoint-requirement.

% subsection: Catalog public release
\subsection{Catalog public release}
The target catalogue over the SGC footprint will be publicly released in mid-2017, as a Value-Added Catalog from the SDSS DR14 release (\href{http://www.sdss.org/dr14/data\_access/vac/}{http://www.sdss.org/dr14/data\_access/vac/}).
It will contain the DECaLS \texttt{brickname}, the (R.A., Dec.) coordinates, the $grz$-band magnitudes and magnitude errors, along with the systematic weights fitted in Section \ref{sec:syst_model}.

%=================================
% first months of obs.
%=================================
\section{First months of observations} \label{sec:firstobs}

We describe here preliminary results from the first months of observations ($57656 \le \texttt{MJD} \le 57787$).
The general procedure for the spectroscopic observations are described in detail in the eBOSS overview paper \citep{dawson16}: we only report here a brief summary, and items specific to the ELG program.
The spectroscopic observations are conducted with the BOSS spectrograph at the 2.5 m aperture Sloan Foundation Telescope at Apache Point Observatory in New Mexico \cite{gunn06}.
From the first data we are able to determine the actual efficiency, which we define as the percentage of observed ELG targets having a reliable $\zspec$ estimation (Eq.\ref{eq:zqzcont}) and $0.7<\zspec<1.1$.

% tiling, plate design, exposure time
\subsection{Tiling, plate design, and exposure time} \label{sec:obs_design}
\begin{figure*}
 	\includegraphics[width=0.95\linewidth]{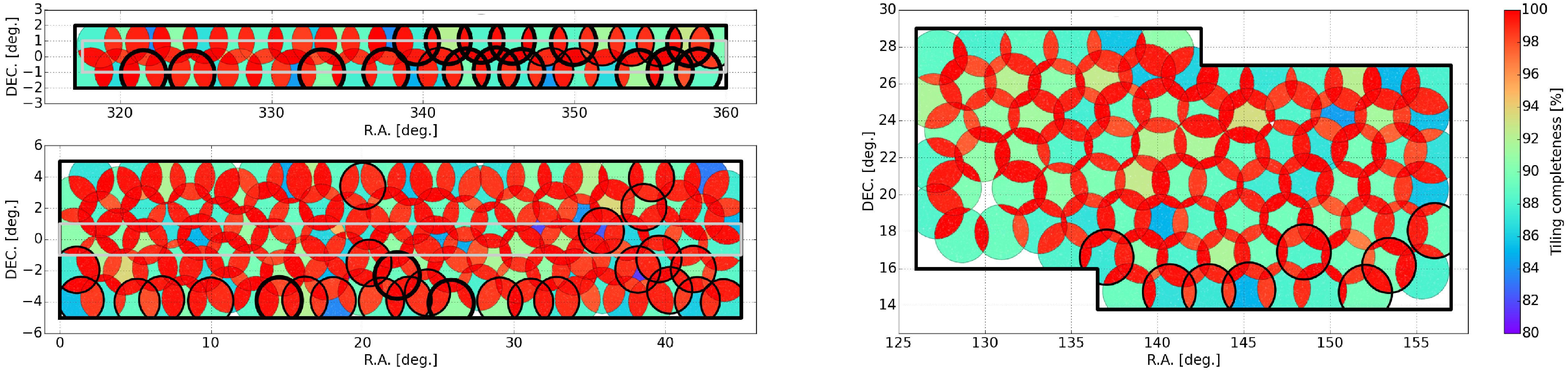}\\
 	\caption{Tiling completeness for the ELG plates.
The full SGC tiling is presented is the top left (chunk \texttt{eboss21}, 46 plates) and bottom left (chunk \texttt{eboss22}, 121 plates) panels.
The currently tiled NGC area is presented in the right panel (chunk \texttt{eboss23}, 87 plates).
The percentage of photometric targets being assigned a fiber is colour-coded for each sector defined by the plates overlaps.
Observed plates are circled in black; plates observed between $57656\le\texttt{MJD}\le57711$ with exposure times longer than nominal are circled in thick black; the gray rectangle in the left panels shows the CS82 imaging survey footprint.}
 	\label{fig:sgccompleteness}
 \end{figure*}

% tiling
Once the target catalog is built from the photometric catalog, one has to define the plate tiling and which targets are assigned a fiber.
For the SGC, the tiling has been designed in two \textit{chunks}, 46 plates at $317^\circ<\textnormal{R.A.}<360.0^\circ$ (\texttt{eboss21}) and 121 plates at $0^\circ<\textnormal{R.A.}<45^\circ$ (\texttt{eboss22}).
For the moment, only $\sim$400 deg$^2$ of the NGC footprint is tiled (\texttt{eboss23}, 87 plates): the rest of the footprint will be tiled in the second semester of 2017, based on the DECaLS imaging available at that time.
Figure \ref{fig:sgccompleteness} shows the plate tiling, along with the tiling completeness, defined as the fraction of targets that were assigned a fiber.
For a typical ELG observed plate, the 1000 fibers available are used as follows: $\sim$100 fibers are used for calibration targets (as for all eBOSS plates), $\sim$50 fibers are dedicated for the Time Domain Spectroscopic Survey \citep[TDSS;][]{morganson15,dawson16} program, and $\sim$850 fibers are dedicated to ELG targets.
For both footprints, the tiling completeness is $\sim$98\% in plate overlapping regions and $\sim$87\% where there is only one plate covering.
The overall tiling efficiency is 95.1\% in the SGC and 91.5\% in the NGC.

% Figure: tiling + completeness

% plate design
During the plate design, the assigned targets celestial coordinates are converted to the coordinate system of the telescope focal plane.
The ELG targets are centered on the focal plane at a position (\texttt{LAMBDA\_EFF}) corresponding to the focus of 7500 \AA~ light: this choice optimises the signal, as it broadly corresponds to the location of the \oii~ emission at redshift 1.

\begin{figure}
 	\includegraphics[width=0.95\columnwidth]{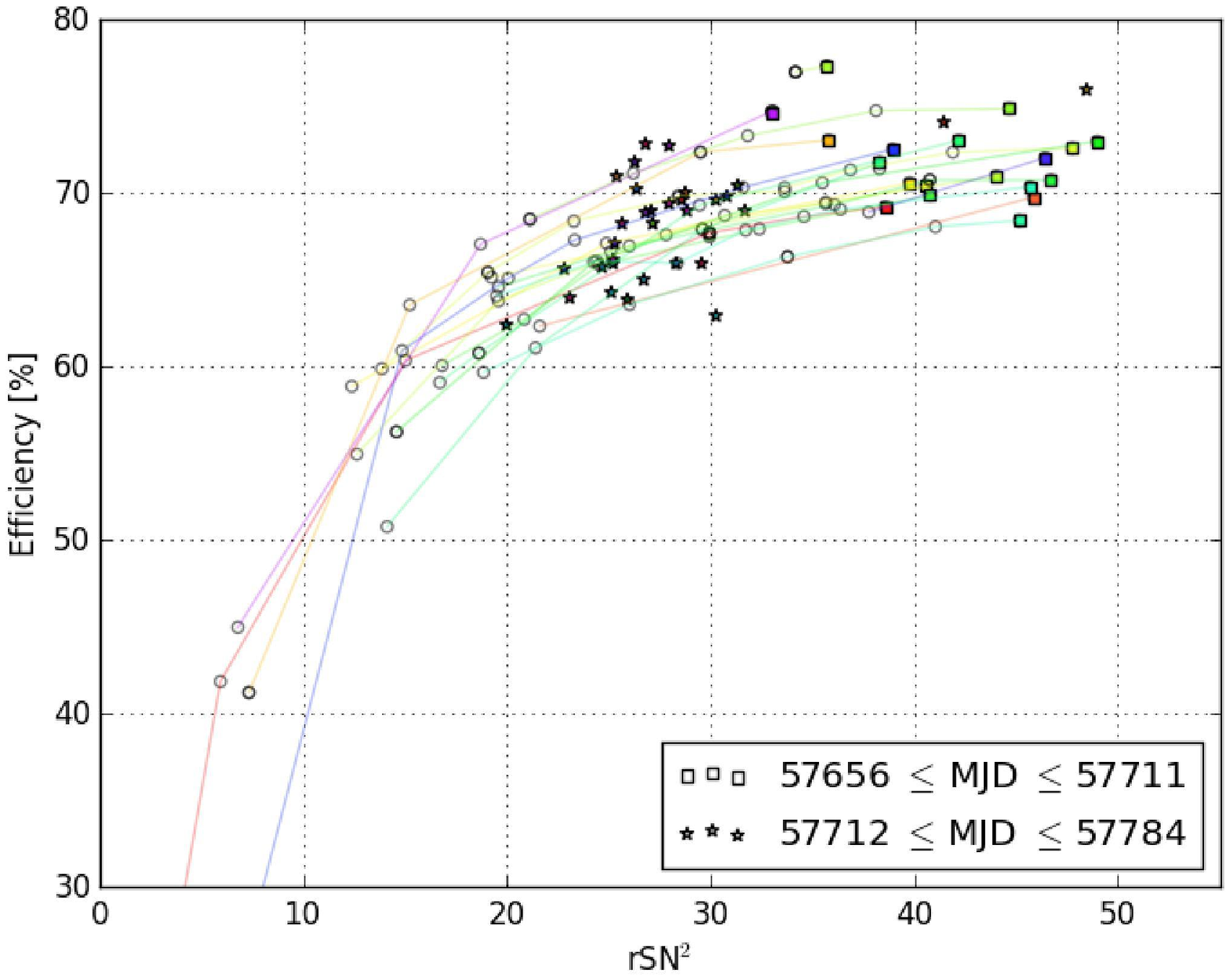}
 	\caption{Efficiency as a function of \texttt{rSN$^2$} for the ELG plates observed in the SGC with longer exposure time ($57656\le\texttt{MJD}\le57711$).
For each plate, the final reduction with all observed individual exposures is displayed as a filled square.
A given plate can have several reductions with different exposures subsets: in this case, those are displayed as empty circles and linked by a thin line of a given colour to the final reduction square.
Plates observed after having set the nominal \texttt{rSN$^2$} ($57712\le\texttt{MJD}\le57749$) are displayed as stars.
}
 	\label{fig:effvsRSN2}
 \end{figure}

% exposure time
The exposure time is dynamically adapted every 15 minutes at the telescope to obtain a homogeneous sample with a redshift success rate as constant as possible. 
After each exposure, the median squared signal-to-noise ratio in the red camera (\texttt{rSN$^2$}; masked from the sky OH lines) is measured: the observers then decide wether or not the observed plate requires more exposure to reach the nominal \texttt{rSN$^2$}.
During the first month of operations ($57656\le\texttt{MJD}\le57711$, 19 plates), the plates were exposed longer, so that those data can be used to optimise the exposure time.
To sample at most the explored parameters space, we re-reduce the data for some plates, including only a subsample of the individual exposures.
Figure \ref{fig:effvsRSN2} illustrates the obtained results.
From this plot it is clear that, when the \texttt{rSN$^2$} is low, the pipeline is unable to provide a reliable redshift at an efficient rate.
For example, the efficiency is only $\sim$50\% when a plate is exposed to a depth of $\texttt{rSN$^2$}=10$.
In addition, the efficiency of the pipeline in producing reliable redshifts begins to approach a plateau in the range of $20<\texttt{rSN$^2$}<30$.
For example, the efficiency only increases from 63\% at $\texttt{rSN$^2$}=20$ to 70\% at $\texttt{rSN$^2$}=30$.
Given that \texttt{rSN$^2$} scales linearly with exposure time, more tracers are classified by investing the time toward a new plate than to increase the signal by this amount.
Accounting for those measurements and for the expected available time to run the ELG program, we established a threshold of $\texttt{rSN$^2$}>22$ for all ELG exposures, which broadly corresponds to an average of 4.5 15 minutes exposures per plate.

% Figure: efficiency vs. RSN2

% zspec reliability
\subsection{$\zspec$ reliability \label{sec:zspecreliability}}

The results presented in this paper used version \texttt{v5\_10\_2}  of the \textsc{idlspec2d} data reduction pipeline to extract and flux-calibrate the ELG spectra.
This version of the eBOSS pipeline includes several improvements over version \texttt{v5\_9\_0} that was used to compute the spectra in DR13 \citep{sdss-collaboration16}, including direct corrections for differential atmospheric refraction as documented in \citet{jensen16} and a new methodology for co-addition of multiple exposures as documented in \citet{hutchinson16}. 
A nearly identical version of the pipeline will be used for all data taken through May 11, 2016 to be released as DR14.  Although these ELG spectra will not be in that public sample, they will be publicly released in 2019 with a new version of the data reduction pipeline that is expected to have a better sky subtraction algorithm.

Spectroscopic redshifts estimations are then a posteriori flagged reliable, using values of signal-to-noise ratio in emission lines (zQ flag) and in the continuum (zCont flag).
We have slightly refined the method discussed \citet{comparat16a} as follows.
The emission line flags are the same, but the continuum flag changed.

For the continuum flag, we consider the spectrum in $3500<\lambda [\textnormal{\AA}] / (1+\zspec)<3800$  and apply a 15 \AA~ \oii~ mask from $3726-15<\lambda [\textnormal{\AA}] / (1+\zspec)<3728+15$.
We set $z\textnormal{Cont}=0$ (1 and 2.5, respectively) if the ratio of the median flux observed to the standard deviation of the flux is in $\leq 3$ (in $]3,8]$ and $\geq 8$, respectively); in this, way the selection criterion of reliable redshifts with the flags remains simillar to the one defined in the Eq.(2) of \citet{comparat16a}:
\begin{equation}
\rm zQ\geq2\;  or \; (zQ\geq1\;  and\;  zCont>0)\;  or \; (zQ\geq0\;  and \;  zCont\geq2.5). 
\label{eq:zqzcont}
\end{equation}
This update was correlated with inspection of two plates that allow an estimation of the catastrophic assignment rate of $1.6\%$.

The \texttt{redmonster} method \citep{hutchinson16}, which currently provides eBOSS LRG redshift, will eventually be ported to ELG to improve the redshift efficiency, thanks to the use of more physically meaningful spectral templates.

% n(z)
\subsection{Redshift distribution and selection efficiency} \label{sec:nz}

\begin{figure}
 	\includegraphics[width=0.95\columnwidth]{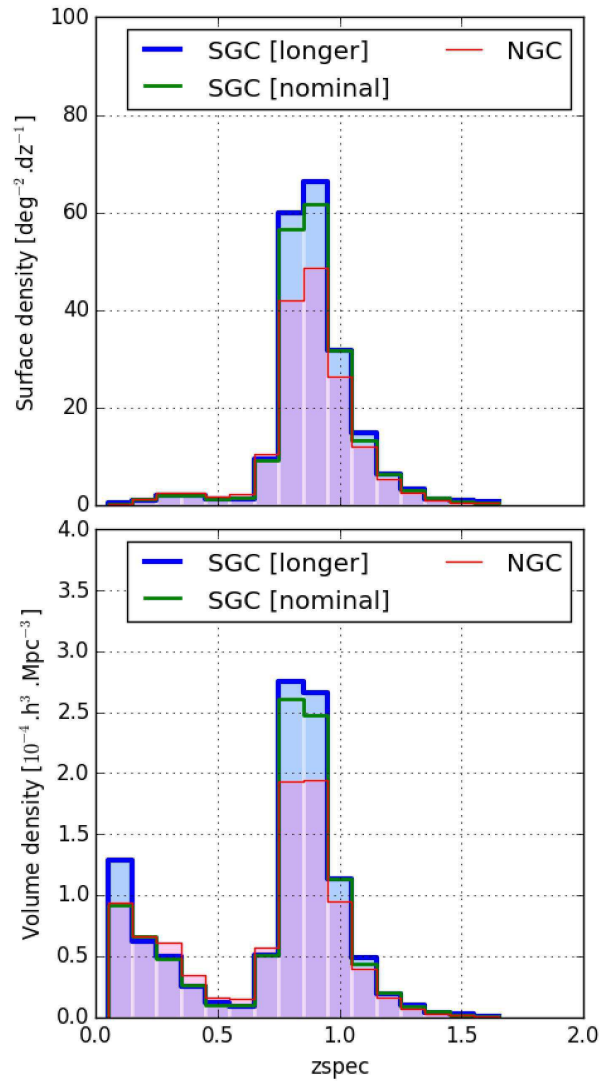}
 	\caption{ELG $n(z)$ for the ELG plates observed between $57656 \le \texttt{MJD} \le 57787$.
We consider only objects with a reliable $\zspec$, i.e. passing the Eq.(\ref{eq:zqzcont}).
For the SGC, we split the sample between the plates observed with $57656\le\texttt{MJD}\le57711$ (longer exposure times), and the plates observed with $57712\le\texttt{MJD}\le57749$ (nominal, shorter exposure times).
\textit{Top panel}: distribution per surface density.
\textit{Bottom panel}: distribution per volume density.
%; the solid and dashed lines represent the required density to overcome the shot noise for the predicted dark matter power spectrum, with a bias value of 1.0 and 1.4, respectively.
}
 	\label{fig:nz}
\end{figure}

Figure \ref{fig:nz} and Tables \ref{tab:nzperarea} and \ref{tab:nzpervolume} detail the redshift distribution $n(z)$ when considering only observed targets with a reliable $\zspec$ (Eq.\ref{eq:zqzcont}).
We compute the surface densities (Table \ref{tab:nzperarea}) correcting for the tiling incompleteness and using the plate covered areas (i.e. not accounting for the masked regions).
We normalise the total target density to the mean density over chunks \texttt{eboss21} and \texttt{eboss22} for the SGC, and over chunk \texttt{eboss23} for the NGC.
The volume densities (Table \ref{tab:nzpervolume}) are computed from the surface densities reported in Table \ref{tab:nzperarea}, assuming a 95\% tiling completeness, and our standard cosmology ($H_0=70$ km.s$^{-1}$.Mpc$^{-1}$, $\Omega_{\rm m}=0.30$, and $\Omega_\Lambda=0.70$).

For the SGC, we split the sample between plates observed with a longer exposure time ($57656\le\texttt{MJD}\le57711$, 19 plates) and plates observed with a nominal exposure time ($57712\le\texttt{MJD}\le57749$, 32 plates).
The former have better statistics than the latter, as less galaxies are rejected because of unreliable $\zspec$ measurement.
The sample with longer exposure times has 71.9\% of the observed targets with a reliable $\zspec$ with $0.7<\zspec<1.1$, versus 68.0\% for the sample with nominal exposure times.
The $57656\le\texttt{MJD}\le57711$ sample has 19.3\% of unreliable $\zspec$, whereas the $57712\le\texttt{MJD}\le57749$ sample has 15.2\% unreliable $\zspec$.

When comparing the $n(z)$ between the SGC and the NGC, we see that the two selections have overall a similar redshift distribution shape (median $\zspec$ of 0.84) and are efficient at removing stars and $\zspec<0.6$ objects.
However, the NGC sample has a lower efficiency, with 63.1\% of the observed targets with a reliable $\zspec$ with $0.7<\zspec<1.1$.

Though those efficiencies are lower than the goal of 74\%, there are several ways to increase them:
1) consider $0.6<\zspec<1.1$ instead of $0.7<\zspec<1.1$: this adds $\sim$5\% to the efficiency, without changing the percentage of catastrophic $\zspec$;
2) refine the redshift quality flags zQ and zCont: visual inspection has shown that about $\sim$50\% of the rejected objects with $0.7<\zspec<1.1$ have a reliable $\zspec$ \citep{comparat16a}: including these objects in the final sample would add$\sim$4\% to the efficiency;
3) improve the pipeline reduction (e.g., using \texttt{redmonster}, improving the sky subtraction), as we currently reject 15-20\% of the observations because of the $\zspec$ reliability criterion.

% Figure: n(z)

% Table: n(z) per area
\begin{table}
\centering
\begin{tabular}{lccc}
\hline\hline
Redshift 	& SGC & SGC & NGC\\
	 	& longer & nominal &  \\
\hline
Unreliable &  36.4	& 46.4	& 43.2\\
Star  &   1.2	& 1.2	& 1.4\\
$0.0 < z < 0.1$	& 0.3	& 0.2	& 0.3\\
$0.1 < z < 0.2$	& 1.1	& 1.1	& 1.1\\
$0.2 < z < 0.3$	& 2.1	& 2.0	& 2.6\\
$0.3 < z < 0.4$	& 1.9	& 1.9	& 2.6\\
$0.4 < z < 0.5$	& 1.3	& 1.1	& 1.7\\
$0.5 < z < 0.6$	& 1.3	& 1.4	& 2.2\\
$0.6 < z < 0.7$	& 9.3	& 9.2	& 10.3\\
$0.7<z<0.8$	& \textbf{59.8}	& \textbf{56.6}	& \textbf{42.0}\\
$0.8<z<0.9$	& \textbf{66.2}	& \textbf{61.6}	& \textbf{48.5}\\
$0.9<z<1.0$	& \textbf{31.6}	& \textbf{31.6}	& \textbf{26.3}\\
$1.0<z<1.1$	& \textbf{14.9}	& \textbf{13.4}	& \textbf{12.0}\\
$1.1 < z < 1.2$	& 6.3	& 6.4	& 5.4\\
$1.2 < z < 1.3$	& 3.2	& 2.9	& 2.5\\
$1.3 < z < 1.4$	& 1.3	& 1.5	& 0.9\\
$1.4 < z < 1.5$	& 0.9	& 0.7	& 0.4\\
$1.5 < z < 10.0$	& 0.7	& 0.6	& 0.5\\
\hline
Total targets  &   240.0	& 240.0	& 204.0\\
Total tracers  &   \textbf{172.6}	& \textbf{163.3}	& \textbf{128.8}\\
\hline
\end{tabular}
\caption{Measured surface density for the ELG plates (in deg$^{-2}$).
The $\zspec$ is considered as reliable if it passes the Eq.(\ref{eq:zqzcont}).
The densities are corrected for tiling incompleteness.
Densities for within the $0.7<\zspec<1.1$ are reported in bold, and their summed density is reported in the last line of the Table.
For the SGC, we split the sample between the plates observed with $57656\le\texttt{MJD}\le57711$ (longer exposure times), and the plates observed with $57712\le\texttt{MJD}\le57749$ (nominal, shorter exposure times).}
\label{tab:nzperarea}
\end{table}

% Cosmological forecast
\subsection{Cosmological forecast} \label{sec:cosmo_forecast}
\begin{figure}
	\includegraphics[width=0.95\columnwidth]{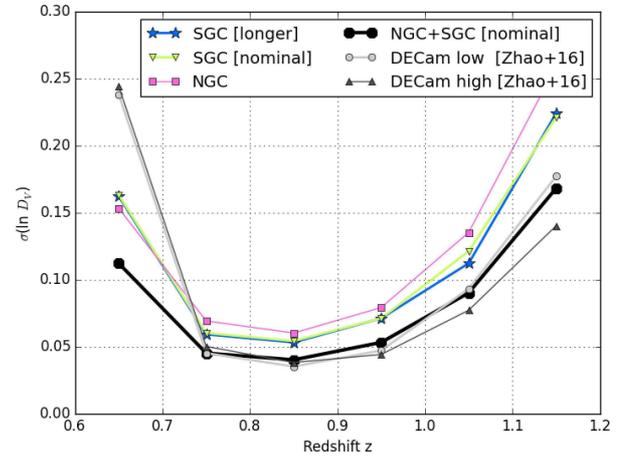}
 	\caption{Predicted $\sigma_{D_V}/D_V$, the 68\% confidence level error of the BAO distance $D_V$, based on the $n(z)$ of Table \ref{tab:nzpervolume}.
For the comparison, we report the values predicted in \citet{zhao16} for the two ELG selections used in that paper, DECam low and DECam high.}
 	\label{fig:cosmo_forecast}
\end{figure}

We compare here the cosmological forecast using the redshift distributions presented in Section \ref{sec:nz}, with that presented in \cite{zhao16}.
Except for the $n(z)$ and the areas, the forecast is performed using exactly the same methods and assumptions as in \cite{zhao16}:
the \textit{Planck} cosmology \citep{planck-collaboration16},
a bias of $b(z) = 1.0G(0)/G(z)$, where $G(z)$ is the linear growth factor at redshift $z$ ($b(z_{\rm eff}) \sim 1.5$ at the effective redshift $z_{\rm eff}=0.84$),
an assumption of 50\% sample reconstruction.
We present in Table \ref{tab:cosmo_forecast} and Figure \ref{fig:cosmo_forecast} the predicted precision of the BAO distance $D_V$.
The quantity $D_V$, defined as $D_V = [d_A^2 (z) cz H^{-1}(z)]^{1/3}$, is a combination of the angular diameter distance $d_A(z)$ and the Hubble parameter $H(z)$.
It is the best constrained distance with the BAO probe for an isotropic distribution of pairs, as it is composed of two transverse dimensions ($d_A^2 (z)$) and one line of sight dimension ($cz H^{-1}(z)$) \citep{eisenstein05}.

% Table: n(z) per volume
\begin{table}
\centering
\begin{tabular}{lccc}
\hline\hline
Redshift 	& SGC & SGC & NGC\\
	 	& longer & nominal &  \\
\hline
$0.0 < z < 0.1$	& 1.281	& 0.915	& 0.939\\
$0.1 < z < 0.2$	& 0.625	& 0.663	& 0.662\\
$0.2 < z < 0.3$	& 0.493	& 0.478	& 0.611\\
$0.3 < z < 0.4$	& 0.255	& 0.256	& 0.344\\
$0.4 < z < 0.5$	& 0.120	& 0.101	& 0.154\\
$0.5 < z < 0.6$	& 0.089	& 0.094	& 0.153\\
$0.6 < z < 0.7$	& 0.509	& 0.504	& 0.566\\
$0.7<z<0.8$	& \textbf{2.753}	& \textbf{2.604}	& \textbf{1.932}\\
$0.8<z<0.9$	& \textbf{2.657}	& \textbf{2.473}	& \textbf{1.947}\\
$0.9<z<1.0$	& \textbf{1.136}	& \textbf{1.137}	& \textbf{0.944}\\
$1.0<z<1.1$	& \textbf{0.489}	& \textbf{0.440}	& \textbf{0.394}\\
$1.1 < z < 1.2$	& 0.191	& 0.195	& 0.164\\
$1.2 < z < 1.3$	& 0.093	& 0.085	& 0.071\\
$1.3 < z < 1.4$	& 0.036	& 0.041	& 0.026\\
$1.4 < z < 1.5$	& 0.023	& 0.019	& 0.011\\
$1.5 < z < 10.0$	& 0.000	& 0.000	& 0.000\\
\hline
\end{tabular}
\caption{Measured volume densities for the ELG plates (in $10^{-4}{\,h^3{\rm Mpc}^{-3}}$).
The densities correspond to the densities per area reported in Table \ref{tab:nzperarea}, assuming a 95\% tiling completeness, and our standard cosmology ($H_0=70$ km.s$^{-1}$.Mpc$^{-1}$, $\Omega_{\rm m}=0.30$, and $\Omega_\Lambda=0.70$)
Densities within $0.7<\zspec<1.1$ are reported in bold.
For the SGC, we split the sample between the plates observed with $57656\le\texttt{MJD}\le57711$ (longer exposure times), and the plates observed with $57712\le\texttt{MJD}\le57749$ (nominal, shorter exposure times).}
\label{tab:nzpervolume}
\end{table}

First we observe that the predicted $\sigma_{D_V}/D_V$ is very similar for the two SGC cases (longer and nominal), and these differ little from the prediction for the NGC.
Figure \ref{fig:cosmo_forecast} shows that the predicted $\sigma_{D_V}/D_V$ for the combined SGC+NGC sample is in broad agreement with the values predicted in \citet{zhao16} for the two then tested ELG selections (DECam low over 1400 deg$^2$ and DECam high over 1100 deg$^2$).
The forecasts at the effective redshift confirm this:
in \citet{zhao16}, both ELG DECam selections provided a constraint of $\sigma_{D_V}/D_V = 0.022$ at the effective redshift;
for the values reported in this paper, the forecast is $\sigma_{D_V}/D_V = 0.023$ at the effective redshift for the combined SGC and NGC samples.
In addition, there is an improvement on the precision of BAO distance in the $0.6<z<0.7$ redshift bin.

% Table: cosmological forecast
\begin{table}
\center
\begin{tabular}{lcccc}
\hline\hline
Redshift 		& SGC 	& SGC 	& NGC	& NGC and SGC\\
			& longer	& nominal	&		& nominal\\
\hline
$0.6<z<0.7$	& 0.162	& 0.163	& 0.153	& 0.112\\
$0.7<z<0.8$	& 0.059	& 0.060	& 0.069	& 0.045\\
$0.8<z<0.9$	& 0.053	& 0.054	& 0.060	& 0.040\\
$0.9<z<1.0$	& 0.071	& 0.071	& 0.079	& 0.053\\
$1.0<z<1.1$	& 0.112	& 0.121	& 0.135	& 0.090\\
$1.1<z<1.2$	& 0.224	& 0.221	& 0.259	& 0.168\\
\hline
$z_{\rm eff}=0.84$ &	0.030	& 0.031	& 0.035	& 0.023\\
\hline
\end{tabular}
\caption{Predicted $\sigma_{D_V}/D_V$, the 68\% confidence level error on the BAO distance $D_V$, based on the $n(z)$ of Table \ref{tab:nzpervolume}.
For the SGC, we split the sample between the plates observed with $57656\le\texttt{MJD}\le57711$ (longer exposure times), and the plates observed with $57712\le\texttt{MJD}\le57749$ (nominal, shorter exposure times).
The assumed areas are 620 deg$^2$ for the SGC and 600 deg$^2$ for the NGC.
The last line displays the forecast at the effective redshift.}
\label{tab:cosmo_forecast}
\end{table}

% Figure: cosmological forecast

%=================================
% Mean properties of the ELGs observed in the SGC
%=================================
\section{Mean properties of the ELGs observed in the SGC} \label{sec:meanproperties}

In this section we characterise the spectral properties, the structural properties, and the stellar mass of the observed ELGs in the SGC.
This analysis is based on the first months of observations ($57656 \le \texttt{MJD} \le 57787$), i.e. on $\sim$43,000 observed ELGs, amongst which $\sim$30,000 have a reliable $\zspec$ estimation with $0.7<\zspec<1.1$.
This sample is already large enough to properly characterise the average properties of our ELGs.

% spectral properties
\subsection{Spectral properties from stacked spectra}

\begin{figure*}
 	\begin{tabular}{ccc}
	 	\multicolumn{3}{c}{\includegraphics[width=1\linewidth]{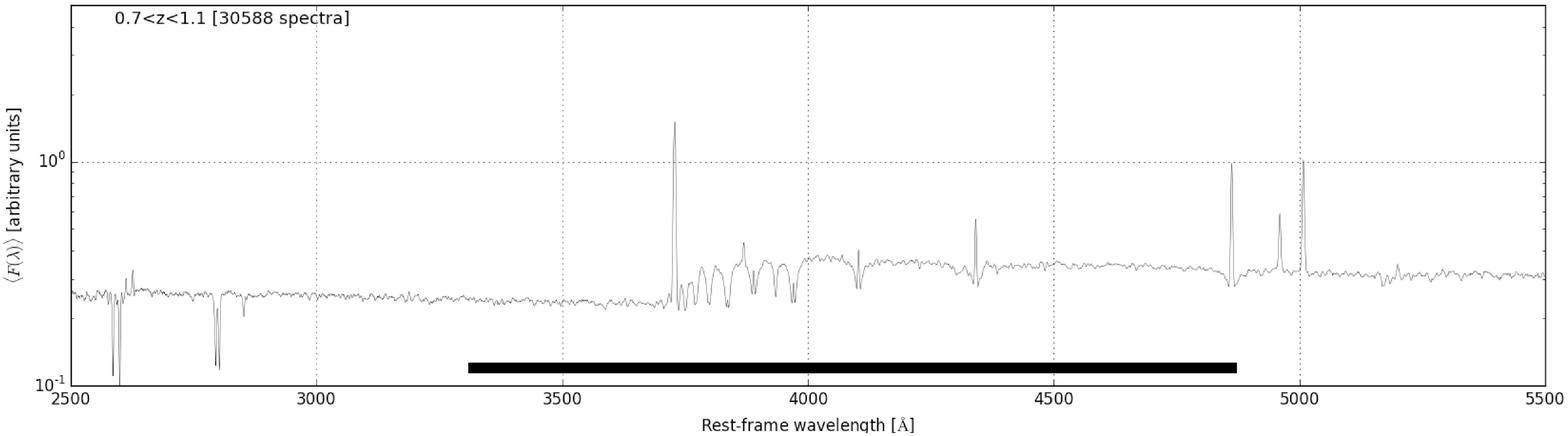}}\\
	 	\includegraphics[width=0.3\linewidth]{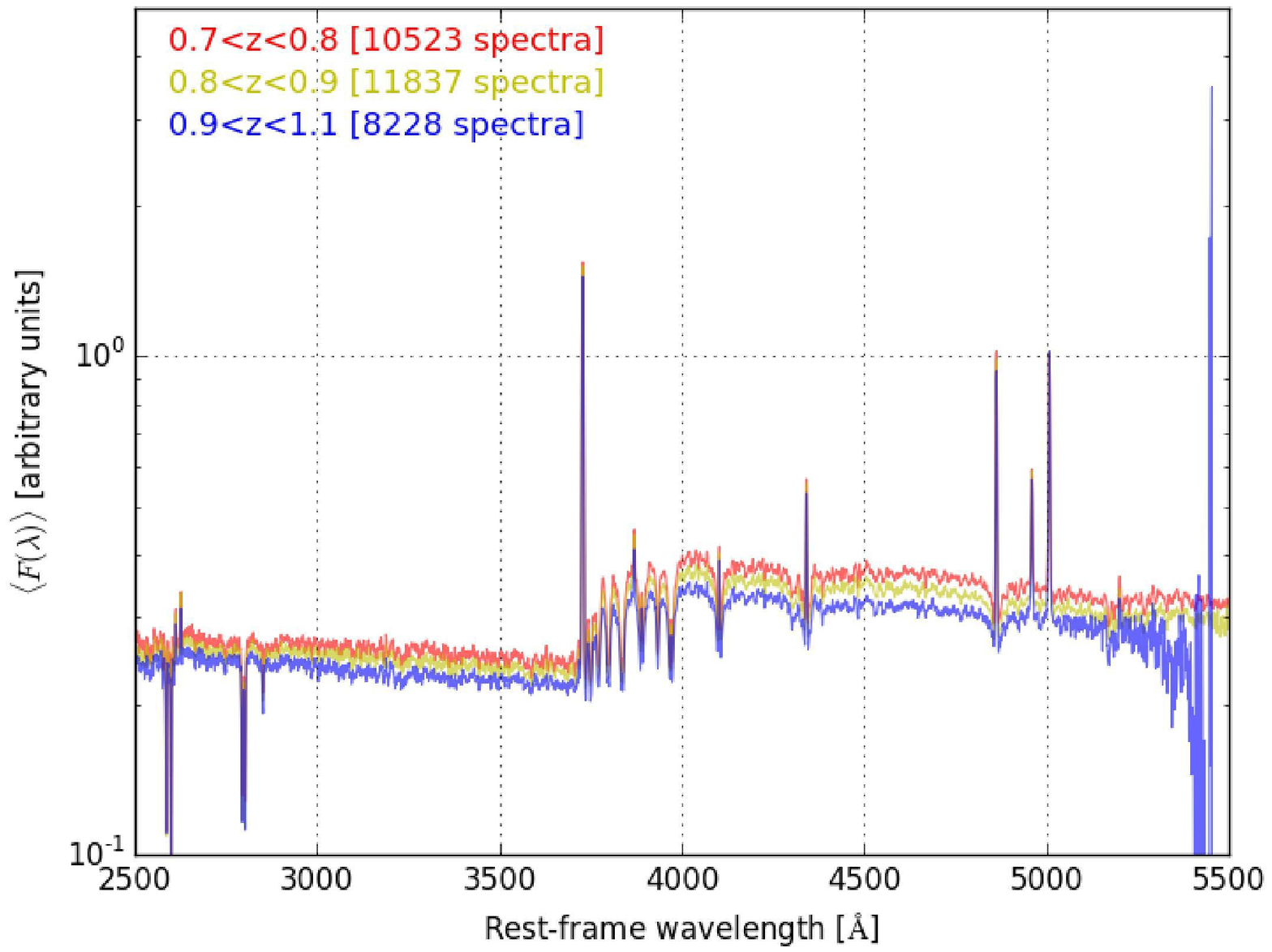}&
	 	\includegraphics[width=0.3\linewidth]{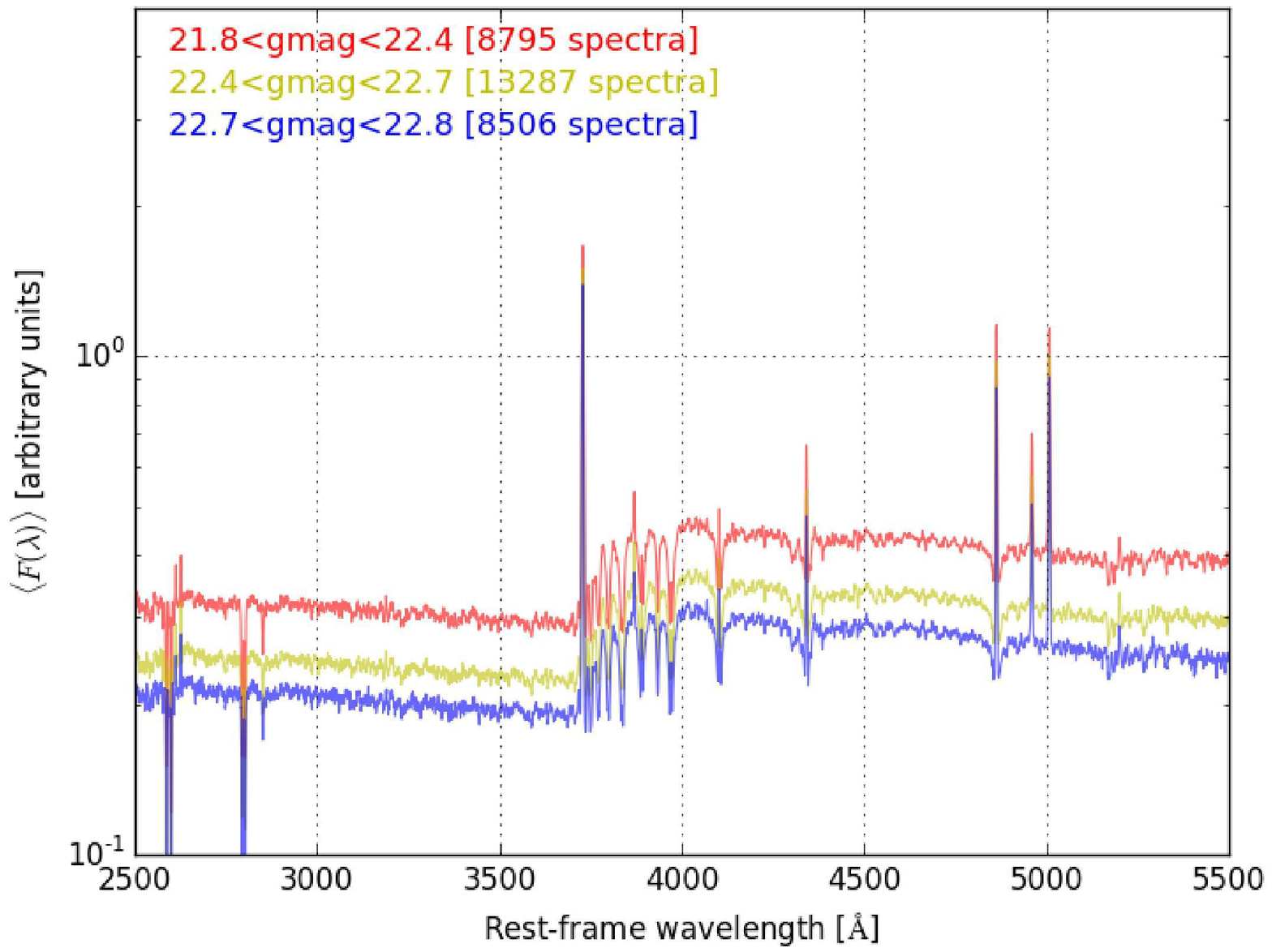}&
	 	\includegraphics[width=0.3\linewidth]{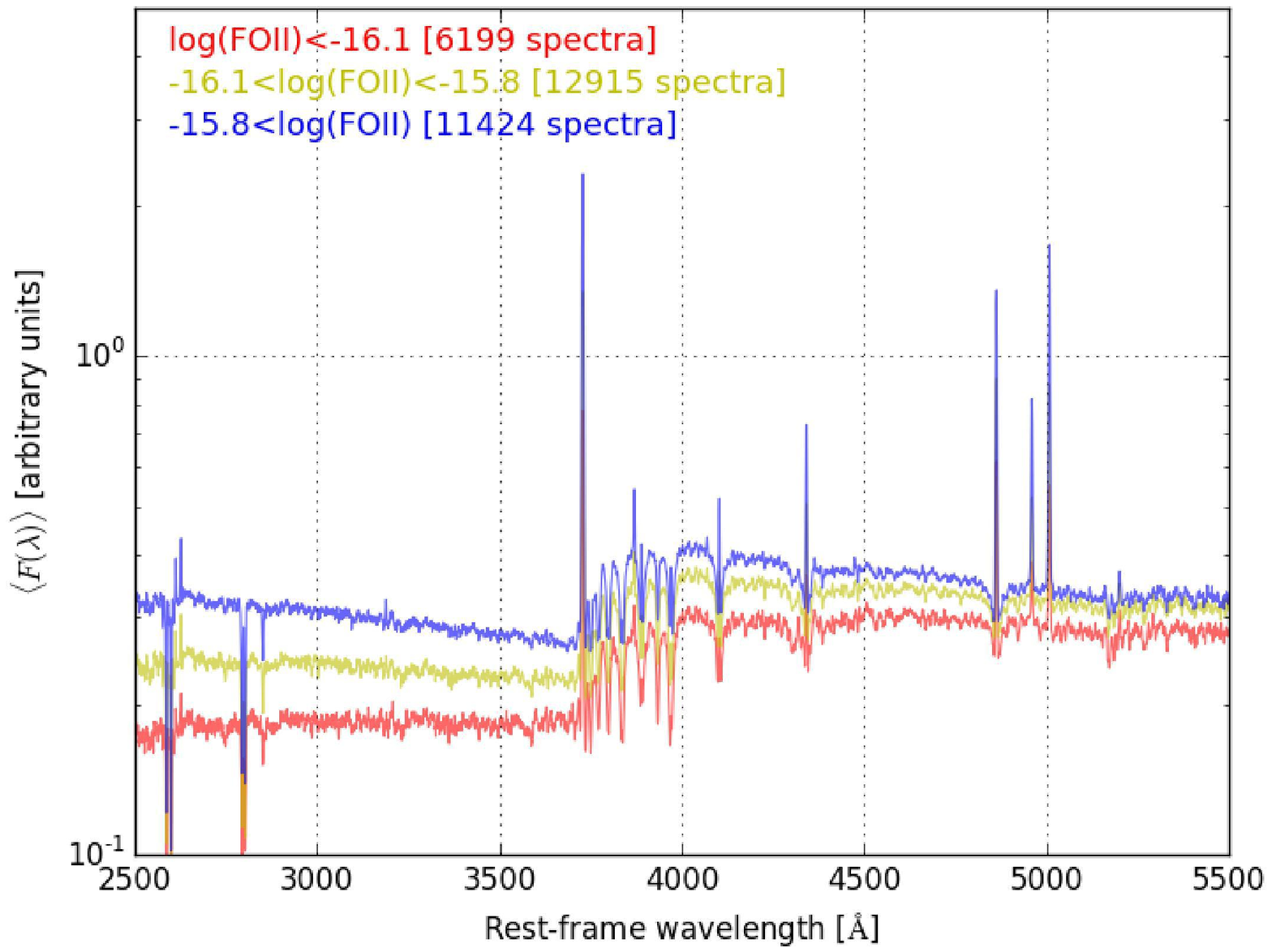}\\
 	\end{tabular}
 	\caption{Median-stacked spectra, built from the currently 51 ELG plates observed in the SGC.
\textit{Top panel}: stacked spectrum obtained using all the ELGs with a reliable $\zspec$ within $0.7<\zspec<1.1$; the thick black line shows the rest-frame wavelength probed by the CS82 $i$-band at $0.7<\zspec<1.1$.
\textit{Bottom panels}: stacked spectra from the same sample, when binning by $\zspec$ (\textit{left}), $g$-band magnitude (\textit{middle}), and $\textnormal{log}_{10}(F_\textnormal{\oii})$ (\textit{right}).}
 	\label{fig:stackspec}
\end{figure*}

The spectral properties are studied through stacked spectra, in order to maximise the signal-to-noise ratio of the typical features.
Indeed, the average signal-to-noise ratio per pixel in the continuum region of the individual ELG spectra is low \citep[see Figure 3 of][]{raichoor16}.
The stacking is done as follows: the spectra are shifted to the rest-frame and then median-stacked.
We display in Figure \ref{fig:stackspec} the stacked spectrum obtained using all the ELGs with a reliable $0.7<\zspec<1.1$ (top panel) and when stacking by redshift bins, $g$-band magnitude bins, and \oii~ flux bins (bottom panels).

The full stacked spectrum (top panel) nicely displays the features typical from star-forming galaxies \citep[e.g.,][]{kennicutt92,moustakas06}.
Those are detailed in \citet{zhu15}, which presents a similar stacked spectrum, though at a lower signal-to-noise ratio, as built from three times less individual spectra.
When looking at the bottom panels, on the one hand, the spectra stacked by redshift bins are similar between them -- the only difference being that the median flux is slightly decreasing with increasing redshift -- indicating that we select ELGs with similar properties through the $0.7<\zspec<1.1$ range.
The same conclusion holds when stacking by $g$-band magnitude bins.
On the other hand, when stacking by \oii~ flux bins, we see differences in the shape of the spectra, as expected: the stellar absorption features around the 4000 \AA~ break are stronger and the near-ultraviolet flux is weaker for low \oii~ emitters.

% Figure: spectra stacks

% structural properties
\subsection{Structural properties from the CS82 imaging}

We here take advantage of the CFHT/MegaCam Stripe 82 Survey survey (CS82; P.I.: J.-P. Kneib; Erben et al., in prep.), which is an $i$-band imaging survey overlapping our ELG SGC footprint (see Figure \ref{fig:sgccompleteness}).
The rest-frame wavelength probed by the $i$-band at redshift 0.7 (0.8,0.9, and 1.1, respectively) is $\sim$4400 \AA~ ($\sim$4150 \AA, $\sim$3950 \AA, and $\sim$3600 \AA, respectively; see Figure \ref{fig:stackspec}).
CS82 covers $\sim$170 deg$^2$ down to $i \sim 24.0$ mag (5$\sigma$ limiting magnitude in a 2\arcsec diameter aperture); its depth and excellent seeing (median value of 0.6\arcsec) makes this dataset ideal to properly estimate the structural properties of the ELGs.
There are $\sim$5000 observed ELGs with a reliable $0.7<\zspec<1.1$ covered by CS82.
We use images processed in a similar fashion to the CFHTLenS dataset \citep{erben13}.

For each ELG, we cut out a 40\arcsec$\times$40\arcsec~ stamp image, masked neighbouring objects, and used the \textsc{Galfit} software \citep[v3.0.5,][]{peng10a} to fit the surface brightness distribution with a \citet{sersic68} profile $I(r) = I_e \times \exp\{-\kappa [(r/r_e)^{1/\nser}-1]\}$, where $I(r)$ is the surface brightness at $r$, and $I_e$ is the surface brightness at the effective radius $r_e$, which is the radius which encloses half of the emitted light.
The S\'{e}rsic index $\nser$ translates the shape of the profile, with a higher value corresponding to a profile more peaked at the centre and with larger wings: $\nser=4$ corresponds to a \citet{de-vaucouleurs48} profile, which is typical of early-type galaxies, while $\nser=1$ corresponds to an exponential profile, typical of late-type galaxies.
\citet{wuyts11a} have shown that, for $0.1 < \zspec < 2.5$, typical passive galaxies have $\nser \sim 4$, while typical star-forming galaxies have $\nser \sim 1$-2.

The initial parameter values (position, magnitude, size axis ratio, position angle) are set to the SExtractor \citep{bertin96} measurements.
The sky is fixed during the fit, to the median value of the (masked) stamp.
For each CFHT pointing, we use $\sim$50 bright, unsaturated ($18 \le i_{\rm AB} \le 21$) SDSS spectroscopic stars \citep{alam15} to create a PSF stamp.
The DECaLS pipeline also provides some size estimation, with fitting a $\nser=1$ and a $\nser=4$ profile.
When restricting to objects classified as exponential by DECaLS and with $|\nser-1|<0.3$ in our measurements, the estimated sizes are in agreement (median of 0.01\arcsec for $\sim$1000 objects); there are not enough objects classified as de Vaucouleurs by DECaLS and with $|\nser-4|<0.3$ in our measurements to do a proper comparison ($<$20 objects).

% stellar mass from SED fitting
\subsection{Stellar masses}
% WISE data
Additionally, we take advantage of the availability of near-infrared photometry in the DECaLS catalogues to estimate the stellar mass of the ELGs.
The DECaLS pipeline also processes the Wide-field Infrared Survey Explorer data \citep[\textit{WISE,}][]{wright10}.
\textit{WISE} -- and its extension \textit{NEOWISE} \citep{mainzer11} -- is an all-sky survey in four bands centered at 3.4, 4.6, 12 and 22 $\mu$m (W1, W2, W3 and W4).
The DECaLS team re-processed the \textit{WISE} images \citep{meisner17} and use them to do forced-photometry \citep{lang16}: the profiles corresponding to the sources detected in the DECam imaging are convolved with the \textit{WISE} PSF and the flux is fitted to the \textit{WISE} imaging.
This results in accurate colour measurements, with accounting for the PSF difference ($\sim$1.5\arcsec for DECam and $\sim$6\arcsec for \textit{WISE}).

% SED fit
We build the spectral energy distribution of the ELGs from the DECam/$grz$- and \textit{WISE}/$W1W2$-bands, and fit it with the \texttt{FAST} code \citep{kriek09}.
We use the default settings: \citet{bruzual03} stellar population models with solar metallicity and a \citet{chabrier03} Initial Mass Function (IMF) to build delayed exponentially declining star-formation history models ($SFR(t) \propto t \times e^{-t/\tau}$, with $8.5 < $log$_{10}(\tau/\textnormal{Gyr})<10$).
During the fit, the redshift is fixed at $\zspec$ and dust is allowed to follow the \citet{kriek13} law.
The model is fitted to the data through a $\chi^2$ minimisation.

% validation with COSMOS Laigle+16
To validate our stellar mass estimate, we use the COSMOS2015 catalogue \citep{laigle16}.
\citet{laigle16} use extremely accurate photometric redshifts and very deep optical and near-infrared imaging in more than 30 bands to estimate the stellar masses with \citet{bruzual03} models with a \citet{chabrier03} IMF, resulting in very well constrained stellar masses.
We select the $\sim$500 galaxies passing our ELG/SGC cuts in the $\sim$2deg$^2$ of the COSMOS field, fix the redshift to the photometric redshift, and apply our fitting procedure.
For the $\sim$300 galaxies with $0.7<z_{\rm phot}<1.1$, we find a difference of 0.05$\pm$0.21 dex between our stellar masses and those from \citet{laigle16}, showing good agreement, given that the photometry is different and the settings are not exactly similar.

% properties overview
\subsection{Properties overview}

\begin{figure*}
	 	\includegraphics[width=1.1\linewidth]{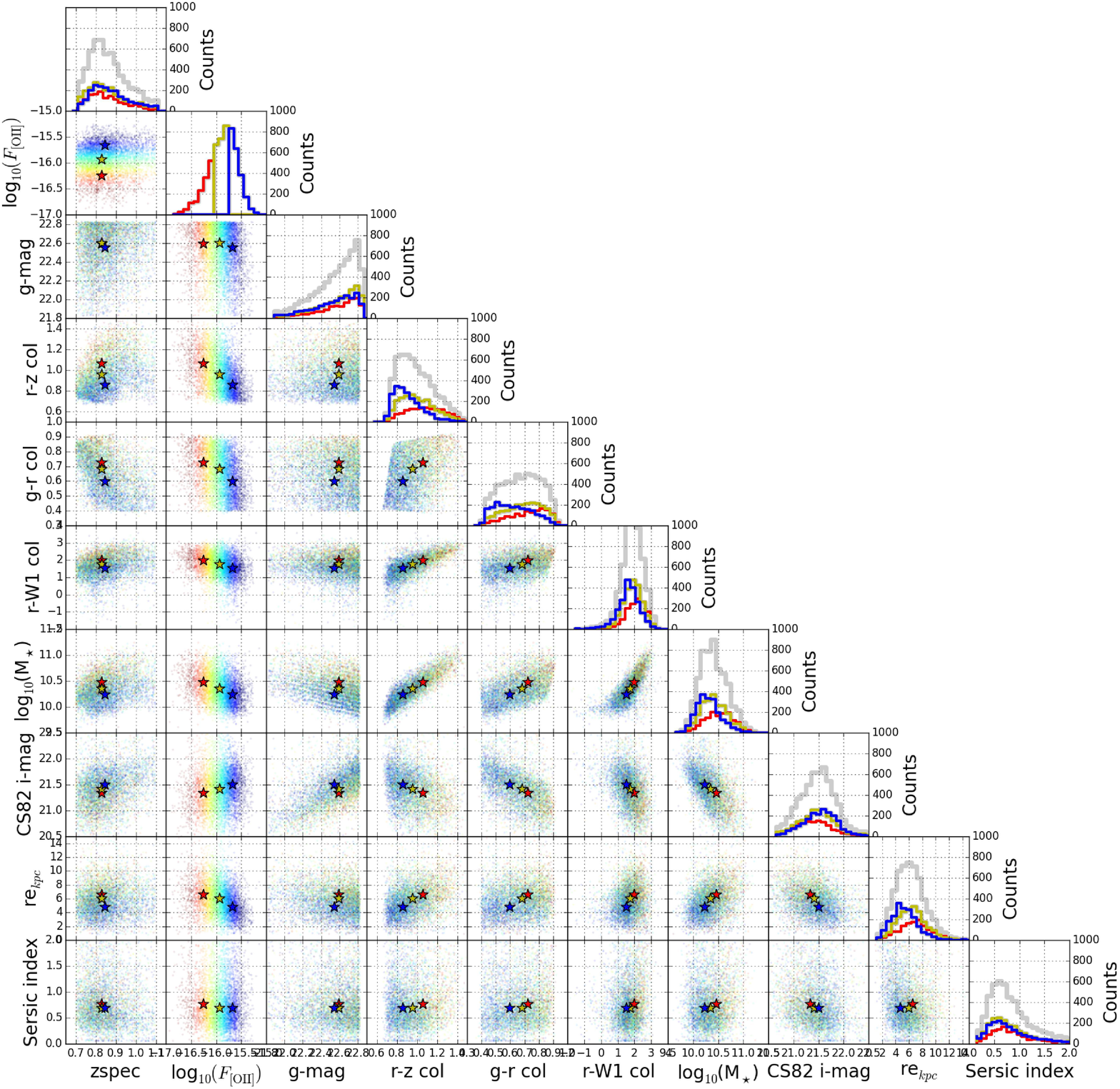}
 	\caption{Matrix plot comparing the various properties of our ELG SGC sample, using objects covered by CS82, with a reliable $\zspec$ and $0.7<\zspec<1.1$ cover ($\sim$5000 objects).
We plot the histogram for this sample (gray), and when splitting it by \oii~ flux bins (red: $\textnormal{log}_{10}(F_\textnormal{\oii})<-16.1$, yellow: $-16.1<\textnormal{log}_{10}(F_\textnormal{\oii})<-15.8$, and blue: $\textnormal{log}_{10}(F_\textnormal{\oii})>-15.8$).
The red, yellow, and blue stars in the scatter plots represent the median values for those bins in \oii~ flux.
In the scatter plots, the individual $\textnormal{log}_{10}(F_\textnormal{\oii})$ is colour-coded, with a colour scale going from -16.5 (red) to -15.5 (blue).}
 	\label{fig:matrixplot}
\end{figure*}

We present an overview of the properties of our ELG sample: photometric, spectroscopic, structural properties, and stellar masses.
We only consider here galaxies with a reliable $\zspec$ within $0.7<\zspec<1.1$ and covered by the CS82 imaging.
The matrix plot in Figure \ref{fig:matrixplot} compares the following quantities: $\zspec$,  log$_{10}(F_\textnormal{\oii})$, $g$-band magnitude, $r-z$, $g-r$, and $r-W1$ colours, the stellar mass log$_{10}(M_\star)$, the CS82 $i$-band magnitude, $r_e$, and $\nser$.
As the log$_{10}(F_\textnormal{\oii})$ is a key quantity for our ELG sample, we also split the data in bins of \oii.
We comment below only some of the information that can be read in this plot.
Additionaly, we provide in Table \ref{tab:meanprop} the median and standard deviation for those quantities; we also provide those numbers when splitting this sample by bins of $\zspec$, $g$-band magnitude, and  log$_{10}$(F\oii).

% gr/rz vs. zpsec/FOII
As already presented above (Figure \ref{fig:grz}), the $r-z$ and $g-r$ colours are correlated with $\zspec$ and log$_{10}(F_\textnormal{\oii})$: this motivated the use of the $grz$-diagram for the ELG selection.
% gmag/imag vs. gr/rz
By construction, as our sample is selected in $g$-band, the $g$-band magnitude is not correlated with the $r-z$ and $g-r$ colours; however a natural outcome of this construction is that the $i$-band magnitude is correlated with these colours, in that ELGs blue in $r-z$ or $g-r$ colours are faint in the $i$-band.
% imag vs. FOII
As a consequence, the $i$-band magnitude is anticorrelated with log$_{10}(F_\textnormal{\oii})$.
% imag vs re
Regarding the structural parameters: our ELGs have on average sizes $r_e = 5.6 \pm 1.93$ and $n_{\rm ser} = 0.7 \pm 1.0$.
The measured Sersic indexes are typical for star-forming galaxies.
We also note a strong correlation between the $i$-band magnitude and the size $r_e$ (fainter objects are smaller).
We also see a trend that strong \oii~ emitters have a smaller $\nser$, however it is difficult to exclude that this is not a consequence of the aforementioned correlations.
% stellar mass
The stellar mass displays some tight correlations with the $r-z$ and $r-W1$ colours.
More massive galaxies will tend to be older, thus having a redder $r-z$ colour, which brackets the 4000 \AA~break; they will also tend to form less stars, as can be seen from the mild correlation between stellar mass and \oii~ flux.
The tight correlation with the $r-W1$ colour is expected, as the \textit{WISE} $W1$ filter probes the rest-frame $H$-band, dominated by emission from the low-mass stars constituting the bulk of a galaxy stellar mass.

% FOII summary
To summarise from the \oii~ emission point of view, in our ELG sample, the strong \oii~ emitters have on average a slightly higher $\zspec$, bluer $r-z$ and $g-r$ colours, smaller stellar masses log$_{10}(M_\star)$, fainter $i$-band magnitudes, smaller sizes, slightly lower $\nser$, and lower stellar masses.

% Table: mean properties
\begin{table*}
	\resizebox{\textwidth}{!}{%
	\begin{tabular}{lccccccccccc}
		\hline
		\hline
		Sample & N	& $\zspec$  & log$_{10}$(F\oii) &	$g$ mag	&  $g-r$ & $r-z$  & $r-W1$ & log$_{10}(M_\star)$ & CS82 $i$ mag	& $r_{\rm e}$	&$n_{\rm sersic}$\\
		        	    & 		& 	& [log$_{10}$(erg.s$^{-1}$.cm$^{-2}$)] &   [mag]	&    [mag]		& [mag]	&    [mag]	& [log$_{10}(M_\odot)$]	& [mag]	&[kpc]		& \\
\hline
All & 5235 & $0.83\pm0.09$ & $-15.88\pm0.27$ & $22.58\pm0.23$ & $0.66\pm0.13$ & $0.94\pm0.17$ & $1.69\pm0.66$ & $10.33\pm0.26$ & $21.44\pm0.38$ & $5.56\pm1.93$ & $0.71\pm1.03$\\
\hline
$0.7<\zspec<0.8$ & 1816 & $0.76\pm0.03$ & $-15.91\pm0.26$ & $22.55\pm0.25$ & $0.74\pm0.11$ & $0.86\pm0.13$ & $1.55\pm0.59$ & $10.26\pm0.22$ & $21.34\pm0.36$ & $5.34\pm1.81$ & $0.74\pm0.98$\\
$0.8<\zspec<0.9$ & 2053 & $0.84\pm0.03$ & $-15.88\pm0.28$ & $22.58\pm0.23$ & $0.64\pm0.13$ & $0.95\pm0.17$ & $1.72\pm0.68$ & $10.33\pm0.26$ & $21.43\pm0.37$ & $5.61\pm1.93$ & $0.69\pm0.93$\\
$0.9<\zspec<1.1$ & 1366 & $0.97\pm0.05$ & $-15.85\pm0.28$ & $22.61\pm0.22$ & $0.57\pm0.12$ & $1.04\pm0.18$ & $1.93\pm0.65$ & $10.45\pm0.27$ & $21.60\pm0.39$ & $5.81\pm2.05$ & $0.70\pm1.22$\\
\hline
$21.825<g\textnormal{-mag}<22.4$ & 1404 & $0.82\pm0.09$ & $-15.84\pm0.29$ & $22.23\pm0.15$ & $0.68\pm0.13$ & $0.92\pm0.17$ & $1.75\pm0.60$ & $10.45\pm0.25$ & $21.10\pm0.37$ & $5.96\pm1.97$ & $0.76\pm1.23$\\
$22.4<g\textnormal{-mag}<22.7$ & 2320 & $0.84\pm0.09$ & $-15.88\pm0.27$ & $22.57\pm0.09$ & $0.66\pm0.13$ & $0.93\pm0.17$ & $1.67\pm0.66$ & $10.31\pm0.25$ & $21.45\pm0.33$ & $5.51\pm1.91$ & $0.70\pm0.95$\\
$22.7<g\textnormal{-mag}<22.825$ & 1511 & $0.84\pm0.10$ & $-15.92\pm0.26$ & $22.77\pm0.04$ & $0.65\pm0.13$ & $0.94\pm0.18$ & $1.66\pm0.70$ & $10.23\pm0.25$ & $21.63\pm0.30$ & $5.29\pm1.86$ & $0.69\pm0.94$\\
\hline
$\textnormal{log}_{10}(F_\textnormal{\oii})<-16.1$ & 1222 & $0.82\pm0.09$ & $-16.24\pm0.17$ & $22.60\pm0.22$ & $0.73\pm0.13$ & $1.06\pm0.18$ & $1.98\pm0.60$ & $10.46\pm0.26$ & $21.36\pm0.40$ & $6.31\pm1.95$ & $0.76\pm1.12$\\
$-16.1<\textnormal{log}_{10}(F_\textnormal{\oii})<-15.8$ & 2045 & $0.83\pm0.09$ & $-15.93\pm0.09$ & $22.60\pm0.23$ & $0.69\pm0.13$ & $0.96\pm0.17$ & $1.74\pm0.63$ & $10.35\pm0.25$ & $21.42\pm0.38$ & $5.91\pm1.86$ & $0.69\pm0.96$\\
$-15.8<\textnormal{log}_{10}(F_\textnormal{\oii})$ & 1968 & $0.84\pm0.09$ & $-15.67\pm0.12$ & $22.55\pm0.24$ & $0.60\pm0.13$ & $0.86\pm0.14$ & $1.52\pm0.67$ & $10.24\pm0.24$ & $21.50\pm0.37$ & $4.73\pm1.76$ & $0.70\pm1.05$\\
\hline
	\end{tabular}}
	\caption{Mean properties for the ELG SGC sample.
All samples are subsamples from the sample which will be used for cosmology (i.e. $0.7<\zspec<1.1$ with a reliable $\zspec$ measurement) overlapping the CS82 imaging survey.
We report the median values and the standard deviation.}
	\label{tab:meanprop}
\end{table*}

% Figure: matrix plot

%=================================
% Conclusion
%=================================
\section{Conclusion} \label{sec:conclusion}
We have presented the final target selection and first observations of the eBOSS/ELG program.
% ELG program
This program will observe $\sim$255,000 ELGs over $\sim$1220 deg$^2$ with the goal of measuring the BAO scale at $z \sim 0.85$ with $a \sim$2\% precision.
Three hundred plates are dedicated to this program, split between the SGC and the NGC, the observation of which will last two years.
Observations started in 2016, September, with 51 plates observed in the SGC and 8 plates in the NGC ($57656 \le \texttt{MJD} \le 57787$) to date.

% ELG requirements
According to previous experience with BOSS and cosmological forecast, the target selection should fulfill the following criteria: 
(1) a surface density $>$170 deg$^{-2}$;
(2) an absolute variation in expected density $<$15\% with respect to imaging depth, Galactic extinction, and stellar density;
(3) an absolute variation in expected density $<$15\% with respect to the estimated uncertainties in the imaging zeropoint;
(4) reliable $\zspec$ measurements, i.e. with a precision better than 300 km.s$^{-1}$;
(5) an ELG sample used for cosmology at $z \sim 0.85$ $>$190,000, i.e. $>$74\% of the observed targets with a reliable $\zspec$ measurement with $0.7<\zspec<1.1$;
(6) $<$1\% of this sample with a catastrophic $\zspec$ measurement (redshift error exceeding 1000 km.s$^{-1}$).

% ELG TS summary
The ELG target selection is based on the DECaLS $grz$-band imaging, requiring: clean photometry,  favouring \oii~ emitters through a cut in the $g$-band magnitude,  and selecting galaxies in the desired redshift range, through a box cut in the $grz$-space.
The ELG footprint is split in two parts, one of 620 deg$^2$ over the SGC and one of 600 deg$^2$ over the NGC.
To take advantage of the deeper DES observations over the SGC, the $grz$-selection box is larger for the SGC, resulting in a target density higher than in the NGC (240 deg$^{-2}$ vs. 200 deg$^{-2}$).

% we pass the requirements...
%% density variation
We led a thorough analysis on the density variation over the footprint, similar to that presented in \citet{delubac17}, and demonstrated that the ELG target selection passes the requirements on the target density variations.
%% catastrophic zspec
We use dedicated $\zspec$ reliability flags ($z$Q, $z$Cont): visual inspections on two plates reduced with the latest pipeline version show that those flags secure a catastrophic failure rate of 1.4\%$\pm$0.5\%.
%% efficiency
Results from the 51 plates already observed in the SGC provide a median $\zspec$ of 0.84 and an efficiency of 68.0\%-71.9\%.
Results from the 8 plates already observed in the NGC provide a median $\zspec$ of 0.84 and an efficiency of 63.1\%.
Overall, the target selection reasonably passes all the requirements, though the efficiency is slightly lower than expected.
The efficiency can be increased by including the $0.6<\zspec<0.7$ redshift bin in the cosmological sample and by pipeline improvements.
% cosmo forecast
The cosmological forecast based on those first months measurements provide $\sigma_{D_V}/D_V = 0.023$, in agreement with the forecast in \citet{zhao16}, which were performed before the start of the ELG observations.

% ELG properties
Lastly, thanks to the current spectroscopic observations, completed with SED fitting using additional near-infrared photometry from the \textit{WISE} satellite and with excellent seeing imaging coverage from the CS82 survey, we have presented a detailed view of the average properties (photometric, spectroscopic, structural properties and stellar masses) of the ELG sample in the SGC.
The typical ELG in the SGC has a stellar mass of log$(M_\star/M_\odot)=10.33 \pm 0.26$, a size of $r_e = 5.6 \pm 1.93$ kpc, and a Sersic index of $n_{\rm ser} = 0.7 \pm 1.0$.
These present the typical features of star-forming galaxies, as seen in a composite spectrum stacking $\sim$30,000 ELG spectra, or with the 2D luminosity profile (low Sersic index).
These observations will be useful in the production of realistic mocks necessary for the cosmological anaysis, and also illustrate the legacy of such a sample for the galaxy evolution studies.
For instance, planned future work includes fitting the stacked spectra with stellar population models, in order to estimate precisely the average properties of those ELGs, including mass, age, and star formation history.

% catalog public release
The target catalogue over the SGC footprint will be publicly released in mid-2017, as a Value-Added Catalog from the SDSS DR14 release (\href{http://www.sdss.org/dr14/data\_access/vac/}{http://www.sdss.org/dr14/data\_access/vac/}).

% experience for future surveys
This ELG selection is paving the way for the future large BAO surveys, such as DESI and 4MOST, in which ELGs will constitue a significant part of the targets.

%=================================
% Bibliography
%=================================
\bibliographystyle{mnras}
\bibliography{ms}

\begin{thebibliography}{}
\makeatletter
\relax
\def\mn@urlcharsother{\let\do\@makeother \do\$\do\&\do\#\do\^\do\_\do\%\do\~}
\def\mn@doi{\begingroup\mn@urlcharsother \@ifnextchar [ {\mn@doi@}
  {\mn@doi@[]}}
\def\mn@doi@[#1]#2{\def\@tempa{#1}\ifx\@tempa\@empty \href
  {http://dx.doi.org/#2} {doi:#2}\else \href {http://dx.doi.org/#2} {#1}\fi
  \endgroup}
\def\mn@eprint#1#2{\mn@eprint@#1:#2::\@nil}
\def\mn@eprint@arXiv#1{\href {http://arxiv.org/abs/#1} {{\tt arXiv:#1}}}
\def\mn@eprint@dblp#1{\href {http://dblp.uni-trier.de/rec/bibtex/#1.xml}
  {dblp:#1}}
\def\mn@eprint@#1:#2:#3:#4\@nil{\def\@tempa {#1}\def\@tempb {#2}\def\@tempc
  {#3}\ifx \@tempc \@empty \let \@tempc \@tempb \let \@tempb \@tempa \fi \ifx
  \@tempb \@empty \def\@tempb {arXiv}\fi \@ifundefined
  {mn@eprint@\@tempb}{\@tempb:\@tempc}{\expandafter \expandafter \csname
  mn@eprint@\@tempb\endcsname \expandafter{\@tempc}}}

\bibitem[\protect\citeauthoryear{{Abazajian} et~al.,}{{Abazajian}
  et~al.}{2009}]{abazajian09}
{Abazajian} K.~N.,  et~al., 2009, \mn@doi [\apjs]
  {10.1088/0067-0049/182/2/543}, \href
  {http://adsabs.harvard.edu/abs/2009ApJS..182..543A} {182, 543}

\bibitem[\protect\citeauthoryear{{Alam} et~al.,}{{Alam} et~al.}{2015}]{alam15}
{Alam} S.,  et~al., 2015, \mn@doi [\apjs] {10.1088/0067-0049/219/1/12}, \href
  {http://adsabs.harvard.edu/abs/2015ApJS..219...12A} {219, 12}

\bibitem[\protect\citeauthoryear{{Bertin} \& {Arnouts}}{{Bertin} \&
  {Arnouts}}{1996}]{bertin96}
{Bertin} E.,  {Arnouts} S.,  1996, \aaps, \href
  {http://adsabs.harvard.edu/abs/1996A%26AS..117..393B} {117, 393}

\bibitem[\protect\citeauthoryear{{Blanton} et~al.,}{{Blanton}
  et~al.}{tted}]{blanton17}
{Blanton} M.~R.,  et~al., 2017, AJ submitted, preprint, \href
  {http://adsabs.harvard.edu/cgi-bin/bib_query?arXiv:1703.00052} {} (\mn@eprint
  {arXiv} {1703.00052v1})

\bibitem[\protect\citeauthoryear{{Bruzual} \& {Charlot}}{{Bruzual} \&
  {Charlot}}{2003}]{bruzual03}
{Bruzual} G.,  {Charlot} S.,  2003, \mn@doi [\mnras]
  {10.1046/j.1365-8711.2003.06897.x}, \href
  {http://adsabs.harvard.edu/abs/2003MNRAS.344.1000B} {344, 1000}

\bibitem[\protect\citeauthoryear{{Chabrier}}{{Chabrier}}{2003}]{chabrier03}
{Chabrier} G.,  2003, \mn@doi [\pasp] {10.1086/376392}, \href
  {http://adsabs.harvard.edu/abs/2003PASP..115..763C} {115, 763}

\bibitem[\protect\citeauthoryear{{Cole} et~al.,}{{Cole} et~al.}{2005}]{cole05}
{Cole} S.,  et~al., 2005, \mn@doi [\mnras] {10.1111/j.1365-2966.2005.09318.x},
  \href {http://adsabs.harvard.edu/abs/2005MNRAS.362..505C} {362, 505}

\bibitem[\protect\citeauthoryear{{Colless} et~al.,}{{Colless}
  et~al.}{2003}]{colless03}
{Colless} M.,  et~al., 2003, ArXiv Astrophysics e-prints, \href
  {http://adsabs.harvard.edu/abs/2003astro.ph..6581C} {}

\bibitem[\protect\citeauthoryear{{Comparat} et~al.,}{{Comparat}
  et~al.}{2013}]{comparat13}
{Comparat} J.,  et~al., 2013, \mn@doi [\mnras] {10.1093/mnras/sts127}, \href
  {http://adsabs.harvard.edu/abs/2013MNRAS.428.1498C} {428, 1498}

\bibitem[\protect\citeauthoryear{{Comparat} et~al.,}{{Comparat}
  et~al.}{2015}]{comparat15}
{Comparat} J.,  et~al., 2015, \aap, \href
  {http://adsabs.harvard.edu/abs/2015arXiv150905045C} {submitted
  [arxiv:1509.05045]}

\bibitem[\protect\citeauthoryear{{Comparat} et~al.,}{{Comparat}
  et~al.}{2016}]{comparat16a}
{Comparat} J.,  et~al., 2016, \mn@doi [\aap] {10.1051/0004-6361/201527377},
  \href {http://adsabs.harvard.edu/abs/2016A%26A...592A.121C} {592, A121}

\bibitem[\protect\citeauthoryear{{Coupon} et~al.,}{{Coupon}
  et~al.}{2009}]{coupon09}
{Coupon} J.,  et~al., 2009, \mn@doi [\aap] {10.1051/0004-6361/200811413}, \href
  {http://adsabs.harvard.edu/abs/2009A%26A...500..981C} {500, 981}

\bibitem[\protect\citeauthoryear{{DESI Collaboration} et~al.,}{{DESI
  Collaboration} et~al.}{2016a}]{desi-collaboration16a}
{DESI Collaboration} et~al., 2016a, preprint, \href
  {http://adsabs.harvard.edu/abs/2016arXiv161100036D} {} (\mn@eprint {arXiv}
  {1611.00036})

\bibitem[\protect\citeauthoryear{{DESI Collaboration} et~al.,}{{DESI
  Collaboration} et~al.}{2016b}]{desi-collaboration16b}
{DESI Collaboration} et~al., 2016b, preprint, \href
  {http://adsabs.harvard.edu/abs/2016arXiv161100037D} {} (\mn@eprint {arXiv}
  {1611.00037})

\bibitem[\protect\citeauthoryear{{Dawson} et~al.,}{{Dawson}
  et~al.}{2013}]{dawson13}
{Dawson} K.~S.,  et~al., 2013, \mn@doi [\aj] {10.1088/0004-6256/145/1/10},
  \href {http://adsabs.harvard.edu/abs/2013AJ....145...10D} {145, 10}

\bibitem[\protect\citeauthoryear{{Dawson} et~al.,}{{Dawson}
  et~al.}{2016}]{dawson16}
{Dawson} K.~S.,  et~al., 2016, \mn@doi [\aj] {10.3847/0004-6256/151/2/44},
  \href {http://adsabs.harvard.edu/abs/2016AJ....151...44D} {151, 44}

\bibitem[\protect\citeauthoryear{{Delubac} et~al.,}{{Delubac}
  et~al.}{2015}]{delubac15}
{Delubac} T.,  et~al., 2015, \mn@doi [\aap] {10.1051/0004-6361/201423969},
  \href {http://adsabs.harvard.edu/abs/2015A%26A...574A..59D} {574, A59}

\bibitem[\protect\citeauthoryear{{Delubac} et~al.,}{{Delubac}
  et~al.}{2017}]{delubac17}
{Delubac} T.,  et~al., 2017, \mn@doi [\mnras] {10.1093/mnras/stw2741}, \href
  {http://adsabs.harvard.edu/abs/2017MNRAS.465.1831D} {465, 1831}

\bibitem[\protect\citeauthoryear{{Drinkwater} et~al.,}{{Drinkwater}
  et~al.}{2010}]{drinkwater10}
{Drinkwater} M.~J.,  et~al., 2010, \mn@doi [\mnras]
  {10.1111/j.1365-2966.2009.15754.x}, \href
  {http://adsabs.harvard.edu/abs/2010MNRAS.401.1429D} {401, 1429}

\bibitem[\protect\citeauthoryear{{Eisenstein} et~al.,}{{Eisenstein}
  et~al.}{2001}]{eisenstein01}
{Eisenstein} D.~J.,  et~al., 2001, \mn@doi [\aj] {10.1086/323717}, \href
  {http://adsabs.harvard.edu/abs/2001AJ....122.2267E} {122, 2267}

\bibitem[\protect\citeauthoryear{{Eisenstein} et~al.,}{{Eisenstein}
  et~al.}{2005}]{eisenstein05}
{Eisenstein} D.~J.,  et~al., 2005, \mn@doi [\apj] {10.1086/466512}, \href
  {http://adsabs.harvard.edu/abs/2005ApJ...633..560E} {633, 560}

\bibitem[\protect\citeauthoryear{{Eisenstein} et~al.,}{{Eisenstein}
  et~al.}{2011}]{eisenstein11}
{Eisenstein} D.~J.,  et~al., 2011, \mn@doi [\aj] {10.1088/0004-6256/142/3/72},
  \href {http://adsabs.harvard.edu/abs/2011AJ....142...72E} {142, 72}

\bibitem[\protect\citeauthoryear{{Erben} et~al.,}{{Erben}
  et~al.}{2013}]{erben13}
{Erben} T.,  et~al., 2013, \mn@doi [\mnras] {10.1093/mnras/stt928}, \href
  {http://adsabs.harvard.edu/abs/2013MNRAS.tmp.1589E} {}

\bibitem[\protect\citeauthoryear{{Finkbeiner} et~al.,}{{Finkbeiner}
  et~al.}{2016}]{finkbeiner16}
{Finkbeiner} D.~P.,  et~al., 2016, \mn@doi [\apj] {10.3847/0004-637X/822/2/66},
  \href {http://adsabs.harvard.edu/abs/2016ApJ...822...66F} {822, 66}

\bibitem[\protect\citeauthoryear{{Flaugher} et~al.,}{{Flaugher}
  et~al.}{2015}]{flaugher15}
{Flaugher} B.,  et~al., 2015, \mn@doi [\aj] {10.1088/0004-6256/150/5/150},
  \href {http://adsabs.harvard.edu/abs/2015AJ....150..150F} {150, 150}

\bibitem[\protect\citeauthoryear{{Gunn} et~al.,}{{Gunn} et~al.}{2006}]{gunn06}
{Gunn} J.~E.,  et~al., 2006, \mn@doi [\aj] {10.1086/500975}, \href
  {http://adsabs.harvard.edu/abs/2006AJ....131.2332G} {131, 2332}

\bibitem[\protect\citeauthoryear{{Guzzo} et~al.,}{{Guzzo}
  et~al.}{2014}]{guzzo14}
{Guzzo} L.,  et~al., 2014, \mn@doi [\aap] {10.1051/0004-6361/201321489}, \href
  {http://adsabs.harvard.edu/abs/2014A%26A...566A.108G} {566, A108}

\bibitem[\protect\citeauthoryear{{Gwyn}}{{Gwyn}}{2012}]{gwyn12}
{Gwyn} S.~D.~J.,  2012, \mn@doi [\aj] {10.1088/0004-6256/143/2/38}, \href
  {http://adsabs.harvard.edu/abs/2012AJ....143...38G} {143, 38}

\bibitem[\protect\citeauthoryear{{Hutchinson} et~al.,}{{Hutchinson}
  et~al.}{2016}]{hutchinson16}
{Hutchinson} T.~A.,  et~al., 2016, preprint, \href
  {http://adsabs.harvard.edu/abs/2016arXiv160702432H} {} (\mn@eprint {arXiv}
  {1607.02432})

\bibitem[\protect\citeauthoryear{{Ilbert} et~al.,}{{Ilbert}
  et~al.}{2006}]{ilbert06}
{Ilbert} O.,  et~al., 2006, ArXiv Astrophysics e-prints, \href
  {http://adsabs.harvard.edu/abs/2006astro.ph..2329I} {}

\bibitem[\protect\citeauthoryear{{Jensen} et~al.,}{{Jensen}
  et~al.}{2016}]{jensen16}
{Jensen} T.~W.,  et~al., 2016, \mn@doi [\apj] {10.3847/1538-4357/833/2/199},
  \href {http://adsabs.harvard.edu/abs/2016ApJ...833..199J} {833, 199}

\bibitem[\protect\citeauthoryear{{Kaiser} et~al.,}{{Kaiser}
  et~al.}{2010}]{kaiser10}
{Kaiser} N.,  et~al., 2010, in Ground-based and Airborne Telescopes III. p.
  77330E, \mn@doi{10.1117/12.859188}

\bibitem[\protect\citeauthoryear{{Kennicutt}}{{Kennicutt}}{1992}]{kennicutt92}
{Kennicutt} Jr. R.~C.,  1992, \mn@doi [\apj] {10.1086/171154}, \href
  {http://adsabs.harvard.edu/abs/1992ApJ...388..310K} {388, 310}

\bibitem[\protect\citeauthoryear{{Kriek} \& {Conroy}}{{Kriek} \&
  {Conroy}}{2013}]{kriek13}
{Kriek} M.,  {Conroy} C.,  2013, \mn@doi [\apjl] {10.1088/2041-8205/775/1/L16},
  \href {http://adsabs.harvard.edu/abs/2013ApJ...775L..16K} {775, L16}

\bibitem[\protect\citeauthoryear{{Kriek}, {van Dokkum}, {Labb{\'e}}, {Franx},
  {Illingworth}, {Marchesini}  \& {Quadri}}{{Kriek} et~al.}{2009}]{kriek09}
{Kriek} M.,  {van Dokkum} P.~G.,  {Labb{\'e}} I.,  {Franx} M.,  {Illingworth}
  G.~D.,  {Marchesini} D.,   {Quadri} R.~F.,  2009, \mn@doi [\apj]
  {10.1088/0004-637X/700/1/221}, \href
  {http://adsabs.harvard.edu/abs/2009ApJ...700..221K} {700, 221}

\bibitem[\protect\citeauthoryear{{Laigle} et~al.,}{{Laigle}
  et~al.}{2016}]{laigle16}
{Laigle} C.,  et~al., 2016, \mn@doi [\apjs] {10.3847/0067-0049/224/2/24}, \href
  {http://adsabs.harvard.edu/abs/2016ApJS..224...24L} {224, 24}

\bibitem[\protect\citeauthoryear{{Lang}, {Hogg}  \& {Schlegel}}{{Lang}
  et~al.}{2016}]{lang16}
{Lang} D.,  {Hogg} D.~W.,   {Schlegel} D.~J.,  2016, \mn@doi [\aj]
  {10.3847/0004-6256/151/2/36}, \href
  {http://adsabs.harvard.edu/abs/2016AJ....151...36L} {151, 36}

\bibitem[\protect\citeauthoryear{{Laureijs} et~al.,}{{Laureijs}
  et~al.}{2011}]{laureijs11}
{Laureijs} R.,  et~al., 2011, preprint, \href
  {http://adsabs.harvard.edu/abs/2011arXiv1110.3193L} {} (\mn@eprint {arXiv}
  {1110.3193})

\bibitem[\protect\citeauthoryear{{Lilly}, {Le Fevre}, {Hammer}  \&
  {Crampton}}{{Lilly} et~al.}{1996}]{lilly96}
{Lilly} S.~J.,  {Le Fevre} O.,  {Hammer} F.,   {Crampton} D.,  1996, \mn@doi
  [\apjl] {10.1086/309975}, \href
  {http://adsabs.harvard.edu/abs/1996ApJ...460L...1L} {460, L1}

\bibitem[\protect\citeauthoryear{{Madau} \& {Dickinson}}{{Madau} \&
  {Dickinson}}{2014}]{madau14}
{Madau} P.,  {Dickinson} M.,  2014, preprint, \href
  {http://adsabs.harvard.edu/abs/2014arXiv1403.0007M} {} (\mn@eprint {arXiv}
  {1403.0007})

\bibitem[\protect\citeauthoryear{{Madau}, {Pozzetti}  \& {Dickinson}}{{Madau}
  et~al.}{1998}]{madau98}
{Madau} P.,  {Pozzetti} L.,   {Dickinson} M.,  1998, \mn@doi [\apj]
  {10.1086/305523}, \href {http://adsabs.harvard.edu/abs/1998ApJ...498..106M}
  {498, 106}

\bibitem[\protect\citeauthoryear{{Mainzer} et~al.,}{{Mainzer}
  et~al.}{2011}]{mainzer11}
{Mainzer} A.,  et~al., 2011, \mn@doi [\apj] {10.1088/0004-637X/731/1/53}, \href
  {http://adsabs.harvard.edu/abs/2011ApJ...731...53M} {731, 53}

\bibitem[\protect\citeauthoryear{{Meisner}, {Lang}  \& {Schlegel}}{{Meisner}
  et~al.}{2017}]{meisner17}
{Meisner} A.~M.,  {Lang} D.,   {Schlegel} D.~J.,  2017, \mn@doi [\aj]
  {10.3847/1538-3881/153/1/38}, \href
  {http://adsabs.harvard.edu/abs/2017AJ....153...38M} {153, 38}

\bibitem[\protect\citeauthoryear{{Morganson} et~al.,}{{Morganson}
  et~al.}{2015}]{morganson15}
{Morganson} E.,  et~al., 2015, \mn@doi [\apj] {10.1088/0004-637X/806/2/244},
  \href {http://adsabs.harvard.edu/abs/2015ApJ...806..244M} {806, 244}

\bibitem[\protect\citeauthoryear{{Moustakas} \& {Kennicutt}}{{Moustakas} \&
  {Kennicutt}}{2006}]{moustakas06}
{Moustakas} J.,  {Kennicutt} Jr. R.~C.,  2006, \mn@doi [\apjs]
  {10.1086/500971}, \href {http://adsabs.harvard.edu/abs/2006ApJS..164...81M}
  {164, 81}

\bibitem[\protect\citeauthoryear{{Moustakas}, {Kennicutt}  \&
  {Tremonti}}{{Moustakas} et~al.}{2006}]{moustakas06a}
{Moustakas} J.,  {Kennicutt} Jr. R.~C.,   {Tremonti} C.~A.,  2006, \mn@doi
  [\apj] {10.1086/500964}, \href
  {http://adsabs.harvard.edu/abs/2006ApJ...642..775M} {642, 775}

\bibitem[\protect\citeauthoryear{{Myers} et~al.,}{{Myers}
  et~al.}{2015}]{myers15}
{Myers} A.~D.,  et~al., 2015, \mn@doi [\apjs] {10.1088/0067-0049/221/2/27},
  \href {http://adsabs.harvard.edu/abs/2015ApJS..221...27M} {221, 27}

\bibitem[\protect\citeauthoryear{{Oke} \& {Gunn}}{{Oke} \&
  {Gunn}}{1983}]{oke83}
{Oke} J.~B.,  {Gunn} J.~E.,  1983, \mn@doi [\apj] {10.1086/160817}, \href
  {http://adsabs.harvard.edu/abs/1983ApJ...266..713O} {266, 713}

\bibitem[\protect\citeauthoryear{{Palanque-Delabrouille}
  et~al.,}{{Palanque-Delabrouille} et~al.}{2016}]{palanque-delabrouille16}
{Palanque-Delabrouille} N.,  et~al., 2016, \mn@doi [\aap]
  {10.1051/0004-6361/201527392}, \href
  {http://adsabs.harvard.edu/abs/2016A%26A...587A..41P} {587, A41}

\bibitem[\protect\citeauthoryear{{Peng}, {Ho}, {Impey}  \& {Rix}}{{Peng}
  et~al.}{2010}]{peng10a}
{Peng} C.~Y.,  {Ho} L.~C.,  {Impey} C.~D.,   {Rix} H.-W.,  2010, \mn@doi [\aj]
  {10.1088/0004-6256/139/6/2097}, \href
  {http://adsabs.harvard.edu/abs/2010AJ....139.2097P} {139, 2097}

\bibitem[\protect\citeauthoryear{{Planck Collaboration} et~al.,}{{Planck
  Collaboration} et~al.}{2016}]{planck-collaboration16}
{Planck Collaboration} et~al., 2016, \mn@doi [\aap]
  {10.1051/0004-6361/201525830}, \href
  {http://adsabs.harvard.edu/abs/2016A%26A...594A..13P} {594, A13}

\bibitem[\protect\citeauthoryear{{Prakash} et~al.,}{{Prakash}
  et~al.}{2016}]{prakash16}
{Prakash} A.,  et~al., 2016, \mn@doi [\apjs] {10.3847/0067-0049/224/2/34},
  \href {http://adsabs.harvard.edu/abs/2016ApJS..224...34P} {224, 34}

\bibitem[\protect\citeauthoryear{{Raichoor} et~al.,}{{Raichoor}
  et~al.}{2016}]{raichoor16}
{Raichoor} A.,  et~al., 2016, \mn@doi [\aap] {10.1051/0004-6361/201526486},
  \href {http://adsabs.harvard.edu/abs/2016A%26A...585A..50R} {585, A50}

\bibitem[\protect\citeauthoryear{{Ross} et~al.,}{{Ross} et~al.}{2011}]{ross11}
{Ross} A.~J.,  et~al., 2011, \mn@doi [\mnras]
  {10.1111/j.1365-2966.2011.19351.x}, \href
  {http://adsabs.harvard.edu/abs/2011MNRAS.417.1350R} {417, 1350}

\bibitem[\protect\citeauthoryear{{Ross} et~al.,}{{Ross} et~al.}{2017}]{ross17}
{Ross} A.~J.,  et~al., 2017, \mn@doi [\mnras] {10.1093/mnras/stw2372}, \href
  {http://adsabs.harvard.edu/abs/2017MNRAS.464.1168R} {464, 1168}

\bibitem[\protect\citeauthoryear{{SDSS Collaboration} et~al.,}{{SDSS
  Collaboration} et~al.}{2016}]{sdss-collaboration16}
{SDSS Collaboration} et~al., 2016, preprint, \href
  {http://adsabs.harvard.edu/abs/2016arXiv160802013S} {} (\mn@eprint {arXiv}
  {1608.02013})

\bibitem[\protect\citeauthoryear{{Schlegel}, {Finkbeiner}  \&
  {Davis}}{{Schlegel} et~al.}{1998}]{schlegel98}
{Schlegel} D.~J.,  {Finkbeiner} D.~P.,   {Davis} M.,  1998, \mn@doi [\apj]
  {10.1086/305772}, \href {http://adsabs.harvard.edu/abs/1998ApJ...500..525S}
  {500, 525}

\bibitem[\protect\citeauthoryear{{Scodeggio} et~al.,}{{Scodeggio}
  et~al.}{2016}]{scodeggio16}
{Scodeggio} M.,  et~al., 2016, preprint, \href
  {http://adsabs.harvard.edu/abs/2016arXiv161107048S} {} (\mn@eprint {arXiv}
  {1611.07048})

\bibitem[\protect\citeauthoryear{{Sersic}}{{Sersic}}{1968}]{sersic68}
{Sersic} J.~L.,  1968, {Atlas de galaxias australes (Observatorio Astronomico,
  Cordoba, Argentina)}

\bibitem[\protect\citeauthoryear{{Smee} et~al.,}{{Smee} et~al.}{2013}]{smee13}
{Smee} S.~A.,  et~al., 2013, \mn@doi [\aj] {10.1088/0004-6256/146/2/32}, \href
  {http://adsabs.harvard.edu/abs/2013AJ....146...32S} {146, 32}

\bibitem[\protect\citeauthoryear{{Sugai} et~al.,}{{Sugai}
  et~al.}{2012}]{sugai12}
{Sugai} H.,  et~al., 2012, in Society of Photo-Optical Instrumentation
  Engineers (SPIE) Conference Series. p.~0 (\mn@eprint {arXiv} {1210.2719}),
  \mn@doi{10.1117/12.926954}

\bibitem[\protect\citeauthoryear{{Takada} et~al.,}{{Takada}
  et~al.}{2014}]{takada14}
{Takada} M.,  et~al., 2014, \mn@doi [\pasj] {10.1093/pasj/pst019}, \href
  {http://adsabs.harvard.edu/abs/2014PASJ...66R...1T} {66, R1}

\bibitem[\protect\citeauthoryear{{Wright} et~al.,}{{Wright}
  et~al.}{2010}]{wright10}
{Wright} E.~L.,  et~al., 2010, \mn@doi [\aj] {10.1088/0004-6256/140/6/1868},
  \href {http://adsabs.harvard.edu/abs/2010AJ....140.1868W} {140, 1868}

\bibitem[\protect\citeauthoryear{{Wuyts} et~al.,}{{Wuyts}
  et~al.}{2011}]{wuyts11a}
{Wuyts} S.,  et~al., 2011, \mn@doi [\apj] {10.1088/0004-637X/742/2/96}, \href
  {http://adsabs.harvard.edu/abs/2011ApJ...742...96W} {742, 96}

\bibitem[\protect\citeauthoryear{{York} et~al.,}{{York} et~al.}{2000}]{york00}
{York} D.~G.,  et~al., 2000, \mn@doi [\aj] {10.1086/301513}, \href
  {http://adsabs.harvard.edu/abs/2000AJ....120.1579Y} {120, 1579}

\bibitem[\protect\citeauthoryear{{Zhao} et~al.,}{{Zhao} et~al.}{2016}]{zhao16}
{Zhao} G.-B.,  et~al., 2016, \mn@doi [\mnras] {10.1093/mnras/stw135}, \href
  {http://adsabs.harvard.edu/abs/2016MNRAS.457.2377Z} {457, 2377}

\bibitem[\protect\citeauthoryear{{Zhu} et~al.,}{{Zhu} et~al.}{2015}]{zhu15}
{Zhu} G.~B.,  et~al., 2015, \mn@doi [\apj] {10.1088/0004-637X/815/1/48}, \href
  {http://adsabs.harvard.edu/abs/2015ApJ...815...48Z} {815, 48}

\bibitem[\protect\citeauthoryear{{de Jong} et~al.,}{{de Jong}
  et~al.}{2014}]{de-jong14}
{de Jong} R.~S.,  et~al., 2014, in Ground-based and Airborne Instrumentation
  for Astronomy V. p. 91470M, \mn@doi{10.1117/12.2055826}

\bibitem[\protect\citeauthoryear{{de Vaucouleurs}}{{de
  Vaucouleurs}}{1948}]{de-vaucouleurs48}
{de Vaucouleurs} G.,  1948, Annales d'Astrophysique, \href
  {http://adsabs.harvard.edu/abs/1948AnAp...11..247D} {11, 247}

\makeatother
\end{thebibliography}

%=================================
% Acknowledgments
%=================================
\section*{Acknowledgments}
% Funding
A.R , T.D. and J.P.K. acknowledges support from the ERC advanced grant LIDA.
W.J.P. acknowledges support from the European Research Council through the Darksurvey grant 614030, and from the UK Science and Technology Facilities Council grant ST/N000668/1 and UK Space Agency grant ST/N00180X/1.
E.J. acknowledges support from the OCEVU Labex (ANR-11-LABX-0060).

% SDSS IV
This paper represents an effort by both the SDSS-IV collaborations.
Funding for SDSS-III was provided by the Alfred
P. Sloan Foundation, the Participating Institutions, the
National Science Foundation, and the U.S. Department
of Energy Office of Science.
Funding for the Sloan Digital Sky Survey IV has been provided by
the Alfred P. Sloan Foundation, the U.S. Department of Energy Office of
Science, and the Participating Institutions. SDSS-IV acknowledges
support and resources from the Center for High-Performance Computing at
the University of Utah. The SDSS web site is www.sdss.org.
SDSS-IV is managed by the Astrophysical Research Consortium for the
Participating Institutions of the SDSS Collaboration including the
Brazilian Participation Group, the Carnegie Institution for Science,
Carnegie Mellon University, the Chilean Participation Group,
the French Participation Group, Harvard-Smithsonian Center for Astrophysics,
Instituto de Astrof\'isica de Canarias, The Johns Hopkins University,
Kavli Institute for the Physics and Mathematics of the Universe (IPMU) /
University of Tokyo, Lawrence Berkeley National Laboratory,
Leibniz Institut f\"ur Astrophysik Potsdam (AIP),
Max-Planck-Institut f\"ur Astronomie (MPIA Heidelberg),
Max-Planck-Institut f\"ur Astrophysik (MPA Garching),
Max-Planck-Institut f\"ur Extraterrestrische Physik (MPE),
National Astronomical Observatory of China, New Mexico State University,
New York University, University of Notre Dame,
Observat\'ario Nacional / MCTI, The Ohio State University,
Pennsylvania State University, Shanghai Astronomical Observatory,
United Kingdom Participation Group,
Universidad Nacional Aut\'onoma de M\'exico, University of Arizona,
University of Colorado Boulder, University of Portsmouth,
University of Utah, University of Virginia, University of Washington,
University of Wisconsin,
Vanderbilt University, and Yale University.

% WISE
This publication makes use of data products from the \textit{Wide-field Infrared Survey Explorer}, which is a joint project of the University of California, Los Angeles, and the Jet Propulsion Laboratory/California Institute of Technology, and NEOWISE, which is a project of the Jet Propulsion Laboratory/California Institute of Technology.
\textit{WISE} and NEOWISE are funded by the National Aeronautics and Space Administration.

% Authors contribution
\section*{Authors contribution}
% ELG work
A.R. led this paper, designed the target selection, analysed the ELG plates, and estimated the structural properties and stellar masses.
J.C. led the $\zspec$ measurement, developed the $\zspec$ confidence flag, the spectra stacking procedure, and installed and ran the DECaLS pipeline on the Utah machines (with J.R.B.).
T.D. led the systematics analysis.
J.P.K, Ch.Y, K.S.D, and W.J.P. supervised the ELG program.
J.C, T.D., K.S.D., and C.G. did the visual inspection.
A.J.R, Y.W, and G.B.Z. did the cosmological forecast.
% eBOSS team
H.J.S. and J.L.T. led the tiling, V.M led the spectroscopic observations, J.B. and J.R.B led the spectroscopic pipeline reduction.
% DECaLS team
A.D, D.L, and D.J.S. led the DECaLS imaging observation and pipeline development.
% ELG review
J.M. and N.P.D. reviewed the ELG program and participated in the ELG program development.
% eBOSS others
E.J., J.A.N., F.P., and G.B.Z also participated in the ELG program development.

%%%%%%%%%%%%%%%%%%%%%%%%%%%%%%%%%%%%%%%%%%%%%%%%%%

% Don't change these lines
\bsp	% typesetting comment
\label{lastpage}
\end{document}